\documentclass[epj]{svjour}
\usepackage{graphicx}
\usepackage{cite}
\newcommand{\as}  {\ensuremath{\alpha_{\mathrm{S}}}}
\newcommand{\asmz}{\ensuremath{\alpha_\mathrm{S}(M_{\mathrm{Z^0}})}}
\newcommand{\Jade}{\mbox{\rm JADE}}

\newcommand{\Opal}{\mbox{\rm OPAL}}

\newcommand{\epem}{\ensuremath{\mathrm{e^+e^-}}}
\newcommand{\stat}{\ensuremath{\mathrm{(stat.)}}}

\newcommand{\expt}{\ensuremath{\mathrm{(exp.)}}}
\newcommand{\had} {\ensuremath{\mathrm{(had.)}}}
\newcommand{\theo}{\ensuremath{\mathrm{(theo.)}}}

\newcommand{\ytwothree}   {{\ensuremath{y^{\rm D}_{23}}}} %pr404
\newcommand{\tma}               {\ensuremath{T_{\mathrm{maj.}}}}
\newcommand{\tmi}               {\ensuremath{T_{\mathrm{min.}}}}
\newcommand{\rs}                {\ensuremath{\sqrt{s}}}
\newcommand{\cp}                {\ensuremath{C}}
\newcommand{\s}                 {\ensuremath{S}}
\newcommand{\obl}                 {\ensuremath{O}}
\newcommand{\bt}                {\ensuremath{B_\mathrm{T}}}
\newcommand{\bw}                {\ensuremath{B_\mathrm{W}}}
\newcommand{\bn}                {\ensuremath{B_\mathrm{N}}}
\newcommand{\mh}                {\ensuremath{M_\mathrm{H}}}
\newcommand{\ml}                {\ensuremath{M_\mathrm{L}}}
\newcommand{\thr}               {\ensuremath{1-T}}
\newcommand{\momn}[2] {\mbox{\ensuremath{\langle#1^{#2}\rangle}}}
\newcommand{\dd}       {\mathrm{d}}
\newcommand{\zzero}     {\ensuremath{\mathrm{Z^0}}}
\newcommand{\asb}               {\ensuremath{\bar{\alpha}_{\mathrm{S}}}}
\newcommand{\asbsq}             {\ensuremath{\bar{\alpha}_{\mathrm{S}}^2}}
\newcommand{\xmu}               {\ensuremath{x_{\mu}}}
\newcommand{\nf}                {\ensuremath{n_{\mathrm{f}}}}
\newcommand{\sigtot} {\ensuremath{\sigma_{\mathrm{tot}}}}
\newcommand{\signull}{\ensuremath{\sigma_{0}}}

\newcommand{\py}                {PYTHIA}

\newcommand{\bbbar}     {\ensuremath{\mathrm{b\bar{b}}}}

\newcommand{\mz}                {\ensuremath{M_{\mathrm{Z^0}}}}
\newcommand{\nlo}{\mbox{NLO}}
\newcommand{\chisqd}    {\ensuremath{\chi^2/\mathrm{d.o.f.}}}
\newcommand{\momone}[1] {\mbox{\ensuremath{\langle#1\rangle}}}

\newcommand{\Petra} {\mbox{\rm PETRA}}

\newcommand{\Lep}{\mbox{LEP}}

\newcommand{\ycut}              {\ensuremath{y_{\mathrm{cut}}}}
\newcommand{\mI}     {{\ensuremath{{\mu_I}}}}
\newcommand{\anulmI} {\ensuremath{\alpha_0(\mu_I)}}
\newcommand{\qcd}    {\mbox{QCD}}
\newcommand{\muR} {\ensuremath{\mu_R}}
\newcommand{\msb}   {\ensuremath{\mathrm{\overline{MS}}}}
\newcommand{\eps}               {\ensuremath{\varepsilon}}
\newcommand{\oaaa}              {\ensuremath{\mathcal{O}(\alpha_\mathrm{S}^3)}}
\newcommand{\tot}   {\ensuremath{\mathrm{(tot.)}}}
\newcommand{\Var}[1]     {{\rm Var}(#1)}

\newcommand{\asmsbmz}{\ensuremath{\alpha_{\mathrm{s}}^{\msb}(M_{\mathrm{Z^0}})}}

\usepackage{color}
\newcommand{\Lepone}{\mbox{\rm LEP 1}}
\newcommand{\Leptwo}{\mbox{\rm LEP 2}}
\newcommand{\bb}    {\ensuremath{\backslash\backslash}}
\newcommand{\Lamqcd}{\ensuremath{\Lambda_{\mbox{\tiny QCD}}}}

\newcommand{\J}[1]{} % wichtiger Kommentar: auch nicht

\newcommand{\abwarten}{\ensuremath{\!\!\!\rightarrow..}}

\begin{document}
\titlerunning{Tests of analytical hadronisation models using event shape moments in          {\epem} annihilation}
\title       {Tests of analytical hadronisation models using event shape moments in \boldmath{\epem} annihilation
    }
\author{C. Pahl\inst{1,}\inst{2} \and S. Bethke\inst{1} \and  O. Biebel\inst{3} \and  S. Kluth\inst{1} \and  J. Schieck\inst{1}}                     % Do not remove
%%
%%\offprints{}          % Insert a name or remove this line
%%
\institute{Max-Planck-Institut f\"ur Physik, F\"ohringer Ring 6, D-80805 Munich, Germany
      \and Excellence Cluster Universe, Technische Universit\"at M\"unchen,
           Boltzmannstr. 2, D-85748 Garching, Germany
      \and LMU M\"unchen, Fakult\"at f\"ur Physik, Am Coulombwall 1, D-85748 Garching, Germany}
\date{Received: date / Revised version: date}
%% The correct dates will be entered by Springer
\abstract{
Predictions of analytical models for hadronisation, namely the
dispersive
model, the shape function and the single dressed gluon approximation, are
compared with moments of hadronic event shape distributions measured in   %ok
  \epem~annihilation at centre-of-mass energies between 14 and 209\,GeV.
  In contrast to Monte Carlo models for hadronisation,
  analytical models require to adjust only two universal parameters,
  the strong coupling and                                                       
  a second quantity parametrising nonperturbative corrections.
  The extracted values of \as\ are consistent with the world average and competitive
  with previous measurements.
%%  Averaging the parameters obtained from several measured moments,
%%  the dispersive model results in
%%  $\asmz = 0.1183\pm0.0056\tot$ and
%%  $\anulmI=0.493\pm0.058\tot$,
%%  while the single dressed gluon prediction in fifth order with
%%  negligible power corrections describes the first moment of thrust  well  
%%  with  
%%  $\asmz=0.1172\pm0.0036\tot$.
  The variance of event shape distributions is compared with predictions given by some of these models.
  Limitations of the models, probably due to unknown
  higher order corrections, are demonstrated and discussed.
%  \[
%     \asmz=0.1172\pm0.0007\stat\pm0.0016\expt^{+0.0031}_{-0.0027}\theo\,.
%  \]
%  %
   %% end of PACS codes
 } %% end of abstract

\maketitle

\section{Introduction}

In previous studies, moments of event shape distributions have been compared to perturbative predictions of 
quantum chromodynamics (QCD, \cite{FritzschGellMann,GrossWilczek1,GrossWilczek2,Politzer}) 
%%quantum chromodynamics (QCD, \cite{FritzschGellMann,GrossWilczek1,Politzer}) 
in next-to-leading order, simulating the
hadronisation process by Monte Carlo models \cite{jadepaper,OPALPR404}.
Alternatively there exist models describing hadronisation analytically.
This paper aims to study these models qualitatively and quantitatively %%and to use them for 
by measurements of the strong coupling and the model parameters.

\sloppy{}The {\it dispersive model}~\cite{DMW} is based on the assumption of a nonpertubatively continued
strong coupling. The {\it shape function}~\cite{KorTaf} additionally describes a 
%%compression (see below) of the
modification of the shape of the perturbatively calculated distribution. The {\it single dressed gluon approximation}~\cite{Gar1,Gar2}
estimates the perturbative
part more completely with reduced perturbative uncertainties of the
prediction. The models test the predicted energy evolution of the strong
coupling\J{Beneke?Dagupta?Salam?zitieren\bb}. Their parameters can be 
determined---one of them the
value of \as\ at some reference energy---and the assumption of universality of these parameters can be probed. 
To date, primarily the distributions themselves and the mean values (first moments) have been 
studied~\cite{ALEPH,L3,DELPHI,JADE-paper}.

This analysis uses data measured by \Jade\ \cite{jadepaper} in the years 1979...1986 at 
six centre-of-mass (cms) energies in the energy range of $Q=14$...44\,GeV, and data measured by 
\Opal~at 12 energy
points over the whole \Lep\ energy range of 91...209\,GeV and combined into 4 energy ranges~\cite{OPALPR404}. The large
energy range covered by the measurements allows to test the employed assumptions 
selectively. 
\J{\tiny Die Resultate bei 91\,GeV beruhen auf Kalibrationsdaten, die waehrend der 
\Leptwo-Periode aufgezeichnet wurden (ab 1996, als \Lep\ ansonsten weit 
oberhalb der \zzero-Masse lief). Sie basieren somit auf der gleichen 
Detektor-Konfiguration (etwas verschieden von derjenigen in der frueheren 
\Lepone\ - Phase, als der Beschleuniger nur nahe dem \zzero-Peak lief) und dem gleichen
Rekonstruktions-Code wie die Hochenergie-Daten. Somit können Resultate ueber einen
gro"sen Energiebereich mit minimalen systematischen Unterschieden zwischen den 
Energien verglichen werden.
Aus dem Vergleich der Daten mit der Theorie ermitteln wir die starke 
Kopplungskonstante \asmz.}

The outline of the paper is as follows.
In Sect.~\ref{theory}, we present the observables 
used in the analysis.
In Sect.~\ref{NPT} we describe the perturbative QCD predictions and introduce the analytical models which describe the
hadronisation process.
In Sect.~\ref{TestsPCs} we discuss predictions for event shape moments at hadron level and compare them with the measurements. 
The predictions for the event shape distribution variance following
from the % first two models 
dis\-per\-si\-ve model and the shape function
are tested as well.
In Sect.~\ref{summary} we summarize and give our conclusions.

%This research was supported by the DFG cluster of excellence `Origin
%and Structure of the Universe'.

\section{Event shape moments}
\label{theory}

Event shape variables are a convenient way to characterise properties of hadronic final states. %events
%%by the distribution of 
They are calculated from particle momenta and energies. For definitions of the variables we refer 
to \cite{OPALPR404}.

In a hadronic event in \epem\ annihilation the virtual vector boson \zzero/$\gamma^*$ generated by annihilation
 of electron
and positron decays into a quark pair q$\bar{\rm q}$.
The quarks may radiate gluons which radiate further gluons or decay into another quark pair. The final state
of this parton shower is called parton level. By the process of hadronisation partons are transferred
into hadrons. The predictions used in this work describe the hadron level. The variables measured
in the experiments have to be corrected for the effects of limited detector acceptance and resolution to probe the hadron level.

%\J{\tiny{}vgl. PR404: The properties of 
%hadronic events may be characterized by a set of
%event shape observables.  These may be used to characterize the
%distribution of particles and thus the topology of an event as
%``pencil-like'', planar, spherical etc.  They can be computed either
%using the measured charged particle tracks and calorimeter clusters,
%or using the true hadrons or partons in simulated events.} 
The event shapes considered here are
thrust {$T$},%:]
\J{\tiny   defined by the expression~\cite{thrust1,thrust2}
  \begin{displaymath}
  T= \max_{\vec{n}}\left(\frac{\sum_i|\vec{p}_i\cdot\vec{n}|}
                    {\sum_i|p_i|}\right)\;\;\;.
  \end{displaymath}
  The thrust axis $\vec{n}_T$ is the direction $\vec{n}$ which
  maximizes the expression in parentheses.  A plane through the origin
  and perpendicular to $\vec{n}_T$ divides the event into two
  hemispheres $H_1$ and $H_2$.}
%\item[
C-parameter \cp,%:]
\J{\tiny   The linearized momentum tensor $\Theta^{\alpha\beta}$ is defined by
  \[
    \Theta^{\alpha\beta}= \frac{\sum_i(p_i^{\alpha}p_i^{\beta})/|\vec{p}_i|}
                               {\sum_i|\vec{p}_i|}\;\;\;,
                           \;\;\;\alpha,\beta= 1,2,3\;\;\;.
  \]
  The three eigenvalues $\lambda_j$ of this tensor define
  \cp~\cite{ert} through
 \[
    \cp= 3(\lambda_1\lambda_2+\lambda_2\lambda_3+\lambda_3\lambda_1)\;\;.
  \]}
heavy jet mass {\mh},%:] 
\J{\tiny The hemisphere 
invariant masses are calculated using  the particles
  in the two hemispheres $H_1$ and $H_2$. We define
  \mh~\cite{jetmass,clavelli81} as the heavier mass, divided by $\rs$.}
jet broadening variables {\bt} and {\bw},%:] 
\J{\tiny   These are defined by computing the quantity
  \[
    B_k= \left(\frac{\sum_{i\in H_k}|\vec{p}_i\times\vec{n}_T|}
                    {2\sum_i|\vec{p}_i|}\right)
  \]
  for each of the two event hemispheres, $H_k$,  defined above.
  The two variables~\cite{nllabtbw} are defined by
  \[
    \bt= B_1+B_2\;,\;\;\mathrm{and}\;\;\;\bw= \max(B_1,B_2)\;\;\;
  \]
  where \bt\ is the total and \bw\ is the wide jet broadening.}
and the transition value {\ytwothree} between 2- and 3-jet final states defined using the Durham jet reconstruction scheme. 
The \as\ determination
in \cite{jadepaper} is based on moments of these variables,
in \cite{OPALPR404}          on distributions and moments of these variables.
Their theo\-re\-ti\-cal description by perturbation theory is 
currently the most advanced~\cite{resummation,NNLOESs,Weinzierl}.
\J{\tiny  Jet algorithms are applied to cluster the large number of particles of
 an hadronic event into a small number of jets, reflecting the parton
 structure of the event. For this analysis we use the Durham
 scheme~\cite{durham}. Defining each particle initially to be a
 jet, a resolution variable $y_{ij}$ is calculated for each
 pair of jets $i$ and $j$: 
\be
  y_{ij}= \frac{2\mathrm{min}(E_i^2,E_j^2)}{E_{\mathrm{vis}}^2}
          (1-\cos\theta_{ij}),
\ee 
 where $E_{i}$ and $E_{j}$ are the energies, $\cos\theta_{ij}$ is the
 angle between the two jets and $E_{\mathrm{vis}}$ is the sum of the
 energies of all visible particles in the event (or the partons in a
 theoretical calculation).  If the smallest value of $y_{ij}$ is less
 than a predefined value \ycut, the pair is replaced by a jet with four
 momentum $p_{ij}^\mu = p_i^\mu + p_j^\mu$, and the clustering starts
 again with $p_{ij}^\mu$ instead of the momenta $p_i^\mu$ and
 $p_j^\mu$.  Clustering ends when the smallest value of $y_{ij}$ is
 larger than \ycut.  The remaining jets are then counted.
 The value of \ycut\ at which
 for an event the transition between a 2-jet and a 3-jet assignment
 occurs is called \ytwothree.}

A generic event shape variable is denoted by the symbol $y$.
Regions dominated by multiple jets give large values of $y$, while
two narrow jets give $y\simeq0$. Thrust $T$ is an exception to this rule.\J{(Durham 
transition value {\ytwothree} wohl doch nicht undefined in this case)} By using 
$y=\thr$ instead, the condition is fulfilled for all event shapes. 
\bw, \ytwothree\ and \mh\ are sensitive to only one suitably chosen 
%%specifically defined 
hemisphere of the event ({\it one-hemisphere variables}), \thr, \cp\ and \bt\ are sensitive to the
whole hadronic event ({\it two-hemisphere variables}).

The $n$th, $n=1,2,\ldots$ moment of the distribution of an event
shape variable $y$ is defined by
\begin{equation}
  \momn{y}{n}=\int_0^{y_{\rm max}} \dd y \, y^n \frac{1}{\sigma_{\rm tot.}}\frac{\dd\sigma}{\dd y} \,,
  \label{defmom}
\end{equation}
where $y_{\rm max}$ is the
kinematically allowed upper limit of the variable and $\sigma_{\rm tot.}$ denotes the total hadronic cross section.

Predictions have been made available for the moments of event shapes. Their evolution
with cms energy allows direct tests of the predicted energy evolution
of the strong coupling \as. \J{vgl. jadepaper:\bb}Furthermore it enables the determination of a single
value of \as\ at a definite energy scale---for example \asmz\ at the rest energy of the 
\zzero\ boson.\J{...---und somit zu obigen 
Verfahren komplementaere Bestimmungen 
des wesentlichen freien Parameters der QCD, der Kopplungs\-staer\-ke \as\,.
Bisher wurden vorrangig die Mittelwerte (ersten Momente) untersucht \cite{ALEPH,L3,DELPHI,JADE-paper}.} The theoretical calculations
always involve a full integration over phase space, which implies that
comparison with data always probes all of the available phase space.
This is in contrast to QCD predictions for distributions; these are
commonly only compared with data---e.g.\ in order to measure \as---in
restricted regions, where the theory is able to describe the data
well, see e.g.~\cite{jadenewas}.\J{\tiny PMF's JADE-paper} Comparisons of QCD predictions for
moments of event shape distributions with data are thus complementary
to tests of the theory using distributions.

\section{Theory} \label{NPT}

The QCD prediction for \momn{y}{n} in next-to-leading-order\footnote{Very recently, perturbative NNLO predictions
for event shape moments became available\cite{NNLOmoments}. 
The nonperturbative models discussed in the following sections
need to be adapted to these new NNLO predictions before a
meaningful study based on NNLO can be done. Therefore, here
we restrict ourselves to consistently using NLO together with the
currently available models.}
%%The non perturbative predictions have to be adapted,
%%and comparing to NNLO is beyond the scope of this work.} %, as in \cite{jadepaper}.} 
(NLO) of the strong coupling $\asb\equiv\as/(2\pi)$
has the form\footnote{The \asbsq\ coefficient is written as ${\cal B}_n-2{\cal A}_n$
because the QCD calculations are normalized to the Born cross section $\sigma_0$,
while the data are normalized to the total hadronic cross section 
$\sigtot=\signull(1+2\asb)$ in LO.}
\begin{equation}
  \momn{y}{n}_{_{\rm NLO}} = {\cal A}_n \,\asb + ({\cal B}_n-2{\cal A}_n) \,\asbsq\,.
\label{eq_qcdmom} 
\end{equation}
The coefficients ${\cal A}_n$ and ${\cal B}_n$ %.  The values of the coefficients ${\cal A}_n$ and ${\cal B}_n$ can be 
are obtained by numerical integration of the QCD matrix
elements using the program EVENT2~\cite{event2}. These predictions were also used in~\cite{OPALPR404,jadepaper}.  

The coupling \asb, and therefore the QCD prediction depends on the renormalisation scale $\muR$, see
e.g.~\cite{ert}. The prediction is changed by this dependence as shown in \cite{jadepaper}. For clearer 
notation the renormalisation scale factor is defined as
$\xmu\equiv\muR/Q$; a truncated fixed order QCD calculation such
as~(\ref{eq_qcdmom}) will then depend on \xmu.  

  Infrared and collinear stability are essential for a
  perturbative description of the
  parton level. These properties, however, do not suffice for a perturbative description of
  the hadron level as hadronisation takes place at low energy scales, where the
  perturbative description breaks down.
\J{schon in etwa gesagt:
  Traditionell werden Hadron- und Partonniveau durch phänomenologische Monte Carlo
  Modelle zueinander in Beziehung 
  gesetzt (siehe den vorhergehenden Abschnitt\J{ ref{MCs}}), in jüngerer Zeit aber 
  auch durch\J{\tiny physikalischer motivierte---mag STK nicht..} analytische Rechnungen.}
  The evolution of partons to hadrons can be approximated by analytical calculations.
  \J{\tiny Schon früh wurde gesehen, dass experimentell gemessene Mittelwerte und
  klassifizierte Verteilungen von Ereignisformvariablen von den perturbativen 
  \nlo-Vorhersagen um Korrekturen
  in inversen Potenzen der Schwerpunktsenergie abweichen \cite{Barreiro:1985db}.}
  These analytical calculations are generally motivated from the transition from perturbative 
  to non-perturbative regime.\J{siehe die folgenden Unterabschnitte} 
  \J{As opposed to a taylor series, }A perturbation series in quantum field theory is a 
  divergent series, see e.g.~\cite{Beneke}.\J{Beneke, Braun find ichs zwar besser erklaert, aber da gibts SPIRES kein
  gutes BibTeX} To obtain finite results, regularisations have to be
  applied.
  \J{Es können
  auch Terme identifiziert werden, deren Reihe alleine schon divergiert---die sogenannten Renormalons, 
  welche Beiträge $\propto\beta_0^n$ zu den 
  perturbativen Koeffizienten liefern.}
  From suitable prescriptions non-perturbative terms are found, which typically
  scale with inverse powers of the cms energy~\cite{KorTaf,Gar1,DMW}.\J{

  Physikalisch lassen sich Terme $\propto 1/Q$ identifizieren mit der Abstrahlung 
  eines weichen Gluons von zwei harten Partonen unter einem grossen Winkel; 
  Terme $\propto \as/Q$
  entsprechen einer Abstrahlung von einem System dreier harter Partonen.}

%%\end{document}
  \subsection{Dispersive model\J{ nach Dokshitzer et al.\tiny gute Bemerkungen 
  in Gar-1?2?\bb | Rev. Dok...\bb | RESUMMED?\bb | ``inclusive'' thrust?\bb}}
  
  \J{\tiny Kopplung ohne Landau-Pol mittles Dispersions=\bb](\bb-Relation
  \[
  \as(Q^2) = \int_0^\infty d m^2 \frac{Q^2}{(m^2+Q^2)^2}\alpha_{\rm eff}(m^2)   
  \]
  mit Gluonvirtualität $m$\,, effektiver Kopplung=\bb ]( \bb$\alpha_{\rm eff}$\,.}
  This model~\cite{DMW,DWthrustmean,DokTalk} is based on the assumption of a non-perturbatively defined strong coupling
  $\as(Q^2)$ which remains finite at and below the Landau pole \Lamqcd. The Landau pole is
  the scale where the usual perturbative coupling diverges.
  \J{Die virtuellen Zustände eines Gluons (verschiedene Partonkonfigurationen) können 
  mit einer virtuellen Gluonmasse und 
  \"uber eine Dis\-persionsrelation mit dieser\J{?\bb} Kopplung verknüpft werden.

  Die Kopplung $\as(Q^2)$ wird verstanden als Summe 
  der üblichen perturbativen Kopplung sowie eines nichtperturbativen Beitrags,
  \begin{displaymath}
  \as(Q^2)=\as^{\rm pt.}(Q^2) + \as^{\rm npt.}(Q^2)\,.
  \end{displaymath}}
  The {\it matching scale} $Q=\mu_I$\,, marking the border between perturbative and 
  non-perturbative region, is not uniquely defined. Usually it is taken as\J{\tiny ?\bb 
  ``on the...'' ch. 6!? ``nicht wirklich frei!?'' PMF:\bb | OHab: ``arbitrary''? sicher im 
  pt. Bereich; vgl. Landau-Pols$\simeq$0.2GeV} $\mu_I \simeq 2\,GeV$.  

  As the non-perturbative coupling\J{ $\as^{\rm npt.}(Q^2)$} cannot be calculated,
  it is parametrised universally in a simple way by the zeroth moment of the
  extended\J{:\bb,Ref.s-aus OHab-253,[71,249]?\bb; \mI\ hier erkl"aren?\bb} coupling over 
  the non-perturbative region,
  \J{\tiny

  Ein Integral der perturbativen Kopplung mit Einbezug kleiner Energieskalen divergiert,
  \[
  \int_0^\mI dQ \, \as^{\rm pt.}(Q^2) \; \rightarrow \; \infty\,,
  \]
  ein analoges der nichtperturbativen Kopplung wird entsprechend negativ,
  \[
  \int_0^\mI dQ \, \as^{\rm npt.}(Q^2) \; \rightarrow \; - \infty\,,
  \]
  so daß ein derartiges Integral über die Kopplung $\as(Q^2)$ einen endlichen
  Wert besitzt, was diese Parametrisierung ermöglicht:}
  \begin{equation}
  \anulmI = \frac{1}{\mu_I} \int_0^{\mu_I} \! dQ \, \as(Q^2) \,.  \label{DefanulmI}
  \end{equation}
  \J{\tiny OHab: kt statt Q---korrekt?\bb\\
     STK weiter unten\bb und [17]\bb: $\tau$-Zerfall bestätigt Existenz von $\as^{NP}$}

  \J{\bb Das Modell von Dokshitzer, Webber und Marchesini (DMW) leitet\bb nichtperturbative
  Terme aus der Untersuchung von Infrarot-Renormalons (STKs [14]: Altarelli '82) ab.
  \bb some resummation here? \bb}
%%    [OHAB]: welches sind seine wichtigen Referenzen: [71]?\bb[73]?\bb[249]?\bb ]( zitieren}
  In first\J{\tiny in zweiter was? )[ bissi 
  ala PMF\bb} approximation, non-perturbative corrections generate a simple shift of the 
  perturbative differential distribution
  ${d \sigma_{_{\rm NLO}}}/{dy}$ of the event shape variables \thr\J{ 1-Hemi Thrust, Ref. Gar?\bb}, 
  \cp, \bt, \bw\ and\footnote{The theoretical calculations are based on the
variable $\mh^2$ because of the problems with the NLO description of \mh\ as discussed
in~\cite{OPALPR404,jadepaper,ICHEP08}.} $\mh^2$ when relating parton to hadron level,\J{\bt\,, \bw\,: squeezed}
\begin{eqnarray}
  \frac{d \sigma_{\rm had.}}{dy} & =& \frac{d \sigma_{_{\rm NLO}}}{dy}(y-a_y \cdot {\cal P}) \;\;\;\;\;  \label{DMWvertVersch}
\end{eqnarray}
  This prediction is\J{\bb} valid only if the value of the event shape variable $y$ is not too
  large ($y\ll 1$), and the cms energy $Q$ not too small %\J{(im Vergleich zur \qcd-Skala $\Lambda$)} 
  ($Q\gg{\Lamqcd}/{y}$).\J{extremer 2Jet-Bereich also ausgenommen}
  Only the numerical factor $a_y$ depends on the event shape variable $y$, see 
  Table~\ref{ay}. However ${\cal P}$ depends\J{\tiny{}STK:[19]\bb} on the hard scale 
  $Q$ but is universal for the event shape variables \thr, \cp, 
  \bt\,, \bw\ and $\mh^2$,\J{---die nichtperturbative Kopplung wird als fundamentaler 
Parameter in nichtperturbativen 
Rechnungen angenommen analog zu $\as^{pt.}$ in perturbativen.\J{---lieber oben?\bb}} and has 
the form \cite{DWthrustmean,DokTalk}\J{\bb{}die 2 Ref.s einfach aus OHab übernommen; checken\bb}
\begin{eqnarray}
{\cal P}&=&\frac{4\,C_F}{\pi^2}\cdot{\cal M}\cdot\left\{\alpha_0-\left[\as(\mu_R^2) \right.\right.\label{calP} \nopagebreak \\
  & & \left.\left.  + 2\,\beta_0\, \as^2(\muR^2)\left(\ln\frac{\muR}{\mu_I}+1+\frac{K}{4\pi\beta_0}\right)+{\cal O}(\as^3) \right]\right\} \times \frac{\mu_I}{Q} \nonumber       
\end{eqnarray} 
In the $\msb$ renormalisation scheme the constant $K$
has the value
\begin{equation}
  K = C_A\left(\frac{67}{18}-\frac{\pi^2}{6}\right)-\frac{5}{9}\nf
\end{equation}
with $\nf=5$ at the studied energies, and the beta function coefficient is $\beta_0=23/(12\pi)$.
\J{Der wesentliche 
  nichtperturbative Parameter in ${\cal P}$ ist also $\alpha_0$ $\equiv$ OHab-(136); (135)=NLO+pc\\
$\alpha_0 - \alpha_s$ wg. matching!?\bb)\abwarten %%\\
STK:vgl. Eur Phys J. C22,1 $\rightarrow$ = JADE-paper!?\bb} 
 The {\it color factors} have the values $C_F=4/3$, $C_A=3$~\cite{ESW}, the 
{\it Milan factor}
%%\K{Name woher?\bb Referenz!!} 
${\cal M}$\J{bisher..Doktalk\bb; {\it on the universality...}-23!\bb} is known in two loops,
${\cal M} = 1.49 \pm 20\%$\J{Ref\bb} (for flavour number 
$\nf=3$ at the relevant low scales).
The cited uncertainty results \cite{DokTalk} from the estimation of the next order contribution,
${\cal M}_{_{\rm NNLO}}={\cal M}_{_{\rm NLO}}\cdot(1+{\cal O}({\as}/{\pi}))$.\J{wobei \as\ an 
der entsprechenden kleinen Energieskala etwa den Wert Eins hat}
%\end{document}

\J{hier?\bb oben, allgemeiner?\bb Die Proportionalität der nichtperturbativen 
Korrektur ${\cal P}$ zu einer Potenz der Schwerpunktsenergie (oben der minus
Ersten) führt zum Namen {\it Energiepotenzkorrektur}.}
%%\begin{table}[h]
%%\begin{center}
%%\begin{tabular}{| c | c |}
%%\hline
%%  event shape variable $y$ & $a_y$                                             \\ \hline
%% \thr       & $2$                                                              \\              
%% \cp        & $3 \pi$                                                          \\
%% \bt        & $1$\J{(\bt\,, \bw\ upgedatet lt. OHAB genauso)}                  \\
%% \bw        & $1/2$                                                            \\  
%% \ytwothree & $0$\J{(erst\bb ${\cal O}(\frac{\ln Q...1}{Q^2})$\bb)\abwarten }  \\   
%% $\mh^2$    & $1$                                                              \\ \hline
%%\end{tabular}
%%\end{center}
%%\caption{Coefficients $a_y$ of  
%%  power correction $\propto {1}/{Q}$ of event shape variables in the dispersive
%%  model\J{lt. OHab, nicht gecheckt:} \cite{UniversalityRescued,Universality}.\J{\tiny 
%%  quer statt hoch?\bb)\abwarten}}\label{ay}
%%\end{table}
\begin{table}[h]
\caption{Coefficients $a_y$ of  
  power correction $\propto {1}/{Q}$ of event shape variables in the dispersive
  model\J{lt. OHab, nicht gecheckt:} \cite{UniversalityRescued,Universality}}\label{ay}
\begin{center}
\begin{tabular}{  l   l   l   l   l   l   l  }
\hline\noalign{\smallskip}
%\begin{tabular}{| c | c   c   c   c   c   c |}
{\parbox{2.cm}{  event shape \\
                  variable $y$ }} &
%%  event shape &\\
%% variable $y$ &
 \thr       &          
 \cp        &
 \bt        &
 \bw        &  
 \ytwothree &   
 $\mh^2$       \\ 
\noalign{\smallskip}\hline\noalign{\smallskip}
  $a_y$     
            & $2$                   
            & $3 \pi$  
            & $1$      
            & $1/2$      
            & $0$         
            & $1$      \\ 
\noalign{\smallskip}\hline
\end{tabular}
\end{center}
\end{table}
\J{Die Verschiebung der differentiellen Verteilung nach Formel (\ref{DMWvertVersch}) ist 
in gewissem Ausmass auch experimentell 
bestätigt \cite{pedrophd,JADE-paper}.\tiny Allerdings liegt auf der Hand,
daß sie nicht vollständig korrekt ist: Die Verteilung nimmt ihr zufolge auch
positive Werte im  unphysikalischen Bereich einer negativen 
Ereignisformvariablen an. Wird sie hier einfach auf Null gesetzt, so erfüllt sie
noch nicht das zum Gewährleisten der Stetigkeit nötige Kriterium des Verschwindens 
an der Phasenraumgrenze.}

Applying the dispersive model calculation~(\ref{DMWvertVersch}) of the normalized event shape distribution 
 in definition~(\ref{defmom}) of the moment of order $n$ and 
naively neglecting the integration over the unphysical range of negative variable values, 
gives %%at cms energy $Q\equiv\rs$\,:\J{$Q\equiv\rs$ \"uberall!?\bb)\abwarten}
\begin{eqnarray*}
  \langle y^n\rangle&=&\int\limits_0^1\!dy\,y^n\cdot\frac{d\sigma}{dy}(y)
  \approx\int\limits_0^1\!dy\,(y+a_y{\cal P})^n\cdot\frac{d\sigma_{_{\rm NLO}}}{dy}(y)\,.\\\nonumber
\end{eqnarray*}
%\end{document}
%%${\cal P}$ is given in~(\ref{calP}).
{The predictions for the moments on hadron level become:
 \begin{eqnarray}
     \langle y ^1 \rangle & =     & \langle y^1   \rangle_{_{\rm NLO}} + a_y {\cal P}    \J{(+{\cal O}(\as/Q)}               \label{Mom1DMW}      \\
      \langle y ^2 \rangle & =     & \langle y^2 \rangle_{_{\rm NLO}} + 2 \langle y^1  \rangle_{_{\rm NLO}}\cdot a_y\,{\cal P} + (a_y\, {\cal P})^2  \J{(+{\cal O}(\as/Q^2)}\label{Mom2DMW}      \\ %f\"ur n=2\\
      \langle y ^3 \rangle & =     & \langle y^3 \rangle_{_{\rm NLO}} + 3 \langle y^2\rangle_{_{\rm NLO}}\cdot a_y\,{\cal P}
                                 + 3 \langle y^1   \rangle_{_{\rm NLO}}\cdot (a_y\,{\cal P})^2 \nonumber \\
                           & &   + (a_y\, {\cal P})^3                                                                                                       \\%\;\;\;, f\"ur n=3\\
      \langle y ^4 \rangle & =     & \langle y^4 \rangle_{_{\rm NLO}} + 4 \langle y^3\rangle_{_{\rm NLO}}\cdot a_y\,{\cal P}
                                 + 6 \langle y^2 \rangle_{_{\rm NLO}}\cdot (a_y\,{\cal P})^2 \nonumber \\
                           & &   + 4 \langle y^1   \rangle_{_{\rm NLO}}\cdot (a_y\,{\cal P})^3 
                                 +                                 (a_y\,{\cal P})^4 \nopagebreak                                                           \\ 
      \langle y ^5 \rangle & =     & \langle y^5 \rangle_{_{\rm NLO}} + 5 \langle y^4\rangle_{_{\rm NLO}}\cdot a_y\,{\cal P}
                                 +10 \langle y^3 \rangle_{_{\rm NLO}}\cdot (a_y\,{\cal P})^2\nonumber \\
                            & &  +10 \langle y^2 \rangle_{_{\rm NLO}}\cdot (a_y\,{\cal P})^3     
                                 + 5 \langle y^1   \rangle_{_{\rm NLO}}\cdot (a_y\,{\cal P})^4\nonumber\\  
                            & &      +                                 (a_y\,{\cal P})^5\,.          \label{Mom5DMW}
\end{eqnarray} 
%%We employ calculations for $\momn{y}{n}_{_{\rm NLO}}$ in second order perturbation theory.
%\J{\tiny The prediction consists of a purely perturbative and a purely nonperturbative term for the 
%first moment---$\momn{B_{T,W}}{1}$ squeezed auch gemischt, s.u.---further mixed terms for the
%higher moments and stronger suppression with cms energy. \tiny nat\"urlich 
%k\"onnte es weitere 
%Korrekturen:\bb geben, die in diesem einfachen Modell vernachl\"assigt 
%sind)\abwarten\footnote{\J{Equation~(\ref{Mom1DMW}) is given in the literature
%\cite{DWthrustmean}. 
%The ansatz basing\J{besser?\bb} (\ref{Mom2DMW}) ... (\ref{Mom5DMW}) %%stellen eine (naive) Erweiterung dar.
%is used also in
%\cite{Webber:1997zj}\J{Webber, Renormalon phenomena}, \J{jedoch weniger
%detailliert ausgef\"uhrt.---in dieser Form} and tested in 
%\cite{L3}.}},\footnote{\J{\tiny Tats\"achlich existierten
%die Vorhersagen f\"ur die Mittelwerte vor jenen f\"ur Verteilungen
%\cite{DW}; (im Fall von Thrust---wegen der Universalit\"at der Energiepotenzkorrekturen 
%\cite{Universality,UniversalityRescued} gilt
%dies aber allgemein---verwirrend 
%wir argumentieren gewissermassen r\"uckw\"arts.}}}}     

Previous studies~\cite{MovillaFernandez:OB259} indicate that the parameters \asmz\ and \anulmI\ when fitted to \bt\ and \bw\ distributions via~(\ref{DMWvertVersch}) 
are not compatible with values derived from \thr\ and \cp. %%\J{\tiny Jetbreiten besonders sensibel auf 2Jetprobs warum?\bb )[ ihre Form:\bb} \cite{MovillaFernandez:OB259}. 
%Insufficiency of a pure distributions shift showed up strongest\J{Doktalk: 
%plot\bb} \cite{JADE-paper} studying
%the distributions of the event shape variables \bt\ and \bw.\J{\tiny der Rückstoß bei
%Gluon-Abstrahlung in der gegenüberliegenden Hemisphäre wurde anfänglich 
%vernachlässigt---sagt OB; Zitat dazu?\bb)\abwarten}
Therefore improved predictions for these distributions were given. %%, which %%additionally 
They also describe a compression of 
the distribution peak, i.e. a narrowing in the two jet region~\cite{DokTalk,revisiting}.
%Therefore the predictions
%were improved for these two variables by a compression of the distribution in the two jet 
%region\J{Die vollst\"andigsten Formeln:} \cite{OHab};\J{\tiny STK sagte: Salam, Dokshitzer!} 
The non-perturbative factor ${\cal P}$ in the case of \momn{\bt}{1}
and \momn{\bw}{1} is replaced by~\cite{OHab} 
\begin{eqnarray}
  {\cal P}_{\momn{\bt}{1}} &=& {\cal P} \cdot \left( \frac{\pi}{2\sqrt{C_F\, \widehat{\alpha}_s(1+K \widehat{\alpha}_s/(2\pi))}} \right.\nonumber\\
                           & & \hspace{2.5cm}+\left.\frac{3}{4} - \frac{2\pi\beta_0}{3C_F} +\eta_0 \right)\,, 
\label{PBT1sque}\end{eqnarray}
rsp.
\begin{eqnarray}
 {\cal P}_{\momn{\bw}{1}} &=& {\cal P} \cdot \left( \frac{\pi}{2\sqrt{2 C_F\, \widehat{\alpha}_s(1+K \widehat{\alpha}_s/(2\pi))}} \right.\nonumber\\
                           & & \hspace{2.5cm}+\left. \frac{3}{4} - \frac{ \pi\beta_0}{3C_F} +\eta_0 \right)\,,
\label{PBW1sque}\end{eqnarray}
with a rescaled coupling $\widehat{\alpha}_s(Q^2) \equiv \as({\rm e}^{-3/2}Q^2)$
and a constant $\eta_0\simeq -0.6137$\,.\J{\tiny Gardis Erkl\"arungen:\bb}
%
%\J{\bf Energiepotenzkorrekturen f\"ur \momn{(\ytwothree)}{n}}

No power correction coefficient has been calculated in the dispersive model for 
the variable \ytwothree.\J{und die Momente dieser Variablen}
\J{Figure~\ref{NPTBWMH2} shows}  The purely perturbative prediction
describes the first moments of \ytwothree\ well \cite{JADE-paper,OHab}\J{JADE-paper,OHab}---therefore we also
compare it with the higher moments.
 
The dispersive model gives predictions for several  
observables and contains only 
universal free parameters\linebreak \asmz\ and $\anulmI$. 
%%The universality of \asmz\ and $\anulmI$ is tested with data in Sect.~\ref{ModellDoksh}.

  \subsection{Shape function\J{{\color{red}\bb{}STK:ask..cite Hoang\bb}\tiny -Rev. Dok?\bb\\
bei Tests: u.a. motiviert aus DMW-Versagen 2. Momente!?---wohl KorTaf;
 Wg. (durch mich explizit gemachter) pt. Probs Antikorrelation $\lambda$/\as 
 eingebaut?\bb)\abwarten}}\label{Theo:ShapeFct}

  Korchemsky and Tafat \cite{KorTaf} describe properties of the event shape variables \thr, 
  \cp\ and $\mh^2$ not included in \nlo\ perturbation theory by a so called 
  {\it shape function}, which does not depend on the variable nor the cms energy.
  This is more general than the dispersive  model, as not only a shift of the perturbative
  prediction is predicted but also a compression of the distribution peak.
    
  The prediction is deduced from studying the two jet region (i.e. $y \ll 1$) in the
  distribution of the event shape variable $y$.\J{$\Rightarrow$ höhere 
  Momente (2Hemi-)ESs nicht gut beschrieben verständlich)
  Diskussion in ``Tests EPK's genügt\bb} 
  The prediction for the differential distribution is\J{spezielle=(4.1,2,3): 
  lohnt?---Unterschied t, C daraus klar??\bb}
  \begin{equation} 
  \frac{1}{\sigma}\frac{d\sigma(Q)}{dy} = \int_0^{Q\cdot y} d\eps f_y(\eps) \frac{d\sigma_{_{\rm NLO}}}{dy}(y-\eps/Q)\,,
  \label{convolution} 
\end{equation} 
  with a non-perturbative function $f_y(\eps)$, dependent on one scale parameter $\epsilon$.
  This function is derived from the shape function $f(\eps_L,\eps_R)$ \cite{KorTaf}, which
  depends on two scale parameters $\eps_L$, $\eps_R$ for the two hemispheres of the event.
\J{\tiny einer observablenunabhängigen 
nichtperturbativen Distribution,\bb welche den Energiefluß in ``linke'' und 
``rechte'' Hemisphäre im hadronischen Endzustand beschreibt. Sie resummiert u.a. auch weiche
Gluonen. Für den Fall der Ereignisformvariablen 
$y=\mh^2$ läßt sich $f_y$ in aufwendigerer Form ebenfalls angeben, wir verweisen 
auf---s.fct. geht auch ein\bb} 
By the compression of the distribution the validity of the prediction is extended
compared to the dispersive model to $y\simeq \Lamqcd/Q$.\J{Es können also auch sehr 
kleine Variablenwerte beschrieben werden (oder niedrigere Schwerpunktsenergien).}

\J{\tiny The shape function receives contributions of two different types 
\begin{displaymath} 
  f(\eps_L,\eps_R)=f_{\hbox{incl}}(\eps_L)f_{\hbox{incl}}(\eps_R)+\delta f_{\hbox{non-incl}}(\eps_L,\eps_R) 
\end{displaymath} 
The first one is the inclusive contribution from gluons produced and decaying 
into the same hemisphere. (This kind of event is partially taken into account by 
IR renormalons models.) The second type of contribution corresponds to 
non-inclusive events, that is when gluon is produced in one hemisphere but decays 
into particles flowing into another hemisphere. 

(nichtso wichtig (und ich denke andersrum!): The non-inclusive contribution 
describes the cross talk between two hemispheres and es wird erwartet, that this effect 
is important for non-inclusive variables like the heavy-jet mass.)}

%%\J{Die Shape function kann nicht berechnet werden, deshalb---vgl. STK, soft fuction:
%%vielleicht doch!?\bb}A physically motivated ansatz is choosen for the shape 
%%function,\J{, welcher drei freie Parameter enthält.} 
%%\J{In what follows we shall rely on a particular ansatz for the shape function
%%$f(\varepsilon_R,\varepsilon_L)$ which agrees with general properties of
%%nonperturbative QCD distributions and has been used in previous studies of power
%%corrections to the thrust distributions [10=KS-shape=\bb]. Namely, one expects that
%%for small values of $\varepsilon_R$ and $\varepsilon_L$ the shape function
%%should vanish as a power of the energy. Similarly,
%%$f(\varepsilon_R,\varepsilon_L)$ should rapidly vanish as $\varepsilon_R$ or
%%$\varepsilon_L$ becomes large. Taking into account these properties together with
%%((4.5))=\bb one chooses the following expression
%%}
%%\begin{eqnarray}
%%f(\varepsilon_R,\varepsilon_L)=&&\frac{{\cal N}(a,b)}{\Lambda^2}
%%\left({\frac{\varepsilon_R\,\varepsilon_L}{\Lambda^2}}\right)^{a-1}\cdot\nonumber\\
%%&&\hspace{1.cm}\exp\left({-\frac{\varepsilon_R^2+\varepsilon_L^2+2\,b\,\varepsilon_R\,\varepsilon_L}{\Lambda^2}}\right)
%%.
%%\end{eqnarray}
%%It depends on two dimensionless parameters $a$ and $b$, and on a 
%%scale $\Lambda$.
%%The factor ${\cal N}(a,b)$ is fixed by normalisation.\J{- condition ((3.10)) Diese 
%%Form der Shape Function wurde in Folge auch physikalisch weiter gehender 
%%interpretiert \cite{Kor01}.}
%%
%%Alternatively the ansatz 

The shape function can be parametrised~\cite{KorTaf} by its ``first moment'',
\begin{equation}
\lambda_1 = \int d\varepsilon_R \int d\varepsilon_L\, (\varepsilon_R+\varepsilon_L)
f(\varepsilon_R,\varepsilon_L) \equiv \momone{\varepsilon_R+\varepsilon_L}
\J{= \int d\varepsilon\, \varepsilon f_t(\varepsilon)} \,,
\end{equation}
its ``second moment'',
\begin{equation}
  \lambda_2 = \momn{(\varepsilon_R+\varepsilon_L)}{2}\,,
\end{equation}
and a $Q$ dependent function $\delta\lambda_2(Q)$.
Predictions for the moments of event shape variables can be derived, and  
$\lambda_1$ and $\lambda_2$ 
can be fitted to the data. %%, see Sect.~\ref{RechngKor}.

From prediction~(\ref{convolution}) for the distribution of the %%event shape 
variables \thr, \cp\ and 
$\mh^2$, predictions for the mean values follow~\cite{KorTaf} by integration\J{ vgl.\bb Text(DMW)},\J{mit der---zugeh\"origen??\bb warum verschiedene Vorhersagen, wenn alles universell?\bb-paper..---Shape function; ``am Energiepunkt $Q$''---noetig?\bb}\J{\tiny Implizit geht die Annahme ein, dass die
Momente durch Zweijetkonfigurationen dominiert sind. - naja, wurden daraus abgeleitet, k\"onnten schon weiter gelten...---hab ich in Theorie!(?\bb)} 
\begin{eqnarray}
  \momn{(1-T)}{1}     &=& \momn{(1-T)}{1}_{_{\rm NLO}} + \frac{\lambda_1}{Q} \J{\tiny+ {\cal O}(\frac{1}{Q^2})                                                             \neq DMW} \label{momonethrKor}\\
  \momn{C}{1}       &=& \momn{C}{1}_{_{\rm NLO}} 
                                                    + \frac{3\pi}{2} \frac{\lambda_1}{Q} \left[1-5.73 \frac{\alpha_s(Q^2)}{2\pi}\J{\tiny+{\cal O}(\as^2)}\right]
                                                    \J{\tiny+ {\cal O}(\frac{1}{Q^2})= DMW-2.61 \asb!?\bb} \label{momonecpKor}\\
  \momn{M_H}{2} &=& \momn{M_H}{2}_{_{\rm NLO}} + \frac{\lambda_1}{2Q}\J{\tiny+{\cal O}(\frac{1}{Q^2})\neq DMW} \label{momnmh2Kor}\,
\end{eqnarray}

\J{obige ${\cal O}(..) \in$ KorTaf, untige nicht!---Im Gegensatz zum dispersiven Modell tritt hier also bereits f\"ur den Mittelwert des 
 C-Parameters ein gemischt perturbativer/nichtperturbativer Term auf,
 mit Ursprung in der Abstrahlung weicher Gluonen von einer Konfiguration mehrerer 
 harter Partonen.}
Analogously for the second moments one finds~\cite{KorTaf},
\begin{eqnarray}
 \momn{(\thr)}{2}     &=& \momn{(\thr)}{2}_{_{\rm NLO}} + 2\frac{\lambda_1}{Q}\momone{\thr}_{_{\rm NLO}}+\frac{\lambda_2}{Q^2} \J{+{\cal O}(\frac{1}{Q^3})} \label{momnthr2Kor}\\
 \momn{C}{2}     &=& \momone{C^2}_{_{\rm NLO}} + \frac{3\pi}2\frac{\lambda_1}{Q} 
                  \left[2\,\momone{C}_{_{\rm NLO}}- 4.30\,\frac{\as(Q^2)}{2\pi}\J{+{\cal O}(\as^2)}\right] \nonumber \\
                 & & +\frac{9\pi^2}4\frac{\lambda_2}{Q^2}\left[1 - 11.46\,\frac{\as(Q^2)}{2\pi}
                               \J{+{\cal O}(\as^2)}\right] 
                               \J{+{\cal O}(\frac{1}{Q^3})}\label{momncp2Kor}\\       
%& &\mbox{\J{... (5.10 c) ausf\"uhren: }}\nonumber\\
 \momn{\mh}{4} &=&\momn{\mh}{4}_{_{\rm NLO}} + \frac{\lambda_1}{Q}\momn{M_H}{2}_{_{\rm NLO}}+\frac{\lambda_2+\delta\lambda_2(Q)}{4Q^2} \J{+{\cal O}(\frac{1}{Q^3}); =DMW+{\cal O}(Q^{-2})!?\bb} \label{momnmh4}\label{momnmh4Kor}
\end{eqnarray}
%%Perturbation theory again is used in two orders.
In this model the more strongly suppressed power corrections have an independent coefficient.
The coefficient $\lambda_1$ is interpreted as the first ``moment'' of the shape function, $\lambda_2$ 
the second, and $\delta\lambda_2$ as\J{oben?\bb} a contribution accounting for the one-hemisphere 
character of $\mh^2$. Therefore these are
universal scales.%%\J{\tiny
%%Da die Form der Shape function aus der Untersuchung des Zweijetbereichs der Verteilung
%%abgeleitet wurde und nur die ersten zwei Momente von diesem Bereich dominiert sind
%%(vgl. Abbildung ref{Moment-Veranschaulichung}; \thr\ wird 1/3 f\"ur drei isotrop 
%%abgestrahlte Partonen \cite{BeStdmodel}), werden keine Vorhersagen f\"ur noch h\"ohere 
%%Momente gegeben.}
%%%%%%%%%%%%%%%%%%%%%%%%%%%%%%%%%%%%%%%%%%%%%%%%%%%%%%%%%%%%%%%%%%%%%%%%%%%%%%%%%%%%%%%%%%%%%%%%%%%%%%%%%%%%%%%%%%%%%%%%%
\subsection{Variance of event shape distributions}

In~\cite{Webber:1997zj} the variance of event shape distributions on hadron level was found to be described
in the dispersive model perturbatively without significant power corrections. This would open the
possibility of an accurate \asmz\ determination. 

A theory predicting first and second moments of an event shape variable $y$ also gives a prediction
for the variance Var$(y)=\momn{y}{2}-\momone{y}^2$, \J{\subsubsection{Dispersive Model}}
and so a simple prediction for the variance of event shape variables on hadron level is 
deduced~\cite{Webber:1997zj}\J{habich, die Referenz habich einfach OB geglaubt, solltich mal checken\bb} 
  by employing the predictions~(\ref{Mom1DMW}) and 
  (\ref{Mom2DMW}),\J{des dispersiven Modells ... in die Darstellung (ref{varcalc}) f\"ur die 
  Varianz der Ereignisformvariablen $y$}

    \begin{eqnarray}
     \mathrm{Var}(y) %& =& \langle y^2 \rangle - \langle y \rangle^2                                      \\
%%            & =& \langle y^2 \rangle_{_{\rm NLO}} + 2 \langle y \rangle_{_{\rm NLO}} \cdot a_y{\cal P} 
            %%                 +(a_y{\cal P})^2  +{\cal O}(\frac{\as}{Q^2})\\   
            %&  &- (\langle y \rangle_{_{\rm NLO}} + a_y{\cal P}+{\cal O}(\frac{\as^2}{Q^2}))^2                                                                 \\
%            &  & - (\langle y \rangle_{_{\rm NLO}} + a_y{\cal P})^2                                     \\
            & =& \langle y^2 \rangle_{_{\rm NLO}} - \langle y \rangle_{_{\rm NLO}}^2  \,.\label{DMWvar}
    \end{eqnarray}
  In the dispersive model one obtains a purely perturbative expression for
  the variance, up to strongly suppressed corrections ${\cal O}({\as}/{Q^2})$.
%%\J{frei von nichtperturbativen Termen $\propto$ ${1}/{Q}$, ${\as}/{Q}$, oder 
%%  ${1}/{Q^2}$\,. 
%%  Sollte diese Beschreibung zutreffen,
%%  so erg\"abe sich die M\"oglichkeit einer sehr reinen Bestimmung der starken
%%  Kopplung. Terme ${\as}/{Q^2}$: hab alte solche
%%     Fits!! D\"urfte aber nur einen sehr grossen Koeffizienten geben u. zu steilen Verlauf!}  

%%  \J{\subsection{Shape function}\label{VarKor}}
  Predictions for the event shape variance
  can also be derived from the shape function. %%\J{\tiny nach Korchemsky et al.}
  The first and second moment predictions~(\ref{momonethrKor}) and (\ref{momnthr2Kor}) 
  give the identical prediction in case of thrust, %%\J{\footnote{\J{\tiny Zur Angabe fehlender Terme mittels ${\cal O}(...)$ sei bemerkt: $1/Q$
  %%f\"allt im Niedrigenergiebereich schneller ab als jede Potenz von \as\,, entsprechend
  %%etwa $1/Q^2$ schneller als $\as^2/Q$, etc.}}}
  \begin{eqnarray*}
   \Var{\thr} %&=& \momn{(\thr)}{2} - \momone{\thr}^2 \nonumber\\
              %&=& \momn{(\thr)}{2} - \left( \momone{1-T}_{_{\rm NLO}} + \frac{\lambda_1}{Q} + {\cal O}(\frac{1}{Q^2}) \right)^2  \nonumber\\
              %&=& \momn{(\thr)}{2}_{_{\rm NLO}} + 2\frac{\lambda_1}{Q}\momone{\thr}_{_{\rm NLO}}+{\cal O}(\frac{1}{Q^2})   \nonumber\\
              %& & - \left( \momone{1-T}_{_{\rm NLO}}^2 + 2\,\frac{\lambda_1}{Q}\,\momone{1-T}_{_{\rm NLO}} + {\cal O}(\frac{1}{Q^2}) \right)  \nonumber\\
              &=& \momn{(\thr)}{2}_{_{\rm NLO}} - \momone{1-T}_{_{\rm NLO}}^2 \,. \nonumber
 \end{eqnarray*}
%% Thus in this model Var(\thr) is described---up to strongly suppressed corrections ${\cal O}({1}/{Q^2})$---purely 
%%perturbatively, too. 
  Analogously this follows from~(\ref{momnmh2Kor}) and 
  (\ref{momnmh4Kor}) for Var($\mh^2$).\J{These predictions have already been discussed in 
  the preceding Subsect.~\ref{VarDMW}.}\sloppy\
  \Var{\cp} follows from~(\ref{momonecpKor}), (\ref{momncp2Kor}) and the perturbative
  LO coefficient \cite{event2}; up to ${\cal O}({1}/{Q^2})$
  \begin{eqnarray}
    \mathrm{Var}(\cp) %&=& \momn{\cp}{2} - \momone{\cp}^2 \nonumber\\
                      %&=& \momone{C^2}_{_{\rm NLO}} + \frac{3\pi}2\frac{\lambda_1}{Q} 
                      %    \left[2\,\momone{C}_{_{\rm NLO}}- 4.30\,\frac{\as(Q^2)}{2\pi}+{\cal O}(\as^2)\right]   +{\cal O}(\frac{1}{Q^3}) \nonumber \\
                      %& & - \left( \momone{C}_{_{\rm NLO}} 
                      %                             + \frac{3\pi}{2} \frac{\lambda_1}{Q} \left[1-5.73 \frac{\alpha_s(Q^2)}{2\pi}+{\cal O}(\as^2)\right]
                      %                              + {\cal O}(\frac{1}{Q^2}) \right)^2 \nonumber \\
                      %&=& \momone{C^2}_{_{\rm NLO}} + \frac{3\pi}2\frac{\lambda_1}{Q} 
                      %    \left[2.066\,\as(Q^2)+{\cal O}(\as^2)\right]   +{\cal O}(\frac{1}{Q^3}) \nonumber \\
                      %& & - \left( 1.375\,\as(Q^2)  + \frac{3\pi}{2} \frac{\lambda_1}{Q} +{\cal O}(\as^2) +{\cal O}(\frac{1}{Q^2}) \right)^2 \nonumber \\
                      &=& \momn{C}{2}_{_{\rm NLO}} - \momone{C}_{_{\rm NLO}}^2-3.23\frac{\lambda_1}{Q}\alpha_s(Q^2) \label{KTvar} \,, 
    \end{eqnarray} 
showing an additional\J{ mixed perturbative / non-perturbative} term $\propto\as/Q$ with coefficient $\lambda_1$ as
in Subsect.~\ref{Theo:ShapeFct}. 
%%%%%%%%%%%%%%%%%%%%%%%%%%%%%%%%%%%%%%%%%%%%%%%%%%%%%%%%%%%%%%%%%%%%%%%%%%%%%%%%%%%%%%%%%%%%%%%%%%%%%%%%%%%%%%%%%%%%%%%%%
 \subsection{Single dressed gluon approximation\J{\tiny nach Gardi et al.\bb Intro [Gar1,2] verstehen\bb}}\label{SDG}

Gardi et al. \cite{Gar1,Gar2} assume the existence of a reordering of the
perturbative series, the so called {\it skeleton expansion} \cite{DSE-paper}---its existence
is proven only for abelian field theory like QED.\J{\tiny Dressed 
mediator vs. dressed vertex verstehen-PhD..erklären\bb; QCD(\nf=0)=QED; Dmed.E=BLM\bb=\bb; 
Bilder übernehmen?\bb}\J{\footnote{\J{\tiny Darin 
gleicht sie an sich jeder \qcd-Rechnung: Die mathematische Existenz einer nicht\-abelschen 
Feldtheorie ist nicht erwiesen \cite{Millenium}.}}}
The first contribution to this expansion is a {\it single dressed gluon} (SDG) which
resums running coupling effects of any order in \as. 
These are renormalons\J{\tiny Bild ala OHABoben bei 1. Erwähnen )[ WWW: Bubblegraphen nur Teil der 
Renormalons!?\bb-haarig!} in the dominant contributions $\propto\beta_0^n$. The single dressed gluon graphs can 
be calculated completely. In this way the perturbative prediction \momn{(\thr)}{n}$_{\rm pt.}$ for a moment 
of thrust can be approximated.
This approximated series already diverges and various regularisations differ by definite powers of the 
cms energy $Q$.\J{\color{red}+genuine NPT:\bb-/in Diss; papers\bb}
These predictions follow:
\begin{eqnarray} 
\langle (1-T)^1 \rangle     &=& \langle 1-T \rangle_{_{\rm pt.}}       + \frac{\nu_1}{Q}                     \nonumber\\
\langle (1-T)^2 \rangle &=& \langle (1-T)^2 \rangle_{_{\rm pt.}}   + \frac{\nu_2}{Q^2} + \frac{\kappa_2}{Q^3}\nonumber        \\
\langle (1-T)^3 \rangle &=& \langle (1-T)^3 \rangle_{_{\rm pt.}}   + \frac{\nu_3}{Q^2} + \frac{\kappa_3}{Q^3}\nonumber        \\
 \langle (1-T)^4 \rangle &=& \langle (1-T)^4 \rangle_{_{\rm pt.}}  + \frac{\nu_4}{Q^2} + \frac{\kappa_4}{Q^5}\nonumber \end{eqnarray} 
In general, the non-perturbative correction is predicted as a sum of two terms with different powers in $Q$
and coefficients $\nu_n$, $\kappa_n$.
The thrust is used in the massless limit,
\begin{equation}
T=\frac{\sum_i \left\vert \vec{p}_i \cdot \vec{n}_T\right\vert}
{\sum _i E_i }=\frac{\sum_i \left\vert \vec{p}_i \cdot \vec{n}_T\right\vert}{Q}\,.\nonumber
\end{equation}
Here the denominator is changed with respect to the standard 
definition~\cite{OPALPR404}\J{OPALPR404}
by ${\sum _i \vert \vec{p}_i \vert } \mapsto {\sum _i E_i }$, 
which does not change the thrust value as long as massless partons are considered.\J{The 
virtual (``massive'') gluon
is understood to fragment eventually into massless partons.

noetig?\bb dann auch massives Gluon einfuehren??\bb} Under this change of definition 
%%JRS: assumption :( 
the thrust value calculated with a
``massive'' virtual gluon is correct, as long as all (massless) partons generated in the
hadronisation end up in the same event hemisphere w.r.t. to the thrust axis.
This so called ``inclusive''%\footnote{inklusiv bezüglich {\it einer} Ereignishemisphäre.}
calculation is valid only if fragmentation of the gluon is approximately collinear.\J{---in 
diesem Fall ist Fragmentation in beide Hemisphären selten. 
Prob der virtuelle Partonlevel mit $m\neq0$---oder der Winkel!?--ESW nachlesen\bb.}
  The perturbative SDG predictions for the moments of thrust are given as perturbation
  series in the coupling $\bar{a}$ \cite{Gar2}, which relates to
  $\as(\muR^2)$ in the \msb\ scheme by a simple   
%%  Reskalierung\J{Gar: Verschiebung-???\bb )[ Antwort}\footnote{Vergleich mit (\ref{asQmu}) liefert in erster Ordnung $\bar{a}(\muR^2)=\as(Q^2)$ mit\J{ $Q^2=e^{5/3}\muR^2$:} $Q=e^{5/6}\muR$\,.
shift of the Landau Pole\footnote{In particular the beta function 
coefficients $\beta_i$ have the same values as in  the \msb\ scheme.},
  \begin{equation}
    \bar{a}(\muR^2) \equiv \frac{\as(\muR^2)/\pi}{1-\frac{5}{3}\,\beta_0\,\as(\muR^2)}\,.\label{bara}
  \end{equation}
%  \J{\bb-h\"ohere..allg. \bb | Gar: Die Kopplung $\bar{a}$ erf\"ullt also die \msb-Renormierungsgruppengleichung\J{(...:\bb...}, 
%  unterscheidet sich aber durch die Definition\J{...:\bb...} von $\Lambda$. \J{...:\bb )[ Antwort Gar}}
%\J{In this section we note $\as^\msb$, to distinguish between these couplings more obviously.}

The perturbative coefficients are given in so called {\it log moments} $d_i$ of the characteristic
thrust function. %% (\ref{Fdef}).
%%,\linebreak $d_0 \equiv {\cal F}_{\momn{(\thr)}{n}}(0)$ in order 0, and 
%%with the abbreviations
%%  \[
%%    {\cal F}_{\momn{t}{m}}(\epsilon)\equiv \int {\cal F}(\epsilon,t)\,t^m\,dt\,,
%%  %\label{char_mom}
%%  \]
%%and\J{ folgendes zusammengeraten:\bb-Lit(WWW)!!,\color{red} sollte ich Gardi fragen\bb}
%%  \begin{displaymath}
%%    \dot{\cal F}_{\momn{t}{m}}(\epsilon)\equiv-\epsilon\,\frac{d}{d\epsilon}\,{\cal F}_{\momn{t}{m}}(\epsilon),
%%  \end{displaymath}
%%the higher orders are defined as
%%  \begin{displaymath}
%%    d_i \equiv \int_0^1  \dot{\cal F}_{\momn{(\thr)}{m}}(\epsilon)\cdot(-\ln \epsilon)^i\,\frac{{\rm d}\epsilon}{\epsilon}\,.
%%  \end{displaymath}
%%  \J{\color{red}oder $\dot{\cal F}\equiv$ Ableitung nach Argument???\bb}
  The first six log moments for \momn{(\thr)}{1}...\momn{(\thr)}{4} were calculated by numerical 
integration~\cite{Gar2}. %% and are given in table~\ref{coef_tab}.
    The perturbative prediction of {\ensuremath{\mathcal{O}(\asb^6)}} in the    
  (obviously incomplete) SDG approximation has the form~\cite{Gardi:privat}
\newpage
\begin{eqnarray} 
&&\momn{(\thr)}{n}_{_{\rm SDG}} \nonumber\\
&=& A_n\cdot\bar{a} + B_n\cdot\bar{a}^2 + C_n\cdot\bar{a}^3 + D_n\cdot\bar{a}^4 + E_n\cdot\bar{a}^5 + F_n\cdot\bar{a}^6\nonumber\\  
&=&  d_0\cdot \bar{a}+  \beta_0 \,d_1\pi \cdot \bar{a}^2\nonumber\\
&                                  +&\left((-\frac{1}{3} d_0 \, \pi^2+d_2) \beta_0^2+\beta_1 \, d_1\right)\pi^2 \cdot \bar{a}^3\nonumber\\
&                                  +&\left((-\pi^2 \,d_1+d_3) \beta_0^3+\frac{5}{2}\,\beta_1\,\beta_0\,d_2+\beta_2\,d_1\right)\pi^3 \cdot \bar{a}^4\nonumber\\
&                                  +&\left((d_4+\frac{1}{5}\,d_0\,\pi^4-2\,\pi^2\,d_2) \beta_0^4+(\frac{13}{3}\,\beta_1\,d_3-\beta_1\,\pi^2\,d_1) \beta_0^2\right.\nonumber\\
&&+3\,\beta_2\,d_2\,\beta_0
\left.+\frac{3}{2}\,\beta_1^2\,d_2+\beta_3\,d_1\right)\pi^4 \cdot \bar{a}^5\nonumber\\
&                                  +&\left((-\frac{10}{3}\,\pi^2\,d_3+\pi^4\,d_1+d_5) \beta_0^5+(\frac{77}{12}\,d_4-\frac{9}{2} \pi^2\,d_2) \beta_1\,\beta_0^3\right.\nonumber\\
&&+(6\,d_3-\pi^2\,d_1) \beta_2\,\beta_0^2%%\nonumber\\
+(\frac{35}{6}\,\beta_1^2\,d_3+\frac{7}{2}\,\beta_3\,d_2) \beta_0\nonumber\\
&&\left.+\beta_4\,d_1+\frac{7}{2}\,\beta_1\,\beta_2\,d_2\right)\pi^5 \cdot \bar{a}^6\,;
\label{SDGformel} \end{eqnarray} 
%%\J{see equation~(3.10) in}
where the $n$-dependent $d_i$ are taken from \cite{Gar2}. %\footnote{
Higher order contributions from the running of the coupling are accounted for up to sixth order. 
Reference~\cite{Gar2} gives the prediction explicitly in ${\cal O}(\bar{a}^3)$ without the term 
$\beta_1 d_1 \pi^2\cdot\bar{a}^3$.
This term 
 with marked effect
%%\J{ (wie f\"ur Effekte in $\beta_1$ \"ublich)} 
on the $\bar{a}^3$-coefficient results from considering the usual \qcd\ beta
function~\cite{Gardi:privat}.
%%\J{ einer allgemeineren Renormierungsgruppengleichung als:\bb in der 
%%ver\"offentlichten Diskussion (mit der \"ublichen QCD Betafunktion).
%%Die Formel in \cite{Gar2} sollte Eingehen h\"oherer Log-Momente der
%%charakteristischen Funktion in die perturbative Entwicklung demonstrieren; 
%%f\"ur ph\"anomenologische Untersuchungen ist die vollst\"andige Betafunktion angebrachter.}
The symbols $\beta_0$...$\beta_3$ denote the usual beta function coefficients~\cite{vanRitbergen:1997va,Czakon}\J{\color{red}ESW; sicher/bis
$\beta_3$!?\bb---andere Ref./selbst definieren\bb}, $\beta_4$ is unknown. 
%We apply the formula with
%\J{\footnote{\J{Man erwartet
%keine starke Abh\"angigkeit der Vorhersage von h\"oheren Betafunktions-Koeffizienten $\beta_i$ 
%mit $i \ge 2$; es gibt auch sog. Padé-N\"aherungen hierf\"ur.}}} 
We set $\beta_4$ to 0. 

%\begin{table}[htb]
%\caption{First six log moments of the characteristic function for the first four moments
%of thrust~\cite{Gar2}.}\label{coef_tab}
%\begin{tabular}{ l  l l l l }
%\hline\noalign{\smallskip}
%&{\momn{(\thr)}{1}}&{\momn{(\thr)}{2}}&{\momn{(\thr)}{3}}&{\momn{(\thr)}{4}}\\
%$d_0$&   0.7888\,   &   0.0713   &   0.0112   &   0.0022  \\
%$d_1$&   3.588\,    &   0.1875   &   0.0160   &   0.000139\\
%$d_2$&   21.06\,    &   0.6296   &   0.0218   &   -0.00903\\ 
%$d_3$&   154.03\,   &   2.5922   &  -0.0020   &   -0.0555 \\
%$d_4$&   1368.38\,  &   12.721   &  -0.3254   &   -0.3108 \\
%$d_5$&   14464.59\, &   72.835   &  -3.1238   &   -1.8546 \\
%\end{tabular}
%\end{table}

\J{Zur leichteren Interpretation l\"asst sich die vollst\"andige numerische Vorhersage (ref{event2moms}) mit Koeffizienten
aus Tabelle ref{abc} umschreiben\footnote{\J{Die Koeffizienten im $\bar{a}$-Schema erh\"alt man 
durch Entwickeln von Gleichung (\ref{bara}) in Potenzen von $\bar{a}$\,.}} \cite{Gar2}, etwa 
f\"ur die ersten zwei Momente von Thrust,
\J{$\beta_0$ umgerechnet; {Einzelbeitr\"age entsprechen was\bb-Gar!!}---wahrsch. nicht sofort
   mit etwa STK vergleichbar, weil dort $\beta_0$ Funktion von $C_A$ etc!}
\begin{eqnarray}
  \momn{(\thr)}{1}        &=& C_F\left[0.7888\,\bar{a} + \left(11.789\,\beta_{0} - 1.1567\,{C_F} - 0.1740 \,{C_A}\right)\,\bar{a}^{2}\right]\nonumber \\
  \momn{(\thr)}{2}        &=& C_F\left[0.0713\,\bar{a} + \left(0.7251\,  \beta_{0} + 0.3073\,  {C_F} + 0.00762\,{C_A}\right)  \bar{a}^{2}\right]\nonumber
%%\\{\large{<}}t^3{\large{>}}&=&C_F\left[0.0112\,\bar{a}(Q^2) + 
%%\left(0.02981\,\beta_{0} + 0.06877\,{C_F} 
%%+ 0.005271\,{C_A}\right)\,\bar{a}^{2}(Q^2)\right]\nonumber\\\nonumber
%%{\large{<}}t^4{\large{>}}&=&C_F\left[0.0022\,\bar{a}(Q^2) + 
%%\left(0.005045\,\beta_{0} + 0.01622\,{C_F} 
%%+ 0.00191\,{C_A}\right)\,\bar{a}^{2} (Q^2)\right]
\end{eqnarray}}

The SDG approximation~(\ref{SDGformel}) is complete in leading order ${\cal O}(\bar{a})$ by 
construction~\cite{Gar2}\J{\bb-7 Mitte}. 
In higher orders this approximation only gives the terms $\propto\beta_0$.\J{ und dieser 
unterscheidet\footnote{\J{Wir bewerten dies als Urteil \"uber die ``inklusive'' N\"aherung innerhalb 
der SDG-N\"aherung und nicht (\"uber die SDG-N\"aherung) innerhalb der vollst\"andigen Berechnung.}} 
sich in zweiter Ordnung wegen der N\"aherung des ``inklusiven'' Thrust von 
seinem exakten Wert um etwa 4\% f\"ur \momn{(\thr)}{1}
und etwa 20\% f\"ur \momn{(\thr)}{2}\,.
(mail1-5:) Zum Vergleich mit Daten sind in Ordnung ${\cal O}(\bar{a}^2)$ die
zu den Gruppen-Strukturkonstanten(Ref.\bb) $C_F$ und $C_A$ proportionalen Terme mit einzubeziehen,
zus\"atzlich zu den zu 
$\beta_0$ proportionalen. Sie sind in \cite{Gar2} Gleichung (3.10) nicht explizit angegeben.
Das jeweilige Log-Moment $d_1$ in Tabelle \ref{coef_tab} ist(\bb) \"uberdies gen\"ahert.}
Therefore we use the numerically calculated \cite{event2} \nlo\ coefficients.\J{ diese liefert 
die vollst\"andigen Entwicklungskoeffizienten---ohne ``inklusive'' N\"aherung} For the third
and higher orders we employ the SDG approximation,\J{\footnote{\J{Hierbei treten keine 
matching-Probleme auf, da die Entwicklung in Terme der Ordnung $\bar{a}^n$ eindeutig ist 
und ein Term davon leicht identifiziert und vollst\"andig ersetzt werden kann.}}}
\begin{eqnarray}
  \langle (1-T)^n \rangle_{_{\rm pt.}} &=& \langle (1-T)^n \rangle_{_{\rm NLO}} + \label{ptSDG}\\
%%                                    & & \hspace{.5cm}\sum_{n=3}^{\infty}\mbox{SDG graphs\J{=\bb} in }{\cal O}(\bar{a}^n)\,. 
                                    & &C_n\cdot\bar{a}^3 + D_n\cdot\bar{a}^4 + E_n\cdot\bar{a}^5 + F_n\cdot\bar{a}^6.\nonumber
\end{eqnarray} 
\J{\color{red}\bb{}Text: vgl. mail Gardi\bb}As the perturbative expansion is an asymptotic (divergent)
series~\cite{Beneke},
the terms are expected to become smaller up to a certain order, and then become
larger again.\J{Bei niedrigen Energien ist die Kopplung gross, Terme mit h\"oheren Potenzen 
der Kopplung tragen relativ st\"arker bei, und somit wird hier diese Ordnung fr\"uher 
erreicht.} 
It is of interest
\begin{itemize}
  \item at which expansion order this happens,            
  \item how the measured coupling depends on the maximum expansion order,              
  \item how the leading power correction $\nu_n/Q^{l_n}$  
        depends on the maximum expansion order,             
  \item how the minimum perturbative term relates to the power correction.
\end{itemize}
\J{\color{red}\bb{}vgl. mail Gardi:\bb}The best approximation of the theory by an asymptotic series is expected when truncating it
near the minimal term---including additional terms does not necessarily result in a better
approximation. Therefore we study the analysis as a function of the truncation order. 

\section{Tests of non-perturbative models}\label{TestsPCs}
  Our tests use event shape moments with statistical and experimental uncertainties measured by \Jade\ and published 
  in~\cite{jadepaper} and 
  analogous \Opal\ data published in~\cite{OPALPR404}, covering in total the energy range of 14 to 209\,GeV.
  In the \Jade\ data $\epem\rightarrow\bbbar$ events have been subtracted on a statistical basis.
  Table~\ref{lumi} gives an overview of the data used.

\begin{table}[h]
\caption{Year of data taking, energy range, integrated luminosity,
average cms energy and the numbers of selected data events 
%using the data version of 5/88
for each \Jade~\cite{jadepaper} or \Opal~\cite{OPALPR404} data sample.
The horizontal lines in the \Opal\ ranges separate the data into the four energy ranges 
used for fit and presentation purposes}
\label{lumi}
\begin{tabular}{ l r@{$\;$...$\;$}l l l r }
\hline\noalign{\smallskip}
year       & \multicolumn{2}{l}{range of  }  & mean $Q$     & luminosity & selected \\
           & \multicolumn{2}{l}{$Q$ in GeV}  &  in GeV      & (pb$^-1$)  & events   \\
\noalign{\smallskip}\hline\noalign{\smallskip}
1981       & 13.0&15.0   & 14.0          &  1.46      & 1783  %& 1722  %&  1588&10
\\
1981       & 21.0&23.0   & 22.0          &  2.41      & 1403  %& 1383  %&  1209&7
\\
1981, 1982 & 33.8&36.0   & 34.6          & 61.7       & 14313 %& 14213 %& 13107&56
\\
1986       & 34.0&36.0   & 35.0          & 92.3       & 20876 %& 20647 %& 19926&83
\\
1985       & 37.3&39.3   & 38.3          &  8.28      & 1585  %& 1584  %&  1450&9
\\
1984, 1985 & 43.4&46.4   & 43.8          & 28.8       & 4376  \\ %& 3896  %&  3919&24
\hline\hline
%\end{tabular}
%\end{table}  
%\begin{table}[h]
%\begin{center}
%\begin{tabular}{| c | r@{$\;$...$\;$}l | c | c | c |}
%\hline
%Year &
%\multicolumn{2}{|c|}{\parbox{2.5cm}{\centering Range of $\sqrt{s}$\\(GeV)}} &
%\multicolumn{1}{|c|}{\parbox{2.5cm}{\centering Mean $\sqrt{s}$\\(GeV)}} &
%\multicolumn{1}{|c|}{\parbox{2.1cm}{\centering \rule{0pt}{0.4cm}Integrated\\luminosity (pb$^{-1}$)\rule[-0.2cm]{0pt}{0pt}}} &
%\multicolumn{1}{|c|}{\parbox{2.5cm}{\centering Number of\\selected events}} & 
%\multicolumn{1}{|c|}{\parbox{2.5cm}{\centering Expected number}} 
%\\ \hline
1996, 2000 &  91.0 & 91.5 &  91.3 &  14.7 & 395695 \\ %%& ---\\
\hline
1995, 1997 & 129.9 & 136.3 & 133.1 &  11.26 & 630   \\ %%& 698\\
\hline
1996       & 161.2 & 161.6 & 161.3 & 10.06 & 281    \\ %%& 275\\
1996       & 170.2 & 172.5 & 172.1 & 10.38 & 218    \\ %%& 232\\
1997       & 180.8 & 184.2 & 182.7 & 57.72 & 1077   \\ %%& 1084\\
\hline
1998       & 188.3 & 189.1 & 188.6 & 185.2 & 3086   \\ %%& 3130\\
1999       & 191.4 & 192.1 & 191.6 & 29.53 & 514    \\ %%& 473\\
1999       & 195.4 & 196.1 & 195.5 & 76.67 & 1137   \\ %%& 1161\\
1999, 2000 & 199.1 & 200.2 & 199.5 & 79.27 & 1090   \\ %%& 1131\\
1999, 2000 & 201.3 & 202.1 & 201.6 & 37.75 & 519    \\ %%& 527\\
2000       & 202.5 & 205.5 & 204.9 & 82.01 & 1130   \\ %%& 1090\\
2000       & 205.5 & 208.9 & 206.6 & 138.8 & 1717   \\ %%& 1804\\
\hline
\end{tabular}
%%\end{center}
%%\caption{The \Opal\ data samples used for the present analysis.}
\end{table}

  JADE and OPAL are similar in construction and many parameters \cite{OPALPR299}.
  Consistent measurements\J{{\it mag SB nicht:} with small systematic differences} can be expected from
  the simultaneous use. The analysis procedures for both data sets were constructed to 
  be similar.
  Systematic variations of the \Jade\ and \Opal\ analyses concern detector event 
  reconstruction, selection cuts, Monte Carlo generators and background.

\subsection{Dispersive model}\label{ModellDoksh} 

\J{\bf Fitprozedur:}
Our test uses the first five moments of \thr, \bt\,, \bw\,, \cp\ and \ytwothree, 
and the second and fourth moment of \mh.\J{\tiny Im 
weiteren verl\"auft der Vergleich wie die Fits der rein perturbativen Vorhersage
an \Jade- und \Opal-Daten einzelner Momente in Unterabschnitt ref{indi-fits}:\bb} We 
compare the prediction with the hadron level data, varying two parameters \asmz\ and \anulmI.
Figs.~\ref{NPTTH}, \ref{NPTBW} show the comparison of data and predictions (\ref{Mom1DMW}...\ref{Mom5DMW}) 
in the case of \thr\ and  
\bw. The \chisqd\ values---calculated from the statistical errors only---vary between 2 and 10, but should 
be regarded as being indicative only.
Experimental differences between \Jade\ and \Opal\ contribute significantly to the 
\chisqd\ values.
At low cms energies the predictions drop off again---this unphysical behaviour (not always visible in the  figures) substantially 
contributes to the high \chisqd\ values.\footnote{The case of \ytwothree\ where  we compared with
the purely perturbative \nlo\ prediction is different.
The energy evolution of the data 
appears consistent with the prediction %with positive coefficients,
but the measured higher moments at 14\,GeV are too low. A suitable power correction would scale with a
higher power of cms energy, and indeed power terms $\ln\,Q/Q^2$ and $1/Q^2$ 
are expected~\cite{DMW} and describe \ytwothree\ distributions better~\cite{pedrophd}. \J{\tiny
In \cite{pedrophd}
wurden verschiedene Energiepotenzkorrekturen mit klassifizierten Verteilungen von \ytwothree\ bei
Schwerpunktsenergien von 14 bis 189\,GeV getestet. Das stellt eine erheblich gr\"ossere 
Datenmenge dar
als die ersten f\"unf Momente, denn sensitiv auf stark unterdr\"uckte Korrekturen ist in 
  diesem Fall fast nur der 
Punkt bei 14\,GeV. 
Korrekturen $1/Q^2$ waren mittels des \chisqd-Wertes nur schwach gegen\"uber Korrekturen $1/Q$ bevorzugt.
Aus den Momenten ist keine bessere Entscheidung zu erwarten.}}\J{\footnote{This evolution is not always visible in
the plot. However the coefficient\J{\color{red}Vorfaktor besser=\bb-LEO!!} of the 
non-perturbative term ${1}/{$Q$}$ is calculated perturbatively in two orders of \as\ by~(\ref{calP}). 
The coefficients of \as\ and $\as^2$ are negative, and this always leads
to the drop of at low enough cms energy.\J{(erinnert sei an 
den G\"ultigkeitsbereich der Vorhersage).
Die Konvergenz dieser St\"orungsreihe ist schlecht; mit dem Fitwert \asmz\ wird der Term
zweiter Ordnung im Fall von \momn{(\thr)}{5}\J{geraten; wars der?\bb-KFaktor](KoeffTab](.xls\bb} bei 
14\,GeV sogar gr\"osser als derjenige erster Ordnung in \as} When calculating the
coefficients in leading order only, the \chisqd\ values of the fits strongly decrease.
\J{Die Theorie kann so aber nicht gut bet\"atigt werden, da} The resulting values of the
assumed universal parameter \anulmI\ however have a large spread then.
\J{{\it Minimal Term - Regularisierung} (Abschneiden
der perturbativen Reihe am kleinsten Summanden f\"ur den jeweiligen 
  Energiepunkt\J{s.\bb--SDG!?}) weist 
die \nlo-Berechnung der 
Energiepotenz-Koeffizienten bis auf den erw\"ahnten Fall stets als die vollst\"andigere aus.}}}
\J{Der Verlauf der h\"oheren Momente des Durham Zweijet-Flipparameters stimmt f\"ur h\"ohere Schwerpunktsenergie mit der 
rein perturbativen Vorhersage \"uberein,
f\"ur niedrige jedoch nicht---dies w\"urde etwa durch eine negative 
Energiepotenzkorrektur berichtigt.\J{aber \chisqd=25..41/16 ca. ok!}}
The moments of the one-hemisphere variables \bw\ and in particular \mh\ are described better
than in the respective comparison with hadronisation correction by Monte Carlo 
models~\cite{jadepaper}.

\J{Min.Ovl. ausf\"uhren: Fkt. \"ofter:\bb-in .xls dokumentiert! nicht (insb. theo. 
Sys.-/verwendet}To estimate the experimental systematic uncertainties, the fits are repeated
based on %%the total errors  and 
the minimum overlap assumption \cite{jadepaper} for combining
the \Opal\ and \Jade\ systematic errors.\footnote{The \chisqd\ values from these fits are substantially 
lower than those from the fits employing the statistical errors.}\J{\bb{}noetig?\bb In jadepaper?\bb:Der erhaltene relative Fehler wird auf das Resultat
    aus dem Fit mit statistischen Fehlern ohne Korrelationen umgerechnet (die 
  Zentralwerte weichen 
   ohnehin\J{\tiny selten?\bb nie signifikant?\bb-check \& dok.\bb)} nur 
  geringf\"ugig voneinander ab).}

We study uncertainties of theory parameters by the following variations
\begin{itemize}
  \item The renormalisation scale uncertainty (i.e. the effect of higher perturbative orders) is estimated 
        by setting $\xmu=0.5$ and $\xmu=2.0$. Our default is always $\xmu=1$. 
  \item The hadronisation uncertainty is estimated by setting
  \begin{itemize}
    \item  ${\cal M} = 1.49 \pm 20\%$, 
    \item  $\mu_I=1$\,GeV and $\mu_I=3$\,GeV. The resulting deviations are included into the
         \asmz\ uncertainty only, as $\alpha_0(\mu_I)$ depends on $\mu_I$ directly by its
         definition~(\ref{DefanulmI}).
  \end{itemize}
\end{itemize}

%%%%%%%%%%%%%%%%%%%%%%%%%%%%%%%%%%%%%%%%%%%%%%%%%%%%%%%%% 
\begin{figure}[htb!] 
     \includegraphics[width=0.50\textwidth]{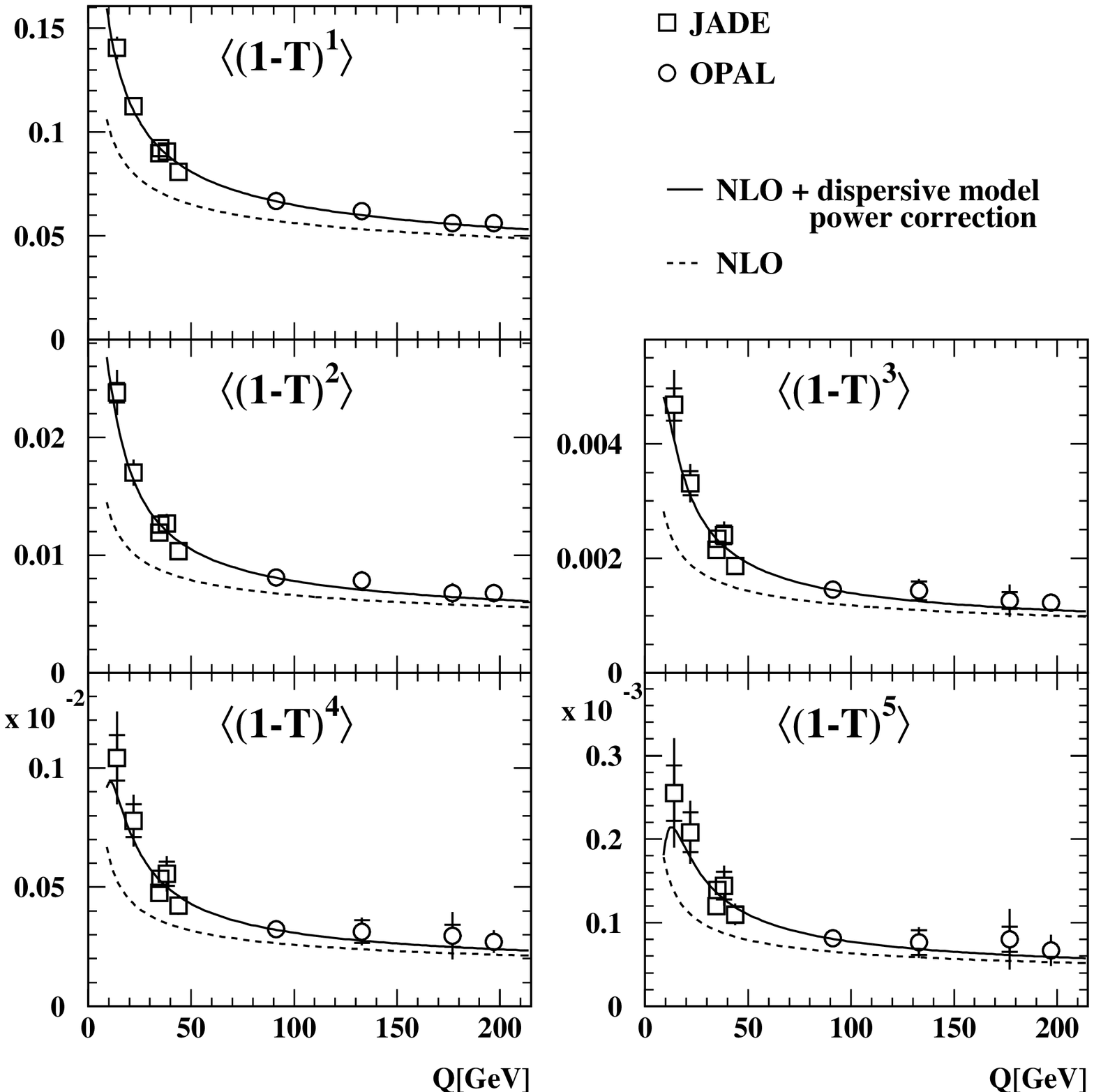}%%\vspace{-13mm}

   \caption{Fits of the dispersive prediction to \Jade\ and \Opal\
            measurements of \thr\ moments.
            The {\it solid line} shows the prediction with fitted values of \asmz\ and
            \anulmI, the {\it dashed line} shows the pure \nlo\ contribution.
            The inner error bars show the statistical uncertainties used in the fit, and
            the outer error bars show the combined statistical and experimental systematic errors
           }\label{NPTTH}
\end{figure}

\begin{figure}[htb!] 
     \includegraphics[width=0.50\textwidth]{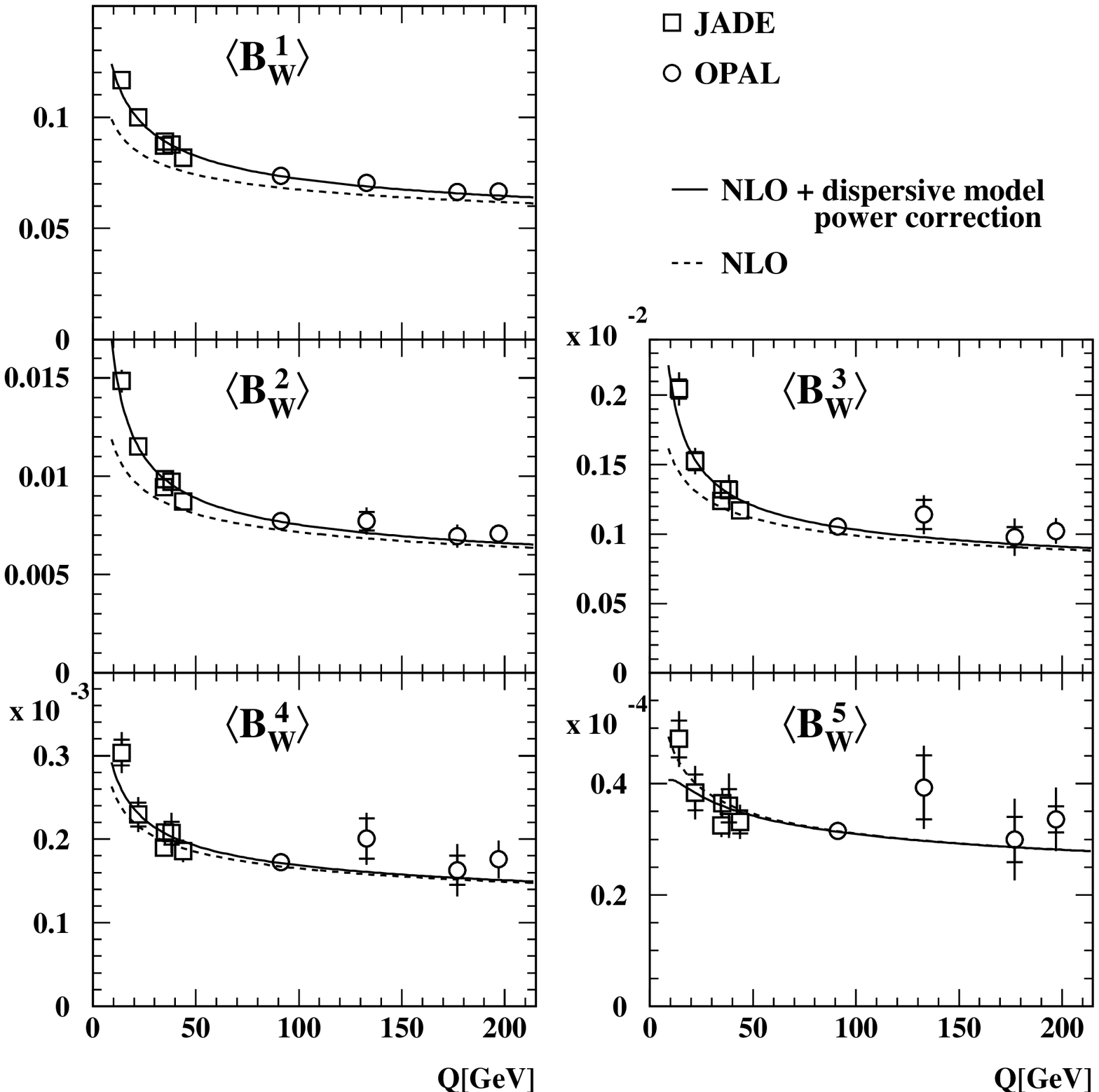}%%\vspace{-13mm}

   \caption{Fits of the dispersive prediction to \Jade\ and \Opal\
            measurements of \bw\ moments.
            The {\it solid line} shows the prediction with fitted values of \asmz\ and
            \anulmI, the {\it dashed line} shows the pure \nlo\ contribution.
            The inner error bars show the statistical uncertainties used in the fit, and
            the outer error bars show the combined statistical and experimental systematic errors
           }\label{NPTBW}
\J{\momn{\thr}{3-5}, \momn{\cp}{2-5}: pc LO f\"allt zu ganz niedrigen $Q$'s weniger stark ab $\Rightarrow$ passt besser.}
\end{figure}

\J{\subsection{Values\J{\& Fehler} of \boldmath{\asmz} and \boldmath{$\alpha_0(\mu_I)$} \J{\tiny vgl. Disk. OB$\sqrt{}$ ( PMF\bb)\abwarten}}}

Tables~\ref{DMW} and~\ref{DMW2} and Fig.~\ref{fitpars-longtabsDMWundminovl} contain the
\asmz\ and $\alpha_0(\mu_I)$ results from the standard measurement of the moments of the
six event shape variables\footnote{We list only results from moments that do not show problems
in the multi jet region dicussed below.}, and from the \xmu, $\mu_I$ and ${\cal M}$ variations.
The values of \asmz\ and \anulmI\ measured from \momn{(\thr)}{1}, \momn{\cp}{1}, \momn{\bt}{1}, \momn{(\ytwothree)}{1}, 
\momn{\mh}{2} are in agreement within errors with previous analyses, see~\cite{OHab,STKrev} and references therein.
The \asmz\ value from \momn{\bw}{1} is lower and the \anulmI\ value higher than in previous analyses because of
the incomplete description of \momn{\bw}{1} in NLO discussed below.
This incompleteness affects the predictions at low centre-of-mass energies, leading to high \anulmI\ values and
low \as\ values. Low energy data have not yet been analysed with a statistics
comparable to ours.

\J{\as: \\}
The \asmz\ results are similar to those in the analysis employing Monte Carlo models
for hadronisiation correction~\cite{OPALPR404,jadepaper}.\J{\tiny sind aber auch kaum 
  besser -schon in PC vs. MC? schon genug besprochen!} In particular the 
\asmz\ values steeply rising with moment order for the variables \thr, \cp\ and \bt\ show the
incomplete description of the multi jet region in \nlo\ as already seen and discussed 
in~\cite{OPALPR404,jadepaper}.\J{\tiny Anstieg 
vergleichbar f\"ur \thr und \bt\,, weniger steil f\"ur \cp.} This is also seen in the 
large renormalisation scale uncertainties.\J{The \momn{(\ytwothree)}{n} and \momn{\mh}{n} results show almost no dependence on the
moment order. \momn{\bt}{n}: \as(n=1), $\alpha_0(n=1)$ liegen deutlich unter unsqueezed; \bw\ kaum so!
Die Werte der starken Kopplung sind nur schwach von Variationen der nichtperturbativen Parameter ${\cal M}$ und $\mu_I$ abh\"angig.

Die Anpassung der alten Rechnung (ohne Verwendung der speziellen Korrekturen 
  (\ref{PBT1sque}) bzw. (\ref{PBW1sque})) f\"ur die Jetbreitenobservablen an unsere Messung von \momn{\bw}{1}\J{Vorherage aus $C^{had}$ kaum unterscheidbar} ergibt
$\asmz=0.1185\pm0.0005$ und $\anulmI=0.563\pm0.018$ (statistische Fehler); aus \momn{\bt}{1}
---Alte Vorherage aus $C^{had}$ bissi h\"oher---erh\"alt man $\asmz=0.1262\pm0.0004$ und $\anulmI=0.486\pm0.015$. Mit 
\chisqd=43/16 f\"ur beide Observablen passen die weniger 
vollst\"andigen Rechnungen sogar besser zu den Daten.---nur weiter disk. falls ich dochmal \chisqd\ aus Kombination hab)\abwarten---
Der Wert von \anulmI\ bleibt nahezu unver\"andert, die gemessenen Kopplungen sind etwas 
h\"oher., was sich f\"ur \momn{\bw}{1 } als Inkonsistenz der Vorhersage interpretieren 
         liesse.---vgl. MC\bb - falschrum, Pedro hatte zu niedrige as, zu hohe a0!! 
Erinnert sei daran, dass die Ver\"anderung der Rechnungen durch Analysen von Verteilungen 
motiviert wurde. Die Mittelwerte wurden bereits durch eine einfache 
Verschiebung der differentiellen Verteilung gut beschrieben 
\cite{OBshiftenough,PMFshiftenough}.            $\alpha_0$:\begin{itemize} 
                                        \item Abh\"angigkeit $\alpha_0(\mu_I)$ von $\xmu$ \"uberhaupt wg.!?\bb )[ Korrelation mit \as\ wg.!?\bb )[ matching\bb  \as/\anulmI 
					\item \momn{\bt}{1} $\xmu$-Var. $\alpha_0$ beide pos. wohl wg. squeezed?\bb )[ vgl. unsqueezed\bb 
                                        \item \momn{\bw}{1}: \xmu-Fehler recht klein!!
                        \end{itemize}

{\bf Universalit\"at von (\asmz\ und?) \anulmI}

  M\"usste innerhalb von stat/exp/\xmu gegeben sein \\
     \tiny schon ungef\"ahr gesagt: Kritisch ist hier (ausser \momn{\bw}{n}, s.o. eher Observable als Momentordnung)

     Bei den Momenten von Thrust, C-Parameter und den Jetbreitenobservablen zeigt 
    sich eine systematische Verschiebung beider Fitparameter mit der Momentordnungszahl, 
  deren (relative) Richtung auch der Antikorrelation der Fitparameter innerhalb eines einzelnen Moments entspricht.\footnote{\J{\tiny In \cite{L3} wurden \"ahnliche Fits an die ersten zwei
   Momente von \thr, \cp, \bt\,, \bw\ und $\mh^2$ durchgef\"uhrt. In den Fits an die zweiten
   Momente wurden die aus den zugeh\"origen ersten Momenten erhaltenen Werte von 
   \asmz\ und \anulmI\ festgehalten und eine st\"arker unterdr\"uckte
   Energiepotenzkorrektur $\propto1/Q^2$ gefittet. F\"ur \momn{(\thr)}{2} und
   \momn{\cp}{2} waren diese
   Beitr\"age dann unerwartet hoch. (die Fits an \momn{\mh}{4} und \momn{\bw}{2} waren 
   nicht klar zu interpretieren) Ein genauerer Vergleich mit unseren Resultaten ist nicht 
   ohne weiteres m\"oglich, da wir die st\"arker unterdr\"uckte Energiepotenzkorrektur aufwendiger
   parametrisieren. Jedoch f\"allt auf, dass der niedrigste Energiepunkt der zweiten Momente
   von \thr, \cp\ und \bt\ von der angepassten Vorhersage stets \"ubersch\"atzt wird---in
   unserer Interpretation: Fehlende perturbative Beitr\"age werden durch 
  Energiepotenzkorrekturen
   ersetzt, welche zu steil verlaufen. Der Wert von \chisqd\ ist auch dort akzeptabel, da
  die radiativen Messungen sehr grosse Unsicherheiten aufweisen.}}

     $\alpha_0$:\\
---- \momn{\cp}{1}: ($\alpha_0=0.42$ etwas tief) \anulmI\ aus \momn{\mh}{2}: 0.58, etwas hoch; .58/.42=140\% (vgl. OB: 20\%) = \bb$\sigma_{sta+exp+\mu_I}$; aus \momn{\bw}{1}: \bb} The \anulmI\ and\J{wenngleich schw\"acher} \asmz\ values from the moments \momn{\bw}{n} are 
not universal but decrease with moment order $n$.\J{\footnote{\J{\tiny Ein Versuch der Erweiterung 
der Beschreibung einer Stauchung der Verteilung auf h\"ohere Momente durch Einsetzen
der nichtperturbativen Faktoren (\ref{PBT1sque}) bzw. (\ref{PBW1sque}) in die Vorhersagen 
(\ref{Mom1DMW}) bis (\ref{Mom5DMW}) \"andert den qualitativen Verlauf beider Fitparameter 
nicht.}}Zumindest hier scheint die naive Ableitung einer Vorhersage f\"ur Momente aus
der Beschreibung der differentiellen Verteilung nicht hinreichend---} We can validate only the
explicitely calculated \momn{\bw}{1} prediction.\J{\tiny Die \chisqd-Werte f\"ur den Fit an diesen Mittelwert sind im Gegensatz zur Analyse mit Monte Carlo -  Hadronisierungskorrektur
zufriedenstellend, vgl. Tabellen \ref{DMW1} und ref{JOdpgminovl}\,.:\bb vgl. 
Fitplots pt,MC/DMW \bb wahrsch. pt Probs von 1/Q gegl\"attet}

Based on the statistical errors, the values of \anulmI\ from the \momn{\bw}{1}, \momn{\mh}{2} and \momn{\mh}{4} fits are significantly
higher\J{\footnote{\J{\tiny Universalit\"at diskutieren wir an dieser Stelle ohne
Einbezug der Unsicherheiten aus dem Milanfaktor, da diese stark korreliert sein d\"urften:
Seine Unsicherheit
wird als Unvollst\"andigkeit der Berechnung einer observablenunabh\"angigen Gr\"osse angesehen
\cite{DokTalk}, nicht etwa als Streuung f\"ur die verschiedenen Ereignisformvariablen.
Diese Universalit\"at ist theoretisch zwar nur auf dem Zweischleifen-Niveau gesichert
\cite{Universality}, sie wird jedoch zum Teil als Eigenschaft des exakten Milanfaktors
betrachtet \cite{DokTalk}.
Der Milanfaktor geht
  f\"ur alle Ereignisformvariablen in dieselbe universelle Energiepotenzkorrektur Formel (\ref{calP}) 
  ein.\J{\tiny(mit unterschiedlichem Vorfaktor, geb ich zu)} 
  Somit entspricht etwa ein erh\"ohter Wert des Milanfaktors stets einem verkleinertem 
  Fitergebnis von \anulmI.\tiny Und viel mehr verwende ich in meiner Diskussion nicht}}}
than those from \momn{(\thr)}{1}, \momn{\cp}{1} and \momn{\bt}{1}.
The \nlo\ description of \bw\ and \mh\ in the two jet region is rather incomplete compared to
\thr, \cp\ and \bt---especially 
at low energy where the 
value of the coupling is high. 
%% \cite{jadepaper,ICHEP08}. This is compensated by a larger power  correction.
%%The noted better description of \bw\ and \mh\ moments is only at the cost of higher
%%\anulmI\ values. 
 At low $Q$ the \nlo\ predictions of the
  \bw, \ytwothree\ and \mh\ distributions are (unphysically) negative 
  in a large range of the two jet region 
  \cite{pedrophd,ICHEP08}. 
This incompleteness of the \nlo\ prediction for the 
%%  and too low to provide a satisfactory description of the data at low c.m. energies. 
moments---in the case of \momn{\bw}{1} the $\as^2$ coefficient is even negative---is 
compensated by a larger power correction.
The perturbative description then contributes less as is seen by the low \asmz\ values\J{(im 
Gegensatz zur Hadronisierungskorrektur mit Monte Carlo - Modellen)}. 
Fits to \Jade\ data of \momn{\bw}{1} alone give 
$\asmz=0.1029\pm0.0016$, $\anulmI=0.728\pm0.017$; to \Opal\ alone they return
$\asmz=0.1242\pm0.0025$, $\anulmI=0.234\pm0.103$ (statistical errors). So at low energies where the coupling is large,
the compensation is stronger, and the power correction is not universal (\Jade\ and \Opal\ agree
better on every other moment).
%and we do not show or use this fit.
% and the incomparable determination of the scale error of \momn{\bw}{1}.
%In case of \momn{\bw}{1}
%the renormalisation scale variations $\xmu=0.5$ and $\xmu=2.0$ both return positive and
%rather small differences. In this case the renormalisation scale uncertainty is not determined in a 
%way comparable with
%the other observables. 
%% \J{\tiny F\"ur Verteilungen von \bw\ und $\mh^2$ werden in \cite{pedrophd} nichtperturbativ 
%%vollst\"andigere Beschreibungen durch zus\"atzliche Terme $\propto (\ln Q)^q/Q^p$\J{trotz 
%%Einschr\"ankung des Fitbereichs!} untersucht.
%%Die Observable \mh\ ist stark von Masseneffekten beeinflusst, 
%%diese werden\bb-SalamWicke; gibt Tests?\bb; gebrochenzahliger Exponent.. 
%%in detailliertere Vorhersagen und Tests mit einbezogen \cite{delphias209,Salam:2001bd}.
%%S\"amtliche Werte von \anulmI\ sind im Vergleich zu den Werten der starken Kopplung wenig von \xmu-Variationen 
%%abh\"angig, welche sich ja zun\"achst auf die perturbative Struktur der Vorhersage auswirken.
%%\J{{\bf Korrelation von \asmz\ und \anulmI}}
%%  Die Werte von \asmz\ und \anulmI\ aus den niedrigsten Momenten sind stark 
%%negativ korreliert
%%  (um etwa -90\%), diese Antikorrelation nimmt f\"ur die h\"oheren Momente meist ab, und betr\"agt nur noch -10\%
%%  f\"ur \momn{\cp}{5}.---warum? wahrscheinlich weil NPT Zweijetbereich wichtiger als 
%%  Multijetbereich. Aus DMW vs. MC nicht\bb klar)\abwarten}
%%

For averaging \asmz\ and \anulmI\ over many variables and moment orders, we exclude
%%We measure the fit parameters from the 
moments that suffer from the 
problems discussed in the preceding two paragraphs:
The deficiencies in the
description of the multi jet region for the two-hemisphere variables have already been seen 
in~\cite{OPALPR404,jadepaper}. 
To select observables with an
apparently converging perturbative prediction, we consider as in~\cite{OPALPR404,jadepaper} only those results for which the NLO term in
equation~(\ref{eq_qcdmom}) is less than half the corresponding LO term (i.e.\ 
$|K|<25$ or $|K\as/2\pi|<0.5$), with $K=\mathcal{B}_n/\mathcal{A}_n$.
%namely \momone{\thr}, \momone{\cp}, \momone{\bt}, \momn{\bw}{n} and
%\momn{(\ytwothree)}{n}, $n=1,\ldots,5$; and \momn{\mh}{n},
%$n=2,\ldots,5$. 
%These are results from %%17 observables in total; or 
%16 observables, %% from \Jade\ and \Opal, 
%exclu\-ding \momn{\bw}{1}.  
  The $K$ values are
shown in Refs.~\cite{OPALPR404,jadepaper}. %%

The higher moments of the jet broadenings are excluded because of
the incomplete~\cite{MovillaFernandez:OB259} description of their distributions
in general. As in \cite{jadepaper} the first moment of \bw\ is excluded because the universality of the fit parameters 
could not be confirmed.

%These are
Thus we combine the results of \asmz\ and \anulmI\ from 
the moments \momn{(\thr)}{1}, 
\momn{\cp}{1}, \momn{\bt}{1}, \momn{(\ytwothree)}{1}...\momn{(\ytwothree)}{5}, 
 and \momn{\mh}{2}, \momn{\mh}{4}.\J{ Die Werte von \asmz\ daraus sind innerhalb der 
totalen Fehler vertr\"aglich; die Werte von \anulmI\ nur unter Einbezug der Unsicherheit 
des Milanfaktors.}
The parameters are consistent within total errors, and the combination procedure 
follows that used in~\cite{OPALPR404,jadepaper}.\footnote{The covariances of the 
statistical errors are estimated by \py\ at hadron level at 91.2~GeV, the covariances
of the experimental systematic uncertainties are taken as\linebreak 
$E_{ij} = \mathrm{Min}\{\sigma_{{\rm exp.},\,i}^2\,,\;\sigma_{{\rm exp.},\,j}^2\}$,
and hadronisation and renormalisation scale uncertainties are found by repeating the analysis with varied parameters.}
The correlations between \asmz\ and \anulmI\ have rather constant values from -0.80 to \mbox{-0.93} and 
thus the combination is done individually for \asmz\ and \anulmI.
The averages are\begin{eqnarray*}
  \asmz   &=&               0.1183  \pm0.0007\stat\pm0.0016\expt\\
          & &                       \pm0.0011\had^{+0.0052}_{-0.0042}(\xmu)\\
          &=&               0.1183  \pm0.0056\tot,\\
%\end{eqnarray*}\newpage\begin{eqnarray*}
  \anulmI &=&               0.493  \pm0.006\stat\pm0.008\expt\\
          & &                      \pm0.050\had^{+0.028}_{-0.014}(\xmu)\\
          &=&               0.493  \pm0.058\tot.           \end{eqnarray*}              

%%\input{DMWnlo}

%%\clearpage
\begin{figure}[htb!]
     \includegraphics[height=8.5cm]{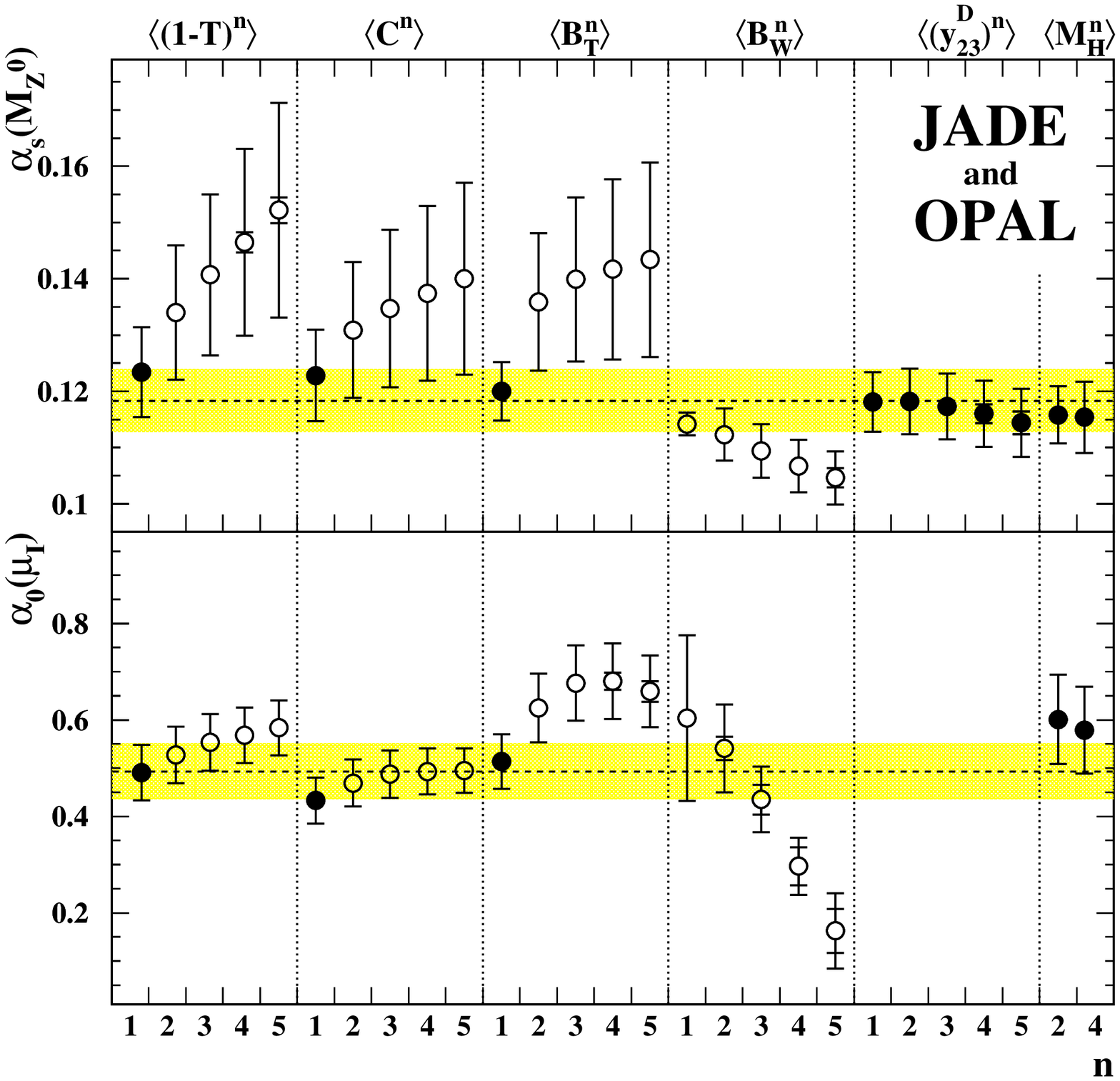}
   \caption{\J{\bb{}{\color{red}update}: ``n'', Legende englisch, ``\ytwothree'', ``ggf.'' einfach 
     weg!?STK!\bb}Measurements of \asmz\ and $\alpha_0(\mu_I)$
     from \J{fits of the dispersive prediction to }moments of six event shape variables 
     at \Petra\ and \Lep\ energies.
  The inner error bars---where visible---show the statistical errors,
  the outer bars show the total errors.
  The  {\it dashed lines} indicate
  the weighted averages, %% described in subsection \ref{ascombs}, 
  the {\it shaded bands} show their errors.
  Only the measurements
  indicated by {\it solid symbols} are used for the averages
%     The error bars show the statistical and the experimental and
%     theoretical uncertainties from varying the renormalisation scale factor $x_\mu$,
%     the Milan factor ${\cal M}$, and in case of \asmz\ also the matching skale $\mu_I$.
%     The dashed lines show a simple                     
%     weighted mean from the measurements indicated by filled 
%     symbols.\J{\tiny lohntnich..keinBock:  \anulmI(\momn{\bw}{1},\momn{\mh}{2}) hoch mit kleinem Fehler 
%               $\Rightarrow$ \anulmI(\momn{(\thr),\cp,\bt}{1}) nicht vertr\"aglich, sondern \bb $\sigma$; 
%               \as: Werte gut vertr\"aglich innerhalb stat/exp/\xmu; JRS will mehr Diskussion!!!---- 
%               Abweichungen aller/der guten in $\sigma$'s und \% \bb}
   }\label{fitpars-longtabsDMWundminovl}
\end{figure}

%%%%%%%%%%%%%%%%%%%%%%%%%%%%%%%%%%%%%%%%%%%%%%%%%%%%%%%% 
%learpage bringtnix, halbe Stunde rumgespielt: ueberlass ic Editoren!!

\subsection{Shape function}\label{RechngKor}

%%\J{glaube [KorTaf] ([TafTalk]?-ex.nur diese$\ge$2000!): DMW passte nicht gut auf h\"ohere Momente war 
%%   wohl u.a. Anlass f\"ur neues NPT-Modell:}
In~\cite{KorTaf} values for the shape function parameters were determined
by comparing a distribution of
%\J{Die Verteilung dieser Ereignisformvariablen ist---vor (4.7) 
%erklaert:\bb---am sensitivsten auf die Form der Shape function. Im Weiteren nahmen die 
%Autoren Universalit\"at der gefundenen Parameter an\J{\asmz\ woher nicht klar... wohl auch 
%daraus, oder Standardwert, jdf. haben die nie Momente gefittet!!} und erstellten hiermit 
%Figuren 3 und 4 in \cite{KorTaf} ohne Fit \cite{Korchemsky:privat}. Gute \"Ubereinstimmung 
%von Daten und T
%heorie wurde nach Augenmass festgestellt.} 
$M_H^2$ at 91\,GeV with the NLO+NLLA
%\J{\footnote{\J{Next to Leading Log Approximation \cite{NLLA}.}}} 
prediction combined with power terms from the shape function
%\J{\tiny mittels Formel 4.3 aus 3.9 (aufschreiben lohnt wohl nicht!?\bb)\abwarten} 
to be $\lambda_1=1.22\,$GeV, $\lambda_2=1.70\,$GeV$^2$ and 
%\J{\tiny umrechnen lohnt nicht weil ich Normierungsunterschiede [TafTalk]/[KorPriv] 
%nicht checke: \tiny $\mapsto 14, 35, 91, 161?, 207\,$GeV, \cite{Korchemsky:privat}}
$\delta\lambda_2$ from $\delta\lambda_2(10$\,GeV$)=1.4\,$(GeV)$^2$ to $\delta\lambda_2(100\,$GeV$)=1.2\,$(GeV)$^2$.

We perform fits to the moments %der Ereignisformvariablen %\linebreak
\momn{(\thr)}{1}, \momn{\cp}{1} and \momn{\mh}{2} with two free parameters \asmz\ 
and $\lambda_1$. Fits to the higher moments \momn{(\thr)}{2}, \momn{\cp}{2} and 
\momn{\mh}{4} are done with an additional parameter. Because of the weak energy
dependence of $\delta\lambda_2$ in~(\ref{momnmh4Kor}) we substitute the numerator
$\lambda_2+\delta\lambda_2(Q)$ by the fit parameter $\overline{\lambda}_2$.
The fits to \momn{\cp}{2} and \momn{\mh}{4} are not sensitive to the 
parameter \J{$\lambda_2$ rsp.} $\overline{\lambda}_2$ while the fit to \momn{(\thr)}{2} 
gives $\overline\lambda_2 = (1.2 \pm 1.3$)\,(GeV)$^2$. %, compatible with zero.
For further analysis we therefore set \J{$\lambda_2=$}$\overline{\lambda}_2=0$.

Fig. \ref{Kor} shows the comparison of the data with the prediction.
The data are fitted well with\J{besser als DMW!?\bb} \chisqd---based on the statistical errors---in the order of one.\J{Die Momente 
der Einhemisph\"arenvariablen \mh\ werden deutlich besser beschrieben als 
im entsprechenden Vergleich mit Monte Carlo - Korrektur f\"ur Hadronisierung.} As before,
the experimental systematic uncertainties are estimated by the minimum overlap assumption, and 
the renormalisation scale uncertainty by setting $\xmu=0.5$ and $\xmu=2.0$\,. 
\begin{figure}
     \includegraphics[width=0.50\textwidth]{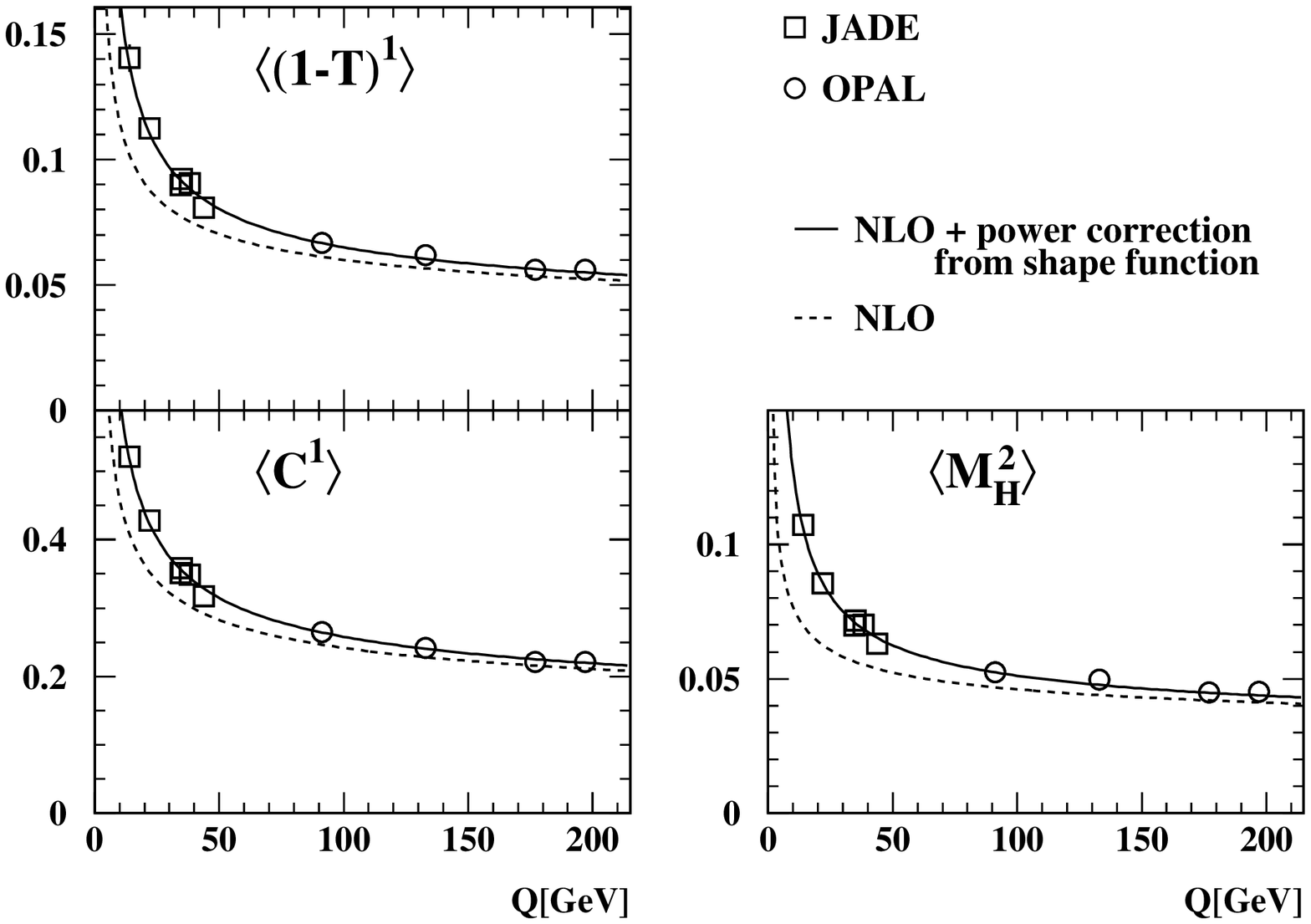} %\nopagebreak
\includegraphics[width=0.50\textwidth]{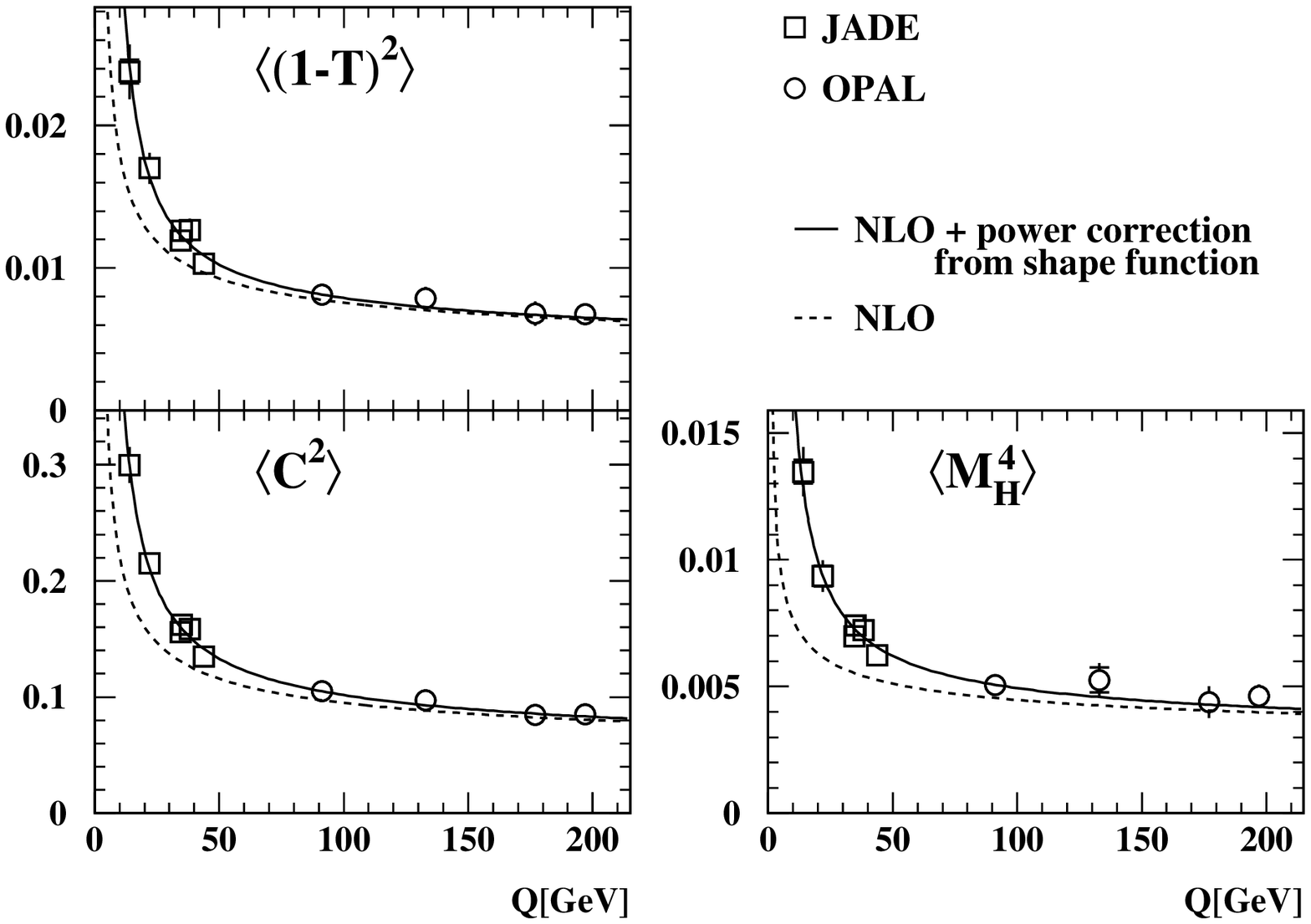}
   \caption{%%\J{{\color{red}update; \chisqd\ eher nicht; ``NLO+shape function power 
            %%correction''}\bb}
            Fits of the shape function prediction with 
            $\overline\lambda_2=\J{\delta\lambda_2=}0$ to 
            \Jade\ and \Opal\ measurements of first and second moments of \thr\ and \cp\ and 
            second and fourth moments of \mh.
            The {\it solid line} shows the fitted prediction %%with fitted values of \asmz\ and $\lambda_1$, 
            and the {\it dashed line} shows the pure \nlo\ contribution. %%prediction with identical value of \asmz.
            The inner error bars show the statistical uncertainties used in the fit and
            the outer error bars show the combined statistical and experimental systematic errors.
            Most of the error bars are smaller than the data points}
%%\vspace{-11.cm}
%%\hspace{1.7cm}{\small $\chi^2/dof.=21.3/16$}\hspace{8.5cm}{\small $\chi^2/dof.=17.1/15$}
%%
%%\vspace{2.65cm}
%%\hspace{1.7cm}{\small $\chi^2/dof.=22.6/16$}\hspace{8.5cm}{\small $\chi^2/dof.=17.3/15$}
%%
%%\vspace{2.6cm}
%%\hspace{1.7cm}{\small $\chi^2/dof.=19.4/16$}\hspace{8.5cm}{\small $\chi^2/dof.=20.6/15$}
%%
%%\vspace{.47cm}
       \label{Kor}
\end{figure}
%%\clearpage
%%\J{\subsubsection{Values of \boldmath{\asmz} and \boldmath{$\lambda_1$}}}
  Table~\ref{KorTab} and Fig.~\ref{fitpars-longtabsKor} contain the results for \asmz\ 
  and $\lambda_1$ from the standard measurement and the systematic variations.
%%\vspace{-.5cm}
\begin{figure}[htb!]  
   \begin{center}
     \includegraphics[height=8.5cm]{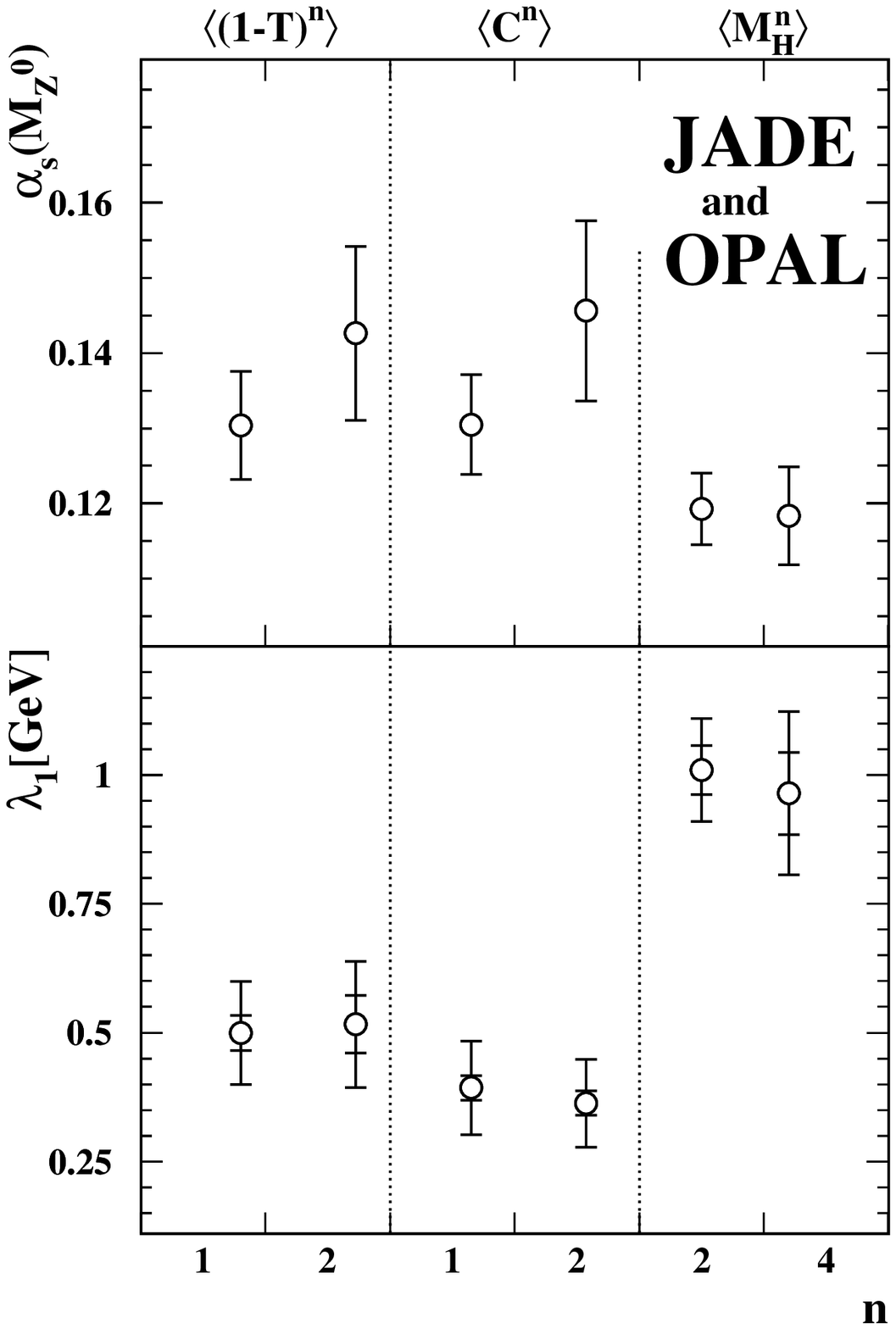} 
   \end{center}
\vspace{-.5cm}
   \caption{\J{fitpars-longtabsKor.eps, \bb {\color{red}update}, engl.; kleiner: Legende rechts oben rein,
    \asmz-Max. ca. .165\bb}
    Measurements of \asmz\J{halbwegs vertr\"aglich..} and $\lambda_1$ from \J{fits of the
    shape function prediction with $\lambda_2=\delta\lambda_2(Q)=0$ to }moments of 
    three event shape variables at \Petra\ and 
    \Lep\ energies.
  The inner error bars---where visible---show the statistical errors,
  the outer bars show the total errors\J{\as\ stark \xmu-abhaengig, $\lambda_1$ fast nicht}}
%%     The error bars show the statistical errors and the experimental and
%%     theoretical uncertainties from varying the renormalisation scale factor $x_\mu$.
     \label{fitpars-longtabsKor}
\end{figure}

%%\input{Kor}%%\clearpage
%%\J{\subsubsection{Universalit\"at von \asmz}}
The values of \asmz\ show the typical increase for the higher moments of \thr\ and \cp.\J{aehneln
jenen aus der Hadronisierungskorrektur mittels MC-Modellen; Abschnitt \bb; sind aber 
auch nicht\bb besser; vgl. PC vs. MC\bb;

Die Variation des Renormalisierungsskalenfaktors sollte sich idealerweise nur auf den perturbativen Teil der Vorhersage 
auswirken.
Die Abh\"angigkeit des nichtperturbativen Parameters $\lambda_1$ von \xmu\ ist aus unserer Messung \J{mit ihrer grossen Statistik} aber im allgemeinen doch
vergleichbar mit seinem statistischen Fehler. F\"ur die Momente der Einhemisph\"arenobservable 
$\mh^2$ ist sie---wie auch schon die Abh\"angigkeit von \asmz\ von diesen Variationen\J{wg. vollst\"andiger, schon in pt., DMW diskutiert?\bb}---
deutlich kleiner.  
(Fazit Kor: Die Momente werden durch den Shape function - Ansatz gut beschrieben,
jedoch ergeben sich aus \momn{(\thr)}{2} und \momn{\cp}{2} hohe Werte von \asmz.)}
%Die Werte von $\lambda_1$ aus \momn{\mh}{n} sind mit jenen aus \momn{(\thr)}{n} und 
%\momn{\cp}{n} nicht vertr\"aglich. 
\J{\subsubsection{Universalit\"at von $\lambda_1$}}
The values of $\lambda_1$ are consistent for the first two moments of every
event shape variable. Furthermore they are consistent for the two-hemisphere variables \thr\ and \cp.
However, the values from \mh\ are not compatible.\J{ auch innerhalb der Gesamtfehler
which is not compatible with a universal shape function.---pt. ok vorausgesetzt!
\as\ fast nur \xmu-abhaengig, \anul\ stark von {\cal M}, \muI}
As the respective predictions are similar, we refer to the high values of \anulmI\ 
from the moments \momn{\mh}{2} and \momn{\mh}{4} in the
dispersive model, see Subsect.~\ref{ModellDoksh}. The large power corrections compensate
for the incomplete perturbative NLO description. Since the values are not consistent and %he theoretical
%uncertainties are not estimated in depth, 
no systematic evaluation of the theory uncertainty is available, we do not combine any of the fit parameters.

\J{\subsubsection{Korrelationen zwischen \asmz\ und $\lambda_1$}
Da sowohl \asmz\ als auch $\lambda_1$ in Formeln (\ref{momonethrKor})\ bis (\ref{momnmh4Kor}) 
in einzelnen Summanden mit positiven Koeffizienten eingehen, sind die Fitwerte von \asmz\ und $\lambda_1$ 
antikorreliert---physikalischer:\bb. Die Antikorrelation ist sehr hoch, sie betr\"agt typischerweise $\rho=-95\%$, siehe Tabelle \ref{KorTab}\,.
---warum $>\rho_{DMW}$?\bb )[ Kor $\mapsto_\delta$ DMW\bb
Taf talk [Ref?\bb] : ``Shape fct. besser als DMW, v.a. h\"ohere... seh ich ? - vgl. \chisqd ]$\mapsto\!\bb$( DMW vs. Kor\bb}

The $\lambda_1$ values from \thr, \cp\ and, less pronounced, from 
\mh\J{beide Momente?\bb} are smaller than the results in \cite{KorTaf}. A substantial 
difference of \cite{KorTaf} with our work is the inclusion of the NLLA approximation in 
the distribution prediction,
which is not available for the moments.

\newpage
  \subsection{Variance of event shape variables}\J{$\Delta$\xmu kann $Q$-Verlauf/aender; fitten lohntnich!!}
%     \tiny (Vgl. allgemeine Diskussion in Abschnitt Momente/Varianz: \bb)\\
%          gute Illustration der Effekte unterdr\"uckten Phasenraums: Man kann sich vorstellen, 
%     was man sich nicht vorstellen kann)\abwarten} 
  From the first and second moments of event shape variables $y$ the variance 
  Var$(y)=\momn{y}{2}-\momone{y}^2$ has been calculated~\cite{hepdata}. 
%  The statistical uncertainties have been determined as\J{\color{red}\bb MC-Test\bb} 
%  \begin{equation}
%    \left( \sigma_{{\rm Var}(y)} \right)^2
%  \sigma^2 = \frac{ \momn{y}{4} - 4\,\momn{y}{3}\momone{y} + 8\,\momn{y}{2}\momone{y}^2 
%             - \momn{y}{2}^2 - 4\,\momone{y}^4 }{N} \,, \label{statvar}
%  \end{equation} and t
 The experimental uncertainties have been determined
 by systematic variations analogously to the moments at hadron level \cite{jadepaper,OPALPR404}. 
 The variance of \cp, \bt, \bw, $\mh$ and $\mh^2$ becomes larger with rising $Q$ while the
 variance of \thr\ becomes smaller, see\footnote{In case of \thr\ this is not seen
very clearly in the plot, but significantly in a straight line fit.} Fig.~\ref{VarWeb} 
and~\cite{hepdata}. Studying other observables defined in~\cite{OPALPR404} we find that the variance of \tma\ and \obl\ becomes larger with 
rising $Q$ while the variance of \tmi, \s, \ml\ and \bn\ becomes smaller, see~\cite{hepdata}.
This behaviour is 
 determined by the
 multi jet region of the distribution \cite{CHPphd}. The evolution is reproduced by Monte Carlo models 
 qualitatively well; however data and models differ by typically up to five standard deviations.

  Fig.~\ref{VarWeb} shows the comparison of the prediction~(\ref{DMWvar}) using the dispersive model and data of \thr, \cp, \bt, \bw,
  \ytwothree\ and $\mh^2$. %; table~\ref{Varianzen} contains the fit results.
  The fits of the free parameter
  \asmz\ use central values of the measurements and the statistical errors.
%%  \J{Die Vorhersagen werden zu h\"oheren Energien stets kleiner, mit Ausnahme derjenigen 
%%  von Var(\bw). Den Energieverlauf der Datenpunkte erkennt man z.T. deutlicher in Abbildung 
%%  ref{vars\_t} mit ihren gr\"oberen Energiebins.} 
  The energy evolution of prediction and data for\linebreak Var(\bw),
%%\J{\footnote{\J{\,Var(\thr) nimmt 
%%  am niedrigsten Energiepunkt von 14\,GeV wieder ab, jedoch nicht statistisch signifikant;
%%  dies wurde bereits in Abschnitt ref{sec_moments}
%%  diskutiert. Es k\"onnte sich eventuell um den Effekt einer stark unterdr\"uckten Energiepotenzkorrektur handeln,
%%  die von den Monte Carlo - Generatoren nicht reproduziert 
%%  wird.\J{
%%%=${\cal O}(\as/Q^2)$
%%}}}} 
%\end{document}
Var(\thr),
%%    \J{ vgl. 
%%  Abbildung ref{vars\_t}}} 
and Var(\ytwothree) coincides qualitatively (in case of Var(\ytwothree) except for the 
  lowest energy points of 14 and 22\,GeV), but not for
  Var(\cp), Var(\bt) and Var($\mh^2$). \J{Energieverlauf k\"onnte durch 
  ${\cal O}(1/Q^2)$-Terme korrigiert werden, kaum aber \asmz: Jene fallen zu schnell ab---
  vgl. \as(\cp)=.0869, Webber vs. 0.0965, Korchemsky} The \chisqd\ values are large
  and the \asmz\ values are not compatible with each other or with established \qcd\ 
  analyses.
%\J{Der folgende Unterabschnitt \ref{VarKor} enth\"alt weitere Diskussion auch 
%  dieser Vorhersagen.
%  In \cite{OHab} wurde die Varianz des C-Parameters bei 91\,GeV aus \Lepone\ - 
%  sowie {\mbox{\rm SLC}} - Messungen von
%  \momn{\cp}{1} und \momn{\cp}{2} berechnet, und hieraus die Kopplung mit statistischem 
%  Fehler bestimmt zu 
%  $\asmz=0.087\pm0.003$. Das gleicht unserem Ergebnis aus 
%  Var(\cp) in den angegebenen Stellen und macht auch unseren kleinen statistischen Fehler
%  glaubw\"urdig 
%  (der statistische Fehler der Varianz konnte in \cite{OHab} lediglich ohne Einbezug 
%  von Korrelationen aus denjenigen der Momente erhalten werden und wurde somit stark 
%  \"ubersch\"atzt).}
%  \input{Varianzen}
 \begin{figure*}
  \begin{tabular}{c c}
     \includegraphics[width=0.485\textwidth]{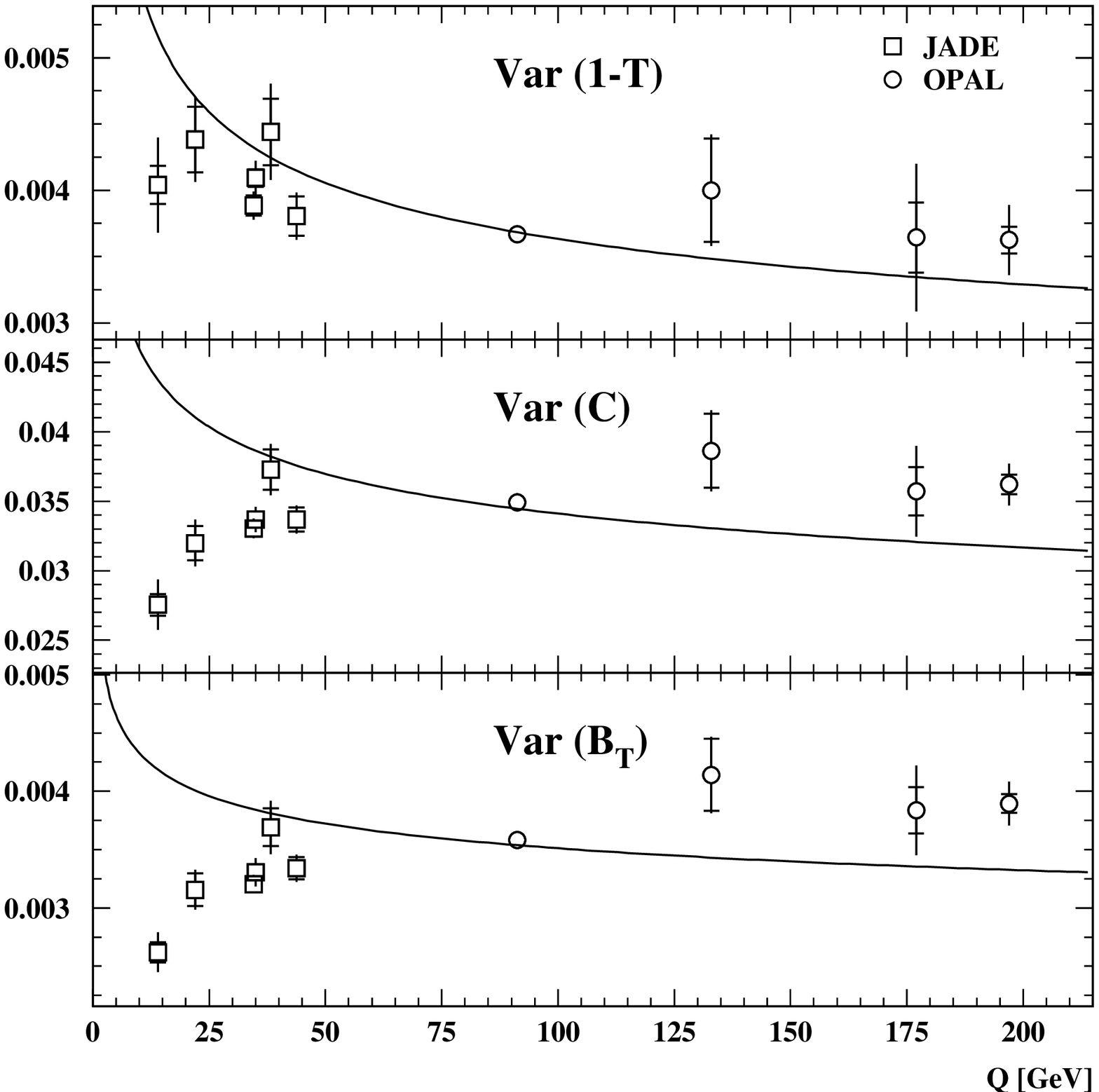} &
     \includegraphics[width=0.485\textwidth]{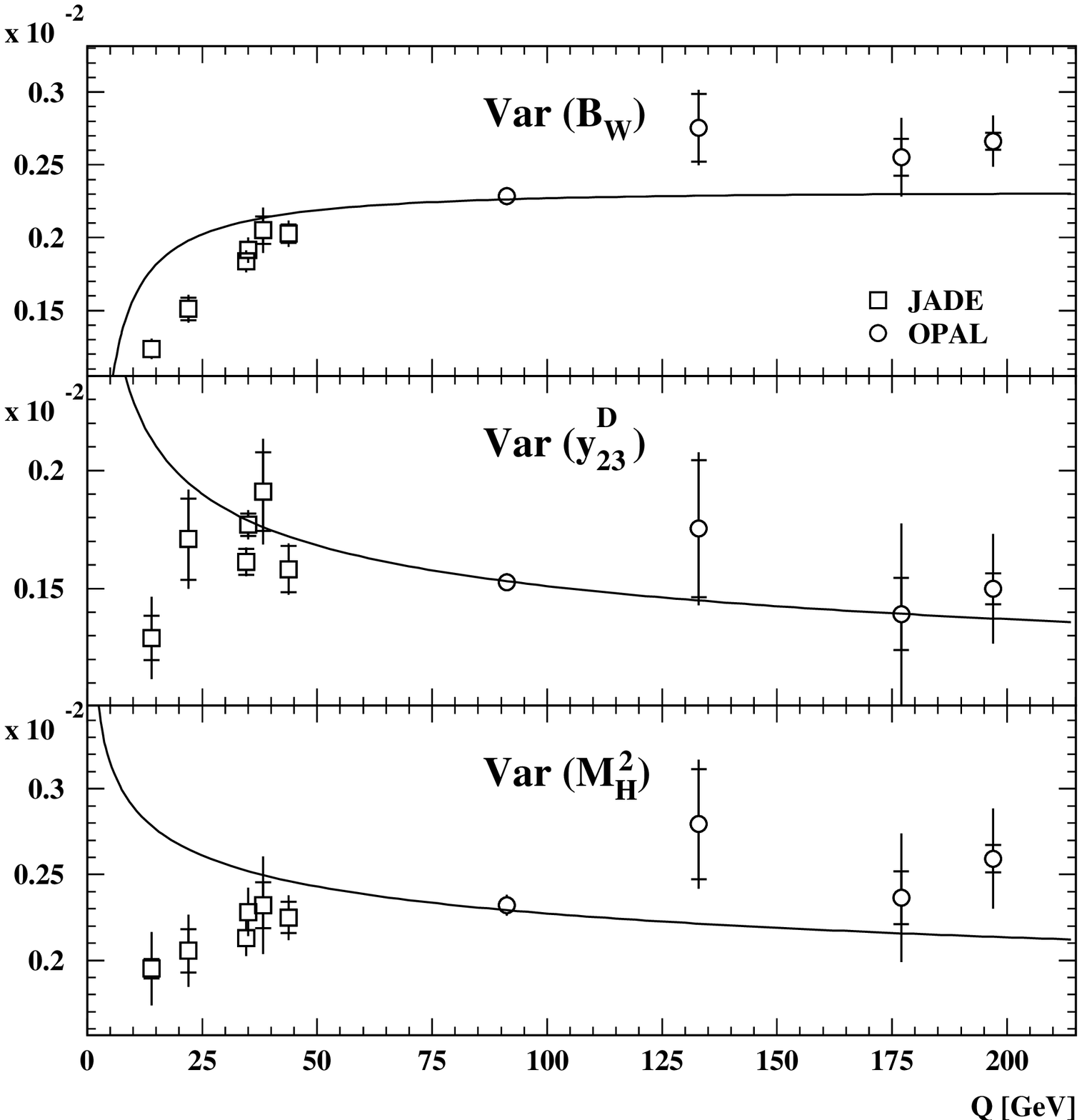}
   \end{tabular} 
    \caption{
\J{{\color{red}update}\bb, ``\ytwothree''\bb}
Comparison of the \nlo\ prediction with \Jade\ and \Opal\ measurements
             of the variance of \thr, \cp, \bt, \bw, \ytwothree\ and $\mh^2$.
            The inner error bars show the statistical uncertainties used in the fit and
            the outer error bars show the combined statistical and experimental systematic errors.
%%             Indicated are the statistical errors used in the fit
%%             (inner bars) and the combined statistical and experimental systematic errors.
             The {\it line} shows the \nlo\ prediction with fitted value of \asmz
%%  \J{Koeefizient 3. Ordnung: \thr: ?; andere $C<$0 !}
}
   \label{VarWeb}
\end{figure*}

The predictions using the shape function differ only for Var(\cp) from the dispersive model.
Fig.~\ref{VarCKor} shows the comparison of this prediction~(\ref{KTvar}) and data.
%, table~\ref{KorVar}contains the results. 
The fit gives $\chisqd=50/8$, $\asmz=0.0963\pm0.0004$ and\linebreak $\lambda_1=(0.824\pm0.025)\,$GeV (statistical errors).
The correlation of \asmz\ and $\lambda_1$ is 0.80\,.
 \begin{figure}[htb!]  %[p]
   \begin{center}
    \includegraphics[height=9.cm]{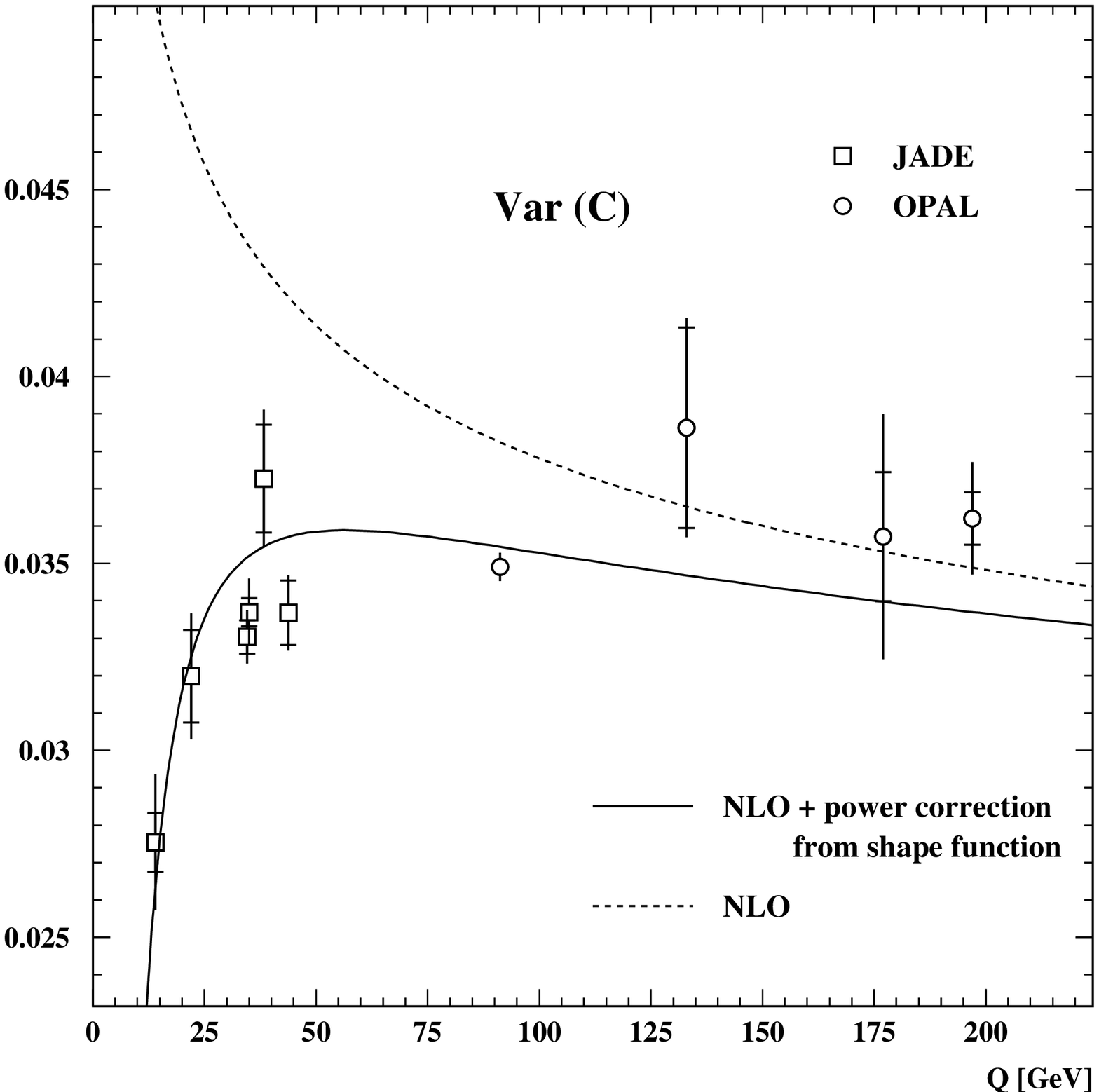}
  \end{center}
     \caption{Comparison of the shape function  
             prediction with \Jade\ and 
             \Opal\ measurements
             of the variance of \cp.
            The inner error bars show the statistical uncertainties used in the fit and
            the outer error bars show the combined statistical and experimental systematic errors.
%%             Indicated are the statistical errors used in the fit
%%             (inner bars) and the combined statistical and experimental systematic errors.
             The {\it solid line} shows the prediction with fitted values of \asmz\ and $\lambda_1$,
             the {\it dashed line} shows the pure \nlo\ contribution}
   \label{VarCKor}\end{figure}
In the shape function description the $Q$\ evolution of the perturbative description deviating
from the data is corrected at low energies by the\J{ coefficient $\lambda_1$ in the negative 
mixt} term\J{ $-3.23\,\as(Q^2)\,$} $\propto\lambda_1$\J{$/{Q}$}. Thus the steep energy 
evolution at low $Q$\ is reproduced successfully.
\J{see Fig.~\ref{VarCKor}. Die erhaltene Kopplung \asmz\ zwingt die Fitkurve ferner in etwa 
durch den \Lepone\ - Punkt mit seiner ausgezeichneten Statistik, was in einem akzeptablen 
Wert von \chisqd\ resultiert.} 
\J{Der hohe Wert von $\lambda_1=0.858\pm0.026$ ist allerdings im 
Rahmen der statistischen Fehler unvertr\"aglich\footnote{\J{Diese Fitergebnisse sind nicht 
inkonsistent, denn die Varianz ist nichtlinear in \momn{\cp}{1}.}} mit den Werten $\lambda_1=0.322\pm0.024$ aus dem
Fit an \momone{\cp}\ oder $\lambda_1=0.306\pm0.023$ aus \momn{\cp}{2}.
Die erhaltene Kopplung $\asmz=0.0951\pm0.0004$ liegt weit unterhalb der Werte aus den Fits 
an die einfachen Momente.:}
However, the fitted values \asmz\ and $\lambda_1$ are not compatible
with the corresponding values from \momn{\cp}{1} or \momn{\cp}{2}.
%%was mit den Daten nicht leicht vereinbar scheint.
%%\footnote{Deutlicher sieht man dies in Abbildung
%%\ref{moms_t} mit ihren gr\"osseren Energiebins; 
%%am signifikantesten f\"ur 
\J{wenn Fits 4 Pkt.e ueberlegen\bb: Im Bereich hoher Schwerpunktsenergien verschwindet 
der Einfluss des gemischten
Terms und die Fitkurve ist---wie die rein perturbative Vorhersage---positiv gekr\"ummt. 
Die Genauigkeit der Messung von $\mathrm{Var}(\cp)$ gen\"ugt nicht zur Diskussion der
Kr\"ummung. %\footnote{
Im Fall von \Var{\bt}, Abbildungen \ref{VarWeb} und ref{vars\_t} erkennt man jedoch: Bei 
hoher Schwerpunktsenergie verschwindet jede Energiepotenzkorrektur und die Kr\"ummung der 
blossen \nlo-Vorhersage widerspricht den Daten. %}
Die naheliegendste Erkl\"arung dieser Diskrepanzen liegt in fehlenden negativen perturbativen Termen
ab der Ordnung $\as^3$ mit nicht zu vernachl\"assigendem Einfluss. Die Varianz scheint
wesentlich von harter Gluonabstrahlung im Multijetbereich beeinflusst, das ist konsistent mit 
unseren Beobachtungen in Abschnitt ref{sec\_moments}: Die Varianz zeigte sich wesentlich 
durch ihren Ausl\"aufer im Multijetbereich des Phasenraum bestimmt, und dieser Bereich
ist durch hohe perturbative Ordnungen dominiert.
{\tiny Zumindest ein Teil dieser Terme liesse sich aus der SDG-N\"aherung gewinnen.}

IN ZUSAMMENFASSUNG!?\bb:
Allgemein haben beliebige Energiepotenzkorrekturen (mit zu erwartenden Skalenparametern
im Bereich einiger GeV) bei \Opal-Energien nur noch wenig Einfluss. Die Genauigkeit
der \Opal-Messpunkte gen\"ugt dann in den meisten F\"allen, um Vorhersagen der 
Varianz
in der Form NLO+Energiepotenzkorrektur
auszuschliessen.}

%%\clearpage 
%%%%%%%%%%%%%%%%%%%%%%%%%%%%%%%%%%%%%%%%%%%%%%%%%%%%%%%%%%%%%%%%%%%%%%%%%%%%%%%%%
%
 \subsection{Single dressed gluon approximation}\label{Gardi}
The hadron level data are compared with the five different predictions which result from
truncating~(\ref{ptSDG}) after order 
${\cal O}(\bar{a}^2)$...${\cal O}(\bar{a}^6)$. The data are described by the predictions well.
Fig.~\ref{Gar5} shows the comparison\footnote{Comparisons
for second to sixth expansion order, and more discussion, can be found in \cite{CHPphd}.}
of data and prediction for %fifth expansion order in $\bar{a}$.
truncating after $\bar{a}^5$. 
The fit results for \as\ and $\nu_i$ are given in Table~\ref{GarTab}. The coefficients 
$\kappa_2$...$\kappa_4$\J{$_5$?\bb$\Rightarrow$ in Diss korrigieren\bb} of the more strongly
suppressed power correction in all instances turn out to be compatible with zero.
\begin{figure}[htb!] 
     \includegraphics[width=0.50\textwidth]{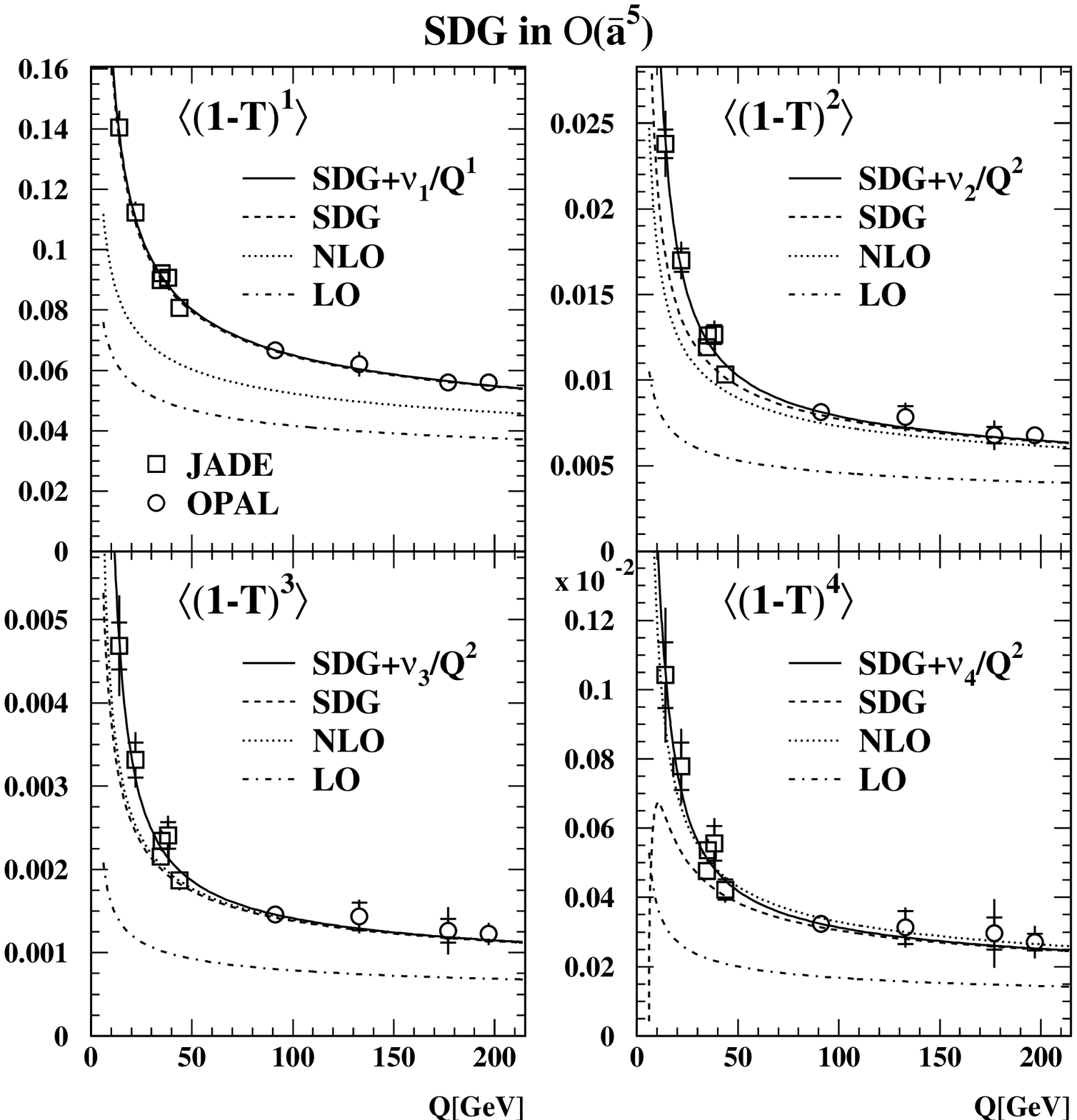} 
   \caption{Fits of the SDG prediction including power correction to \Jade\ and \Opal\ measurements of first to  fourth 
            moment of \thr. The perturbative part of  the prediction has been calculated in
            ${\cal O}(\bar{a}^5)$.
            Superimposed are the perturbative prediction in leading, next-to-leading, and
            maximum order as indicated in the figure.    
            The inner error bars show the statistical uncertainties used in the fit, the outer error bars show
            the combined statistical and experimental systematic errors\J{$y_{max}$ \momn{\thr}{4} 
            geht mit paw nicht besser}}\label{Gar5}
%\vspace{-11.13cm}
%\hspace{6.7cm}{\small $\chi^2/dof.=21.1/16$}\hspace{8.8cm}{\small $\chi^2/dof.=18.7/16$}
%
%\vspace{1.17cm}
%\hspace{6.7cm}{\small $\chi^2/dof.=15.8/16$}\hspace{8.8cm}{\small $\chi^2/dof.=16.0/16$}
%
%\vspace{1.82cm}
%\hspace{6.7cm}{\small $\chi^2/dof.=12.5/16$}\hspace{8.8cm}{\small $\chi^2/dof.=12.5/16$}
%
%\vspace{1.82cm}
%\hspace{6.7cm}{\small $\chi^2/dof.=10.4/16$}\hspace{8.8cm}{\small $\chi^2/dof.=10.4/16$}
%
%\vspace{1.47cm}
%%\J{3., 4. Moment: SDG $<$ 0 disk: pt. Prob noch groesser,  $\bar{a}$(++)(++)\bb}
\end{figure}

For every moment order we study specific properties to find the series describing the data best.
The \chisqd\ values from the various expansion orders do not differ significantly for any moment
order and thus do not discriminate between different expansion orders.  \begin{description}
  \item[\momn{(\thr)}{1}:] For the first thrust moment 
    %% the \chisqd\ value decreases with increasing truncation
    %%order up to $m_{\rm max}=5$, and then increases again. A
    at the order $m_{\rm max}=5$ the power
    correction is compatible with zero, then it becomes negative.\footnote{There is no distinct sign
    of the power term. It can be positive or negative, or change its sign when using
    different regularisations \cite{Gardi:privat}.}
    The values of \asmz\ and the power correction reach their minimum at $m_{\mathrm max}=5$,
    so the convergence\J{ der Reihe} appears best at $m_{\mathrm max}=5$.\J{\footnote{\J{\tiny Das
    ist auch aus der perturbativen Vorhersage ohne Fits an Daten zu erahnen: Der 
 Beitrag $E_1\cdot\bar{a}^5$ liegt mit dem Fitwert von $\as(\mz)$ ab ca. 15\,GeV unter $D_1\cdot\bar{a}^4$\,,
    der n\"achste Term $F_1\cdot\bar{a}^6$ ist erst ab etwa 40\,GeV noch kleiner---wichtige Datenpunkte (mit der st\"arksten Energieabh\"angigkeit) liegen
    aber bereits darunter.}}}
  \item[\momn{(\thr)}{2}:] Including higher perturbative orders the $\as(\mz)$ and $\nu_2$ values 
    from the second thrust moment decrease. %% while \chisqd\ barely changes. 
    The purely perturbative
    description appears continually more accurate, we therefore choose $m_{\rm max}=6$\,.
  \item[\momn{(\thr)}{3}, \momn{(\thr)}{4}:] Fitting the third or fourth thrust moment, to compensate the negative 
    perturbative terms the power correction increases when increasing the truncation order.
    %%the \chisqd\ values remain approximately constant. %Wie im Fall des vierten Moments 
    Thus the \nlo\ description appears more accurate than the SDG approximation, and
    we choose $m_{\mathrm max}=2$.
\end{description}%%\J{
%  F\"ur drittes und viertes Moment von Thrust diskutieren wir also nicht mehr die 
%  eigentliche SDG-N\"aherung: Die Vorhersagen \"ahneln denjenigen aus dem 
%  dispersiven Modell, wobei die Korrekturen aber nur in einer bestimmten Potenz der 
%  Schwerpunktsenergie unterdr\"uckt sind.
  %die Vorhersage f\"ur \momn{(\thr)}{3}\ \"ah\-nelt derjenigen aus
  %dem DMW-Modell, die Energiepotenzkorrektur f\"ur\\ \momn{(\thr)}{4} ist allerdings st\"arker unterdr\"uckt.
%  eher Illustration..}
%\end{document}
%\clearpage %: nochne Leerseite mehr
%\input{GarTab}

Fig.~\ref{plotgarptpc} compares the perturbative terms and the power correction terms.\J{Zum 
Gr\"ossenvergleich zwischen minimalem perturbativem Term und der Energiepotenzkorrektur..}             
\begin{figure}[htb!] 
%%   \hspace{-1.cm}\includegraphics[width=.95\textwidth]{plotgarptpc.eps} 
   \includegraphics[width=.50\textwidth]{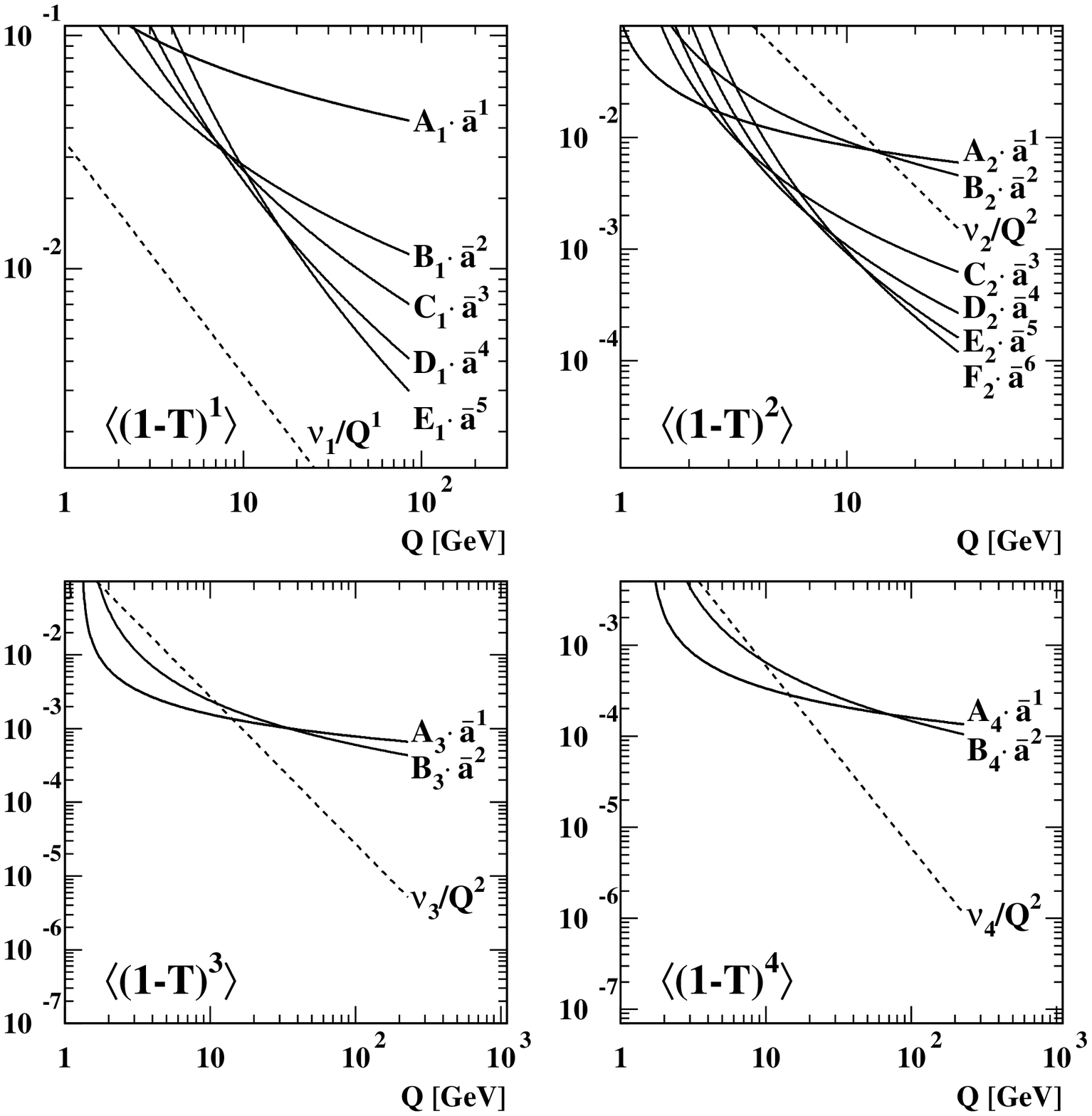} 
   \caption{\J{plotgarNLOSDG=0.eps; {\color{red}update}\bb}$Q$\ evolution of the employed 
  perturbative terms from~(\ref{ptSDG})
     and of  the leading power correction in powers of the cms energy
     $Q$ for the case of first to fourth moment of thrust.
     A$_n$...F$_n$ are the coefficients of the terms in 1st...6th order of the coupling $\bar{a}$ as
     defined in~(\ref{bara}).
     The parameters $\as(\mz)$ 
  \J{---somit $\bar{a}$---}and
            $\nu_i$ are fitted to the data\J{Schrift groesser\bb}}\label{plotgarptpc}
\end{figure}
%\clearpage %%=> ganze Spalte frei
In case of the first moment the power correction is compatible with zero; the purely perturbative
prediction from the SDG approximation already describes the data well.\J{Verschiedene 
Regularisierungen der perturbativen Reihe unterscheiden sich ja in etwa um diese 
Energiepotenzkorrektur, und die Abschneideordnung wurde---wie oben diskutiert, so 
gew\"ahlt, dass m\"oglichst viel Information aus einer Vorhersage fester Ordnung gewonnen 
wird} The power correction is larger for the third and fourth moments, but is still small compared
to the LO and NLO perturbative terms down to $Q=10$\,GeV.
In case of the second moment the higher SDG terms up to ${\cal O}(\bar{a}^6)$ are comparable
with the power correction.\J{STK will: ``NLO'', aber $\neq DMW$\bb)\abwarten}

\J{\subsubsection{\J{Akzeptierte }Werte von \asmsbmz\ und $\lambda_i$}
QUATSCH!?\bb: Im Normalfall ist die Energiepotenzkorrektur positiv und deshalb
mit der starken Kopplung antikorreliert---im Fall von \momn{(\thr)}{1} sogar noch st\"arker als im
dispersiven Modell, f\"ur die h\"oheren Momente analog immer schw\"acher.}
%%\J{alter plot: \begin{figure}[htb!]  
%%   \begin{center}
%%     \includegraphics[width=0.3\textwidth]{fitpars-longtabsGar.eps} 
%%   \end{center}
%%  \caption[\asmz\ und $\lambda_1$ bis $\lambda_4$ aus Fits an die ersten vier Momente von Thrust bei \Petra- und \Lep-Energien.]
%%   {\J{
%%    \asmz \J{=.118/.146/.161/.169}, sowie $-\lambda_1$\,, $\lambda_2$\,, $\lambda_3$ und $\lambda_4$ aus Fits der SDG-Vorhersage mit Energiepotenzkorrekturen\J{ $Q^{-1,-3,-3,-5}$} an die ersten vier Momente von Thrust  bei \Petra- und \Lep-Energien.
%%     Gezeigt sind die statistischen Fehler.}
%%   }
%%\label{fitpars-longtabsGar}
%%\end{figure}}

\subsection{Measuring \boldmath{\asmz} in the SDG approximation}

The \asmz\ values steeply rise with moment order in this model as observed before in
Sect.~\ref{ModellDoksh}, \ref{RechngKor}. 
The problem of describing the higher thrust moments consistently---i.e. with compatible fit parameters---is not solved in this model.
%\J{Der Wert aus der Anpassung an \momn{(\thr)}{2},
%\begin{eqnarray*}
%  \asmz=0.1463 \pm 0.0004\stat 
%\end{eqnarray*}
%ist vergleichbar mit dem entsprechenden Shape function Resultat
%\begin{eqnarray*}
%  \asmz=0.1446\pm0.0014\stat\,.
%\end{eqnarray*}
%S\"amtliche Werte aus h\"oheren Momenten liegen \"uber den 
%entsprechenden Werten aus der Monte Carlo - Korrektur oder dem dispersiven Modell.{\tiny 
%ist oder sollte das in PCvsMC?}
%Aus der Analogie zu den bisher untersuchten Modellen
%k\"onnen wir vermuten, dass die SDG-N\"aherung f\"ur dieses Moment zwar einen Teil der perturbativen 
%Vorhersage h\"oherer Ordnung in \as\ enth\"alt (der Fitwert f\"ur \asmz\ nimmt mit steigender Entwicklungsordnung stets ab), wesentliche Terme aber bereits
%hier fehlen. Dies d\"urfte teilweise eine Folge der Berechnung mittels ``inklusivem'' Thrust
%sein---nach unserer Interpretation der mit der Momentordnung stark ansteigenden Fitergebnisse f\"ur \asmz\ aus Monte Carlo -
%korrigierten Momenten von \thr, \cp\ und \bt\ in Unterabschnitt ref{indi-fits} ist 
%diese Vereinfachung bereits 
%f\"ur \momn{(\thr)}{2} nicht mehr gerechtfertigt. Gardi deutet in \cite{Gar2} 
%die M\"oglichkeit an, die SDG-Rechnung auch f\"ur nicht ``inklusiven'' Thrust durchzuf\"uhren, was
%hier von Interesse w\"are.}
We measure the strong coupling in this model only from \momn{(\thr)}{1}. 

The central value of the measurement and its statistical error are shown in table \ref{GarTab} as case $m_{\max}=5$.
To determine the
experimental systematic uncertainty, the fit is repeated 
employing the minimum overlap assumption for the experimental uncertainties of the data.\J{machen\bb 
(die Zentralwerte weichen 
    ohnehin (selten? nie signifikant?) nur geringf\"ugig voneinander ab). 
  )[ \as-Messung} To estimate the theoretical uncertainty we study the following contributions: 
\begin{itemize}
  \item In order ${\cal O}(\bar{a}^6)$ of the perturbative prediction~(\ref{SDGformel}) 
    the unknown coefficient $\beta_4$ appears. We set it to the simple Pade\J{Appostroph korrekt\bb} 
    estimate\footnote{In this approximation the coefficients $\beta_i$ follow a simple geometric series.}
    $\beta_4=\beta_3^2/\beta_2$ instead of $\beta_4=0$. This does not change any of the four
    cited digits of the \asmz\ fit result.\J{ und somit 
    auch nicht unsere Diskussion, die perturbative Vorhersage der Ordnung ${\cal O}(\as^5)$
    zur Messung zu verwenden.}
%%  \item We estimate the renormalisation scheme dependence of the prediction by varying
%%   $\beta_2 \mapsto 2\cdot\beta_2$ or $\beta_2 \mapsto 0$.
%%   \J{\color{red}hab ich $2 beta_2$ auch in der laufenden Kopplung gemacht? 0 in Vorhersage UND 
%%   Kopplung gibt Absturz!} This changes \asmz\ by\J{\color{red}UPDATE!!} +0.0001 or less;
%%   we do not include this effect further.\J{Die Vorhersage in ${\cal O}(\as^5)$ ist bereits nur noch sehr wenig vom
%%   Renormierungsschema abh\"angig.\footnote{\J{Auf Variation des Renormierungskalenfaktors 
%%   verzichten wir, da die n\"otige Erweiterung von Gleichung (\ref{SDGformel}) sehr aufwendig 
%%   w\"are.}}}  
 \item Instead of three loops we use the four known loops for calculating the $Q$\ evolution 
   of \as. This changes the fit result by\J{\color{red}UPDATE!!} less than 0.0001. 
 \item \J{Zentrale Bedeutung besitzt die G\"ute der N\"aherung des \oaaa-Koeffi\-zi\-en\-ten 
   C$_1$\,.}
   The \J{exakte }${\cal O}(\bar{a}^2)$ coefficient B$_1$ is approximated by SDG within 34\%~\cite{Gar2},
%%\J{vgl. Abbildung \ref{plotgar}} 
   and the approximation is expected to be complete in high order
   \cite{Gar1}.\J{oder wars Gar2?\bb wars DGE statt SDG?\bb (meinen Fit von C$_1$ 
   erw\"ahnen)} Thus we
   vary the coefficient C$_1$ by $+34\%$ or $-34\%$. This results in changes of \asmz\ by 
   $-0.0027$ or $+0.0031$. This variation is consistent with truncating after expansion order
   $m_{\rm max}=3$ or $m_{\rm max}=6$.  
A renormalisation scale variation
   is not studied
   and is expected to cause a small effect in a prediction of fifth order. %%of the same order of magnitude.
 \item As $\nu_1$ is one of the two fit parameters, the hadronisation error is part of the fit error.
\end{itemize}
The result is
\begin{eqnarray*}
  \asmz   &=&  0.1172\pm0.0007\stat\pm0.0016\expt\\
          & &           ^{+0.0031}_{-0.0027}\theo\\ 
          &=&  0.1172\pm0.0036\tot\,.
\end{eqnarray*}

%\J{ALT: \tiny
%Die rein perturbative Vorhersage f\"unfter Ordnung ohne Energiepotenzkorrektur beschreibt die Daten mit \chisqd=19.4/17 und ergibt
%\begin{eqnarray*}
%  \asmz&=&0.1173\pm0.0001{\rm(stat.)}\pm0.0003{\rm(exp.)}\nonumber\\
%       & &\pm0.0034{\rm(Theorie)}\\
%       &=&0.1173\pm0.0034\,. \nonumber
%\end{eqnarray*}
%Wir betrachten dies aber nicht als zentrales Ergebnis, da es stark von experimentellen und
%theoretischen Umst\"anden abh\"angt: 
%Der niedrigste Datenpnkt von 14\,GeV hat bedeutenden Einfluss auf die erhaltene Enegiepotenzkorrektur.
%Ver\"anderungen in der Rekonstruktionsprozedur d\"urften grossen Einfluss auf ihren erhaltenen Wert haben.
%($SDG\pm14\,GeV$\bb)\abwarten)
%Unsere Regularisierung durch Abschneiden nach dem f\"unften Term erfolgte einfach ad hoc.
%Andere Verfahren k\"onnten ebenfalls eine merkliche Energiepotenzkorrektur nach sich 
%ziehen.}This measurement agrees\J{war: remarkably} well with the world average of 
%$\asmz=0.1189\pm0.0010$~\cite{bethke06}.\J{und unseren anderen Messungen?\bb
%Ihre Pr\"azision ist nicht vergleichbar mit den genauesten dazu beitragendenWerten, etwa 
%aus dem Tau-Zerfall, der Z-Lineshape, 4-Jetrate (Aleph: \as=.1170(2)(13))
% oder RGI. Ref.s hab ich inzwischen anderswo..} 
It is our most precise value. %% from one single event shape moment.
%This follows
%from the more advantageous and better motivated estimation of higher 
%perturbative orders as compared with the renormalisation scale variation
%$\xmu=0.5\ldots2.0$.
The variation of the NNLO coefficient C$_1$ by $\pm34\%$ is better motivated than the somewhat arbitrary
variation of the renormalisation scale factor $\xmu=0.5$, 2.0\,. It leads
to a smaller perturbative uncertainty than in the NLO analyses discussed above or in \cite{jadepaper}.

%\J{STK: confusing---As already the third order coefficient is approximated, this
%measurement has to be understood as an \oaa\ analysis too.---und die vorangegangene Diskussion als Illustration zu erwartender Effekte,
%sobald gen\"ugend perturbative Koeffizienten h\"oherer Ordnung bekannt sind.}
In \cite{Gar1} the SDG prediction was fitted to \momone{\thr} data, notably of the \Petra\ experiments 
{\mbox{\rm TASSO}} und {\mbox{\rm MARK~J}} at cms energies down to 12\,GeV. The results were $\asmz=0.110$ und 
$\lambda_1=0.62\,$GeV. In the old \Petra\ measurements $\epem\rightarrow\bbbar$ events have not been subtracted. They enlarge \momone{\thr}
in an effect decreasing with energy faster than $1/Q$. This results in an increase of $\lambda_1$ and 
a decrease of $\asmz$.
\J{\tiny In \cite{Gar1} wurde die Vorhersage an \"altere Daten von 
\momn{(\thr)}{1} angepasst, insbesondere der \Petra-Experimente {\mbox{\rm TASSO}} und 
{\mbox{\rm MARK~J}} bei niedrigen
Schwerpunktsenergien bis hinunter zu 12\,GeV. Dies ergab $\asmz=0.110$ und 
$\lambda_1=0.62\,$GeV. 
Dabei wurde eine andere Regularisierungsvorschrift verwendet.
Ein zus\"atzlicher Grund f\"ur die positive und relativ grosse
Energiepotenzkorrektur d\"urfte die fehlende Subtraktion von b-Ereignissen sein. Diese 
erh\"ohen den Mittelwert von Thrust in einem Effekt, der aber schneller abf\"allt als $1/Q$.
Daraus ergibt sich eine Erh\"ohung des Fitwertes von $\lambda_1$ und eine Erniedrigung von $\asmz$.
Bei Ausschluss der Daten unterhalb von 22\,GeV finden die Autoren dieser Arbeit die Werte
$\asmz=0.111$ und $\lambda_1=0.54\,$GeV. Die numerische Implementierung der SDG-Vorhersage
in \cite{Gar1} unterscheidet sich ebenfalls von der von uns verwendeten aus \cite{Gar2}, mit
Differenzen im Bereich einiger Prozente. Der
sehr niedrige erhaltene Wert der Kopplung wird in \cite{Gar1} hypothetisch \cite{Grunberg:privat} als Effekt der 
Resummierung h\"oherer perturbativer Ordnungen interpretiert.}

\section{Summary}\label{summary}
As an alternative to the description of hadronisation by Monte Carlo models we studied analytical
models of hadronisation.
These descriptions
require 
no laborious tuning of a Monte Carlo model
but
fitting of \as\ and an additional parameter.
In this way, their consistency can be quantified.

We studied the dispersive model, the shape function and the single dressed gluon approximation,
including the NLO predictions.
All models describe the mean values of all studied event shape variables 
  at cms energies $Q=14$...209\,GeV 
well. The dispersive model
and the shape function also describe the higher moments of the one-hemi\-sphere variables well.
The measurements of \asmz\ and \anulmI\ 
in the dispersive model from the mean values of \thr, \cp, \bt, all studied moments of 
\ytwothree\ and \momn{\mh}{2}, \momn{\mh}{4} confirm the universality of these parameters 
within total errors. 
%mit Ausnahme der Ergebnisse aus \momn{\bw}{1}, \momn{\mh}{2} und \momn{\mh}{4}.

Higher moments of the two-hemisphere variables \thr, \bt\ and \cp\ can only be described by higher values of
\asmz, moments of \bw\ and \mh\ only by higher values of the power correction coefficient.
Both deviations apparently compensate for the deficiencies of the perturbative predictions.
%% and violate the universality of the theory parameters.

Averaging the parameters from \momn{(\thr)}{1}, 
\momn{\cp}{1}, \momn{\bt}{1}, \momn{(\ytwothree)}{1}...\momn{(\ytwothree)}{5}, 
 and \momn{\mh}{2}, \momn{\mh}{4} in the dispersive model results in
\begin{eqnarray*}
  \asmz   \J{&=&  0.1174\pm0.0002\stat\pm0.0018\expt\\
          & &  \pm0.0047(\xmu)\pm0.0003(\mu_I)\pm0.0001({\cal M})\\}
          &=&  0.1183\pm0.0056\tot\,,\\
  \anulmI \J{&=&  0.484\pm0.003\stat\pm0.006\expt\\
          & &  \pm0.026(\xmu)\pm0.046({\cal M})\\}
          &=&  0.493\pm0.058\tot\,.                                                  
\end{eqnarray*}
\J{\color{red}vgl.mail\bb:}Event shape moments are barely sensitive to more strongly suppressed power
corrections resulting from the shape function.\J{Interessant wären hier Vorhersagen 
und Messungen von Momenten negativer Ordnung. 
Fits der Shape Function - Vorhersagen an Momente ergeben andere nichtperturbative Skalen 
als jene an klassifizierte Verteilungen.

Interpretationen höherer Momente der Zweihemisphärenvariablen, wie sie in
\cite{L3,KorTaf} gegeben wurden, sind aufgrund unserer bei niedriger Schwerpunktsenergie 
wesentlich genaueren Messung ausgeschlossen.}
In general, a detailed understanding of non-perturbative effects cannot reach any further
than the reliability of the underlying perturbative structure which was studied in~\cite{OPALPR404,jadepaper}.          
\J{Die Varianz der Ereignisformvariablen wird am \Lepone-Punkt bis auf etwa 1\% genau gemessen (totale Fehler).
Die Messung ist an den \Jade-Punkten im allgemeinen genauer als an den \Leptwo-Punkten.}

For some variables the variance shows an unexpected energy evolution.
Of the variables analysed here only the \thr\ distributions are broader at low 
cms energy; the $\mh^2$, \cp, \bt\ and \bw\ distributions
are narrower at low energy.
 The energy evolution of the variances is not
given correctly by the pure \nlo\ prediction. The description by an additional power correction  
from the shape function is not quantitatively correct. 
%%; or more generally by any power correction with scale parameters --as expected-- in the GeV region.
\J{Der Multijetbereich bestimmt stark 
die Breite einer Verteilung -- er ist durch mehrfache Gluonabstrahlung dominiert, 
und die Varianz wird
durch die perturbativen Terme in vorhandener Ordnung und berechnete
Energiepotenzkorrekturen (oder allgemeiner zu erwartende derartige Korrekturen
mit Skalenparametern im Bereich h\"ochstens einiger GeV) nicht hinreichend beschrieben.}

The single dressed gluon approximation %%describes well only the first thrust moment. 
shows the same problems describing the higher thrust moments as the other models.
Fitting the prediction 
for \momn{(\thr)}{1}
including terms up to fifth order gives power corrections compatible with zero 
and 
\begin{eqnarray*}
  \asmz\J{&=&0.1172\pm0.0007\stat \pm0.0016\expt\nonumber\\
       &  &^{+0.0031}_{-0.0027}\nonumber\theo\\}
       &=&0.1172\pm0.0036\tot\,.\nonumber  
\end{eqnarray*}
With a competitive total error of 3\% this is our most precise \as\ measurement. \J{o. schon 
genauso!?\bb:} It agrees\J{war: remarkable well} with the world average of $\asmz=0.1189\pm0.0010$ 
\cite{bethke06}. For this result a Monte Carlo model has only  been used for correcting the
data for \bbbar\ background, taking into account its uncertainty in the
experimental systematic error.
%Ein analytischer Ausdruck zur Energiepotenzkorrektur ging lediglich in die Abschätzung
%der theoretischen Unsicherheit ein. 
%
%Die Information in der perturbativen Reihe lie\3e sich noch vollständiger
%ausnutzen durch Abschneiden am exakt minimalen Term -- vgl. Figur 8 in \cite{Gar1},
%Vorhersage der ``minimal term'' Regularisierung. Dadurch dürfte sich die Konvergenz
%noch verbessern, und in Folge statistischer und experimentell systematischer Fehler 
%noch verkleinern.\J{Das Ergebnis ist so gut wie die zugrundeligende Theorie --}
%Am meisten verbessern lie\3e sich an diesem Ergebnis die Abschätzung der theoretischen
%Unsicherheit. Etwa wird erwartet\J{ \cite{Gar1}!?\bb}, da\3 die SDG-Näherung
%in hohen Ordnungen von \as\ vollständig ist. Aus einer genaueren Quantifizierung dieser
%Aussage könnte man die Qualität der Näherung besser bewerten. 
\J{Der zugrundeliegende Ansatz der Dressed Gluon Expansion wurde bereits auf anderen Gebieten
erfolgreich angewandt, etwa dem Spektrum des Zerfalls von B-Hadronen, siehe
Referenzen \cite{GarB1,GarB2,GarB3,GarB4}.}

Moments are an interesting alternative to distributions:\J{In intro 
schon!?\bb:} Specific parts of phase space are tested selectively and the energy evolution
shows up clearly. This evolution can be studied thoroughly by the combination of
the experiments \Jade\ and \Opal. As in \cite{jadepaper} the most interesting energy range
is provided by the \Jade\ experiment, as the perturbative and non-perturbative effects both scale inversely
with the energy.
%A study of analytical hadronisation models based on moments and variances of event shape distributions shows various
%unexpected phenomena but the strong coupling can be measured quite precisely in one of these models.

{\small
\section*{Acknowledgements}
\par

We would like to thank E. Gardi and G. Korchemsky for explanations of
their work and useful discussions.
This research was supported by the DFG cluster of excellence `Origin
and Structure of the Universe'.
}

\newcommand{\Mvar}{${\cal M}$ variation: }
\newcommand{\Mvard}{${{\cal M}-20\%}$}
\newcommand{\Mvaru}{${{\cal M}+20\%}$}
\newcommand{\hl}{\noalign{\smallskip}\hline\noalign{\smallskip}}
\newcommand{\hhl}{\noalign{\smallskip}\hline\hline\noalign{\smallskip}}
\newcommand{\hhhl}{\noalign{\smallskip}\hline\hline\hline\noalign{\smallskip}}
\newcommand{\GeV}{\rm GeV}

\begin{table*}
\caption{Measurements of \asmz\ and \anulmI\ in the dispersive model from first moments
 of six event shape variables over the full analysed range of c.m.\
 energy, 14...209\,GeV. 
The \chisqd\ values are based on the statistical errors only, see text for
further discussion
%The rather high values of \chisqd\ are discussed in the text.
}\label{DMW}
\begin{center}
\begin{tabular}{ l r@{}lr@{}lr@{}lr@{}lr@{}lr@{}l  }\hline\noalign{\smallskip}%ebensoM+20%
&\multicolumn{2}{c}{ \momn{(\thr)}{1} }&\multicolumn{2}{c}{\momn{\cp}{1}}&\multicolumn{2}{c}{\momn{\bt}{1}}&
\multicolumn{2}{c}{\momn{\bw}{1}}&\multicolumn{2}{c}{ \momn{(\ytwothree)}{1} }\\\hhl
{\bf\boldmath\asmz}& 0.1234 && 0.1228 && 0.1200 && 0.1142 && 0.1181  &\\ \hl
Statistical error  & 0.0006 && 0.0004 && 0.0005 && 0.0006 && 0.0004  &\\ \hl
Experimental syst.          & 0.0015 && 0.0012 && 0.0011 && 0.0015 && 0.0016  &\\ \hl
              \xmu\ variation:  ${\xmu=2.0}$&${ +0.0073}$&&${ +0.0073}$&&${+ 0.0047}$&&${+ 0.0004}$&&${+ 0.0050}$&\\ 
{\color{white}\xmu\ variation:} ${\xmu=0.5}$&${ -0.0062}$&&${ -0.0063}$&&${ -0.0041}$&&${+ 0.0009}$&&${ -0.0038}$&\\ \hl
              \mI\ variation:  ${\mI=1\,\GeV}$ &${ +0.0025}$&&${ +0.0030}$&&${+ 0.0017}$&&${+ 0.0008}$&& ---&\\
{\color{white}\mI\ variation:} ${\mI=3\,\GeV}$ &${ -0.0019}$&&${ -0.0022}$&&${ -0.0013}$&&${ -0.0007}$&& ---&\\ \hl
              \Mvar\ \Mvard &${ +0.0011} $&&${ +0.0014} $&&${+ 0.0008} $&&${+ 0.0004} $&&---&\\
{\color{white}\Mvar} \Mvaru &$ { -0.0011}$&&$ { -0.0013}$&&$ { -0.0007}$&&$ { -0.0004}$&&---&\\ \hl
Theoretical syst. & 0.0078 && 0.0080 && 0.0051 && 0.0012 && 0.0050 &\\ \hhl
{\bf\boldmath \anulmI} & 0.490 && 0.433 && 0.514 && 0.604 && --- &\\ \hl
Statistical error & 0.007 && 0.003 && 0.009 && 0.015 && --- &\\ \hl
Experimental syst. & 0.010 && 0.005 && 0.009 && 0.014 && --- &\\ \hl
              \xmu\ variation:  ${\xmu=2.0}$&${ -0.022}$&&${ -0.024}$&&${+ 0.013}$&&${-0.049}$&&---&\\ 
{\color{white}\xmu\ variation:} ${\xmu=0.5}$&${ +0.031}$&&${ +0.033}$&&${+ 0.011}$&&${+0.147}$&&---&\\ \hl
              \Mvar\ \Mvard &${ + 0.047} $&&${ +0.034} $&&${+ 0.053} $&&${+ 0.086} $&&---&\\ 
{\color{white}\Mvar} \Mvaru &$ { -0.033}$&&$ { -0.025}$&&$ { -0.037}$&&$ { -0.057}$&&---&\\ \hl
Theoretical syst. & 0.056 && 0.048 && 0.055 && 0.170 && --- &\\ \hhl
Correlation \asmz, \anulmI& -0.86 && -0.80 && -0.93 && -0.90 && --- &\\ \hl
\chisqd & 47.8/10&& 66.3/10&& 106/10&& 75.3/10&& 37.7/10&\\ 
\noalign{\smallskip}\hline\end{tabular}\end{center}\end{table*}

\begin{table*}
\caption{Measurements of \asmz\ and \anulmI\ in the dispersive model from second to fifth moments
 of six event shape variables over the full analysed range of c.m.\
 energy, 14...209\,GeV. 
The \chisqd\ values are based on the statistical errors only, see text for
further discussion
%The rather high values of \chisqd\ are discussed in the text.
}\label{DMW2}
\begin{center}
\begin{tabular}{ l r@{}lr@{}lr@{}lr@{}lr@{}lr@{}l  }\hline\noalign{\smallskip}%ebensoM+20%
&\multicolumn{2}{c}{ \momn{(\ytwothree)}{2} }&\multicolumn{2}{c}{\momn{\mh}{2}}&\multicolumn{2}{c}{ \momn{(\ytwothree)}{3} }&\multicolumn{2}{c}{ \momn{(\ytwothree)}{4} }&\multicolumn{2}{c }{\momn{\mh}{4}}  
&\multicolumn{2}{c}{ \momn{(\ytwothree)}{5} }\\\noalign{\smallskip}\hline\hline\noalign{\smallskip}
{\bf\boldmath\asmz}  & 0.1182 && 0.1158 &&  0.1173 && 0.1160 && 0.1154 & & 0.1144 &\\ \hl
Statistical error    & 0.0009 && 0.0007 &&  0.0013 && 0.0017 && 0.0010 & & 0.0020 &\\ \hl
Experimental syst.   & 0.0021 && 0.0024 &&  0.0020 && 0.0025 && 0.0034 & & 0.0033 &\\ \hl
              \xmu\ variation:  ${\xmu=2.0}$&${+ 0.0054}$&&${ +0.0042}$&&${+ 0.0053}$ &&${+ 0.0050}$&&${ +0.0051}$& &${+ 0.0047}$&\\ 
{\color{white}\xmu\ variation:} ${\xmu=0.5}$&${ -0.0042}$&&${ -0.0033}$&&${ -0.0041}$ &&${ -0.0038}$&&${ -0.0042}$& &${ -0.0036}$&\\ \hl
              \mI\ variation:  ${\mI=1\,\GeV}$ &---&&${ + 0.0013}           $&&--- &&---&&${ + 0.0011}           $& & --- &\\ 
{\color{white}\mI\ variation:} ${\mI=3\,\GeV}$ &---&&$            { -0.0010}$&&--- &&---&&$            { -0.0009}$& & --- &\\ \hl
              \Mvar\ \Mvard &---&&${ + 0.0006}           $&&---&&---&&${ + 0.0005}           $& &---&\\ 
{\color{white}\Mvar} \Mvaru &---&&$            { -0.0006}$&&---&&---&&$            { -0.0005}$& &---&\\ \hl
Theoretical syst.    & 0.0054 && 0.0045 && 0.0053 && 0.0050 && 0.0053 & & 0.0047 &\\ \hhl
{\bf\boldmath \anulmI}&  ---      && 0.601  && --- && --- && 0.579 & & --- &\\ \hl
Statistical error     &  ---   && 0.011  && --- && --- && 0.015 & & --- &\\ \hl
Experimental syst.    &  ---   && 0.012  && --- && --- && 0.022 & & --- &\\ \hl
              \xmu\ variation:  ${\xmu=2.0}$&---&&${ -0.022}           $&&---&&---&&${ -0.022}           $& &---&\\ 
{\color{white}\xmu\ variation:} ${\xmu=0.5}$&---&&$          { + 0.037}$&&---&&---&&$          { + 0.034}$& &---&\\ \hl
              \Mvar\ \Mvard &---&&${ + 0.083}          $&&---&&---&&${ + 0.079}          $ &&--- &\\ 
{\color{white}\Mvar} \Mvaru &---&&$           { -0.056}$&&---&&---&&$           { -0.054}$ &&--- &\\ \hl
Theoretical syst.     &  ---   && 0.091 && --- && --- && 0.086 & & --- &\\ \hhl
Correlation \asmz, \anulmI &  ---   && -0.88 && --- && --- && -0.82 & & --- &\\ \hl
\chisqd                              & 18.3/10&& 25.2/10&& 21.0/10&& 18.7/10&& 26.4/10& & 15.2/10&\\
\noalign{\smallskip}\hline\end{tabular}\end{center}\end{table*}

\clearpage
\begin{table}
\caption{Measurements
of \asmz\ and $\lambda_1$\ in the shape function model from the first
%%\J{\chisqd=$\frac{19..22}{16}$ ( : vgl. plot, disk.\bb;\as = :|, :|, :) )} 
and second moments
%%\J{\chisqd=$\frac{17..21}{16}$ ( : vgl. plot, disk.\bb; \as = :((, :((, :)); vgl. pt.\bb , (DMW,Gar))\abwarten} 
of three event shape variables (second and fourth moment in case of \mh) over the full analysed range of c.m.\
energy, 14...209\,GeV
%  The experimental systematics are estimated by the minimum overlap assumption.
}\label{KorTab}
\begin{tabular}{  l   r@{}l r@{}l r@{}l  }
\hline\noalign{\smallskip}
& \multicolumn{2}{c}{$\langle (1-T)^1 \rangle$}
& \multicolumn{2}{c}{$\langle C^1 \rangle$}
& \multicolumn{2}{c }{\momn{\mh}{2}}
 \\ \noalign{\smallskip}\hline\hline\noalign{\smallskip}
{\bf \boldmath \asmz} &$ 0.1304 $&&$ 0.1305 $&&$ 0.1193 $& \\ \noalign{\smallskip}\hline\noalign{\smallskip}
Statistical error     &$ 0.0008 $&&$ 0.0006 $&&$ 0.0008 $& \\ \noalign{\smallskip}\hline\noalign{\smallskip}
Experimental syst.    &$ 0.0018 $&&$ 0.0015 $&&$ 0.0027 $& \\ \noalign{\smallskip}\hline\noalign{\smallskip}
              \xmu\ variation:  ${\xmu=2.0}$&${+ 0.0069} $&&${+ 0.0064} $&&${+ 0.0038} $&\\ 
{\color{white}\xmu\ variation:} ${\xmu=0.5}$&$ { -0.0054}$&&$ {-0.0050}$&&$ { -0.0025}$&\\ 
\noalign{\smallskip}\hline\hline\noalign{\smallskip}
{\bf \boldmath $\lambda_1$[\,GeV]} &$ 0.499 $&&$ 0.393 $&&$ 1.010 $& \\ \noalign{\smallskip}\hline\noalign{\smallskip}
Statistical error &$ 0.034 $&&$ 0.024 $&&$ 0.047 $& \\ \noalign{\smallskip}\hline\noalign{\smallskip}
Experimental syst. &$ 0.065 $&&$ 0.045 $&&$ 0.075 $& \\ \noalign{\smallskip}\hline\noalign{\smallskip}
              \xmu\ variation:  ${\xmu=2.0}$&${+ 0.053} $&&${+ 0.063} $&&${+ 0.044} $& \\ 
{\color{white}\xmu\ variation:} ${\xmu=0.5}$&$ { -0.067}$&&$ { -0.075}$&&$ { -0.045}$& \\ 
\noalign{\smallskip}\hline\hline\noalign{\smallskip}
Correlation \asmz, $\lambda_1$ &$ -0.96 $&&$ -0.96 $&&$ -0.94 $& \\ \noalign{\smallskip}\hline\noalign{\smallskip}
\chisqd &$ 23.3/8 $&&$ 22.6/8 $&&$ 19.6/8 $& \\ \noalign{\smallskip}\hline\hline\hline\noalign{\smallskip}
& \multicolumn{2}{c}{$\langle (1-T)^2 \rangle$}
& \multicolumn{2}{c}{$\langle C^2 \rangle$}
& \multicolumn{2}{c }{\momn{\mh}{4}}
 \\ \noalign{\smallskip}\hline\hline\noalign{\smallskip}
{\bf \boldmath \asmz} &$ 0.1426 $&&$ 0.1456 $&&$ 0.1184 $& \\ \noalign{\smallskip}\hline\noalign{\smallskip}
Statistical error &$ 0.0014 $&&$ 0.0008 $&&$ 0.0012 $& \\ \noalign{\smallskip}\hline\noalign{\smallskip}
Experimental syst. &$ 0.0028 $&&$ 0.0023 $&&$ 0.0041 $& \\ \noalign{\smallskip}\hline\noalign{\smallskip}
              \xmu\ variation:  ${\xmu=2.0}$&${+0.0111} $&&${+ 0.0117} $&&${+ 0.0049} $& \\ 
{\color{white}\xmu\ variation:} ${\xmu=0.5}$&${-0.0091} $&&$ {-0.0097} $&&$ { -0.0037}$& \\ 
\noalign{\smallskip}\hline\hline\noalign{\smallskip}
{\bf \boldmath $\lambda_1$[\,GeV]} &$ 0.516 $&&$ 0.364 $&&$ 0.964 $& \\ \noalign{\smallskip}\hline\noalign{\smallskip}
Statistical error &$ 0.056 $&&$ 0.023 $&&$ 0.079 $& \\ \noalign{\smallskip}\hline\noalign{\smallskip}
Experimental syst. &$ 0.096 $&&$ 0.052 $&&$ 0.133 $& \\ \noalign{\smallskip}\hline\noalign{\smallskip}
              \xmu\ variation:  ${\xmu=2.0}$&${+ 0.044} $&&${+ 0.063} $&&${+ 0.031} $& \\ 
{\color{white}\xmu\ variation:} ${\xmu=0.5}$&$ { -0.052}$&&$ { -0.054}$&&$ { -0.030}$& \\ 
\noalign{\smallskip}\hline\hline\noalign{\smallskip}
Correlation \asmz, $\lambda_1$&$ -0.95 $&&$ -0.91 $&&$ -0.92 $& \\ \noalign{\smallskip}\hline\noalign{\smallskip}
\chisqd &$ 20.9/8 $&&$ 18.3/8 $&&$ 20.2/8 $& \\ \noalign{\smallskip}\hline
\end{tabular}
%[\J{Kor.tex . $\Delta\xmu(\lambda_1)$-Vorzeichen stimmen jetzt wahrscheinlich: analog DMW
% gewählt - dort waren die .txt-Files ok, aber hier lief was schief :( aber neu laufen lassen oder theoretisch überlegen mir zu aufwendig}
%Messungen von \asmz\ und $\lambda_1$\ aus erstem \J{\chisqd=$\frac{19..22}{16}$: vgl. plot, disk.\bb;
%\as = :|, :|, :)} und zweitem Moment\J{\chisqd=$\frac{17..21}{16}$: vgl. plot, disk.\bb;
%\as = :((, :((, :)); vgl. pt.\bb , (DMW,Gar)\abwarten} von drei Ereignisformvariablen "uber
%den gesamten untersuchten Bereich der \Petra-Schwerpunktsener\-gie, sowie "uber den
%gesamten Bereich der \Lep-Schwerpunkts\-ener\-gie.]
\end{table}
 %% Kortab oben bringt Var durceinander :(
\begin{table*}
\caption{Measurements of $\as(\mz)$\ and %%the coefficient 
$\nu_n$ %%of the power correction $1/Q$  
%with statistical errors in the single dressed gluon approximation 
from first %%, $1/Q^2$ from the second...
to fourth moment of \thr\
 over the full analysed range of cms\ 
 energy, 14...209\,GeV. The single dressed gluon approximation is used in maximum order $m_{\rm max}$.
 The errors are statistical. 
 Units of $\nu_n$ are GeV for $n=1$ and (GeV)$^2$ for $n=2$, 3, 4 
}\label{GarTab}
\begin{tabular}{l l l l l l}
%%\begin{tabular}{ l  l  l  r@{}l   r@{}l  r@{}l   r@{}l  r@{}l   r@{}l  r@{}l   r@{}l }
\hline\noalign{\smallskip}
%& $m_i$  
& $m_{\rm max}$
& $\as(\mz)$
& $\nu_n$
%%& \multicolumn{1}{l}{\parbox{1.8cm}{corre\-lation \asmz, $\nu_n$}}
& {\parbox{1.8cm}{corre\-lation \asmz, $\nu_n$}}
& \chisqd
%%& \multicolumn{1}{c}{\parbox{1.cm}{                           $m_{\rm max}$}}
%%& \multicolumn{2}{l}{\parbox{1.4cm}{                                         $\as(\mz)$}}
%%& \multicolumn{2}{l}{\parbox{1.cm}{                                          $\nu_n$}}
%%& \multicolumn{2}{l}{\parbox{1.8cm}{                                         corre\-lation \asmz, $\nu_n$}}
%%& \multicolumn{2}{l}{\parbox{1.3cm}{                                         \chisqd}}
 \\

\noalign{\smallskip}\hline\noalign{\smallskip}
\momn{(\thr)}{1}
  & 2 & 0.1314$\pm$0.0008 & 0.430$\pm$0.036 &-0.96 & 29.1/10\\
  & 3 & 0.1209$\pm$0.0007 & 0.352$\pm$0.039 &-0.97 & 30.0/10\\
  & 4 & 0.1178$\pm$0.0007 & 0.223$\pm$0.045 &-0.97 & 29.3/10\\
  & 5 & 0.1172$\pm$0.0007 & 0.035$\pm$0.058 &-0.98 & 26.4/10\\
  & 6 & 0.1182$\pm$0.0001 &-0.293$\pm$0.010 & 0.88 & 24.0/10\\\noalign{\smallskip}\hline\noalign{\smallskip}
\momn{(\thr)}{2}
  & 2 & 0.1463$\pm$0.0009 & 1.855$\pm$0.455 & 0.73 & 23.8/10\\
  & 3 & 0.1426$\pm$0.0009 & 1.762$\pm$0.454 & 0.74 & 24.0/10\\
  & 4 & 0.1416$\pm$0.0008 & 1.644$\pm$0.458 & 0.73 & 24.2/10\\
  & 5 & 0.1412$\pm$0.0008 & 1.546$\pm$0.460 & 0.73 & 24.3/10\\
  & 6 & 0.1410$\pm$0.0008 & 1.465$\pm$0.470 & 0.73 & 24.4/10\\\noalign{\smallskip}\hline\noalign{\smallskip}
\momn{(\thr)}{3}
  & 2 & 0.1523$\pm$0.0015 & 0.273$\pm$0.136 & 0.73 & 20.7/10\\
  & 3 & 0.1525$\pm$0.0015 & 0.274$\pm$0.137 & 0.73 & 20.7/10\\
  & 4 & 0.1531$\pm$0.0015 & 0.286$\pm$0.137 & 0.73 & 20.7/10\\
  & 5 & 0.1533$\pm$0.0015 & 0.298$\pm$0.136 & 0.73 & 20.6/10\\
  & 6 & 0.1534$\pm$0.0015 & 0.307$\pm$0.135 & 0.74 & 20.6/10\\\noalign{\smallskip}\hline\noalign{\smallskip}
\momn{(\thr)}{4}
  & 2 & 0.1581$\pm$0.0017 & 0.059$\pm$0.024 & 0.55 & 17.6/10\\
  & 3 & 0.1625$\pm$0.0018 & 0.063$\pm$0.024 & 0.55 & 17.5/10\\
  & 4 & 0.1645$\pm$0.0019 & 0.075$\pm$0.023 & 0.56 & 17.4/10\\
  & 5 & 0.1652$\pm$0.0019 & 0.085$\pm$0.022 & 0.58 & 17.3/10\\
  & 6 & 0.1655$\pm$0.0019 & 0.095$\pm$0.021 & 0.60 & 17.3/10\\\noalign{\smallskip}\hline
\end{tabular}
\end{table*}

\clearpage

%%%%%%%%%%%%%%%%%%%%%%%%%%%%%%%%%%%%%%%%%%%%%%%%%%%%%%%%%%%%%%%%%%%%%%%%%%%%%%%%%%%%%%%%%%%%%%
% BibTeX users please use
%% \bibliographystyle{iopart}
 \bibliographystyle{ephja}
 \bibliography{papers}

\begin{thebibliography}{10}

\bibitem{FritzschGellMann}
{H. Fritzsch, M. Gell-Mann, H. Leutwyler}, Phys. Lett. B {\bf 47}, 365 (1973)

\bibitem{GrossWilczek1}
{D. Gross, F. Wilczek}, Phys. Rev. Lett. {\bf 30}, 1343 (1973)

\bibitem{GrossWilczek2}
{D. Gross, F. Wilczek}, Phys. Rev. D {\bf 8}, 3633 (1973)

\bibitem{Politzer}
{H. Politzer}, Phys. Rev. Lett. {\bf 30}, 1346 (1973)

\bibitem{jadepaper}
{C. Pahl, S. Bethke, S. Kluth, J. Schieck}, Eur. Phys. J. C {\bf 60}, 181
  (2009)

\bibitem{OPALPR404}
OPAL Coll., {G. Abbiendi et al}, Eur. Phys. J. C {\bf 40}, 287 (2005)

\bibitem{DMW}
{Y. Dokshitzer, G. Marchesini, B. Webber}, Nucl. Phys. B {\bf 469}, 93 (1996)

\bibitem{KorTaf}
{G. Korchemsky, S. Tafat}, JHEP {\bf 0010}, 010 (2000)

\bibitem{Gar1}
{E. Gardi, G. Grunberg}, JHEP {\bf 9911}, 016 (1999)

\bibitem{Gar2}
{E. Gardi}, JHEP {\bf 0004}, 030 (2000)

\bibitem{ALEPH}
{ALEPH} Coll., {A. Heister et al}, Eur. Phys. J. C {\bf 35}, 457 (2004)

\bibitem{L3}
L3 Coll., {P. Achard et al}, Phys. Rept. {\bf 399}, 71 (2004)

\bibitem{DELPHI}
DELPHI Coll., {P. Abreu et al}, Physics Letters B {\bf 456}, 322 (1999)

\bibitem{JADE-paper}
{JADE} Coll., {P. A. Movilla Fern\'andez, S. Bethke, O. Biebel, S. Kluth et
  al}, Eur. Phys. J. C {\bf 22}, 1 (2001)

\bibitem{resummation}
{M. Dasgupta, G. Salam}, J. Phys. G {\bf 30}, R143 (2004)

\bibitem{NNLOESs}
{A.~Gehrmann-De Ridder, T.~Gehrmann, E.~W.~N.~Glover, G.~Heinrich}, JHEP {\bf
  12}, 094 (2007)

\bibitem{Weinzierl}
{S.~Weinzierl}, Phys. Rev. Lett. {\bf 101}, 162001 (2008)

\bibitem{jadenewas}
JADE Coll., {P. A. Movilla Fern\'andez, O. Biebel, S. Bethke, S. Kluth, P.
  Pfeifenschneider et al}, Eur. Phys. J. C {\bf 1}, 461 (1998)

\bibitem{NNLOmoments}
{A.~Gehrmann-De Ridder, T.~Gehrmann, E.~W.~N.~Glover, G.~Heinrich}, {arXiv
  0903.4658} (2009)

\bibitem{event2}
{S. Catani, M. Seymour}, Phys. Lett. B {\bf 378}, 287 (1996)

\bibitem{ert}
{R. Ellis, D. Ross, A. Terrano}, Nucl. Phys. B {\bf 178}, 421 (1981)

\bibitem{Beneke}
{M. Beneke}, Phys. Rept. {\bf 317}, 1 (1999)

\bibitem{DWthrustmean}
{Y. Dokshitzer, G. Marchesini, B. Webber}, Phys. Lett. B {\bf 352}, 451 (1995)

\bibitem{DokTalk}
{Y. Dokshitzer}, In: High energy physics {ICHEP} 1998, proceedings of the 29th
  international conference, 1998

\bibitem{ICHEP08}
{C. Pahl}, arXiv 0810.3326, {to appear in} High Energy Physics {ICHEP} 2008,
  Proceedings of the 34th International Conference (2008)

\bibitem{ESW}
{R. Ellis, W. Stirling, B. Webber}, Cambridge Monographs on Particle Physics,
  Nuclear Physics and Cosmology {\bf 8} (1996), {C}ambridge University Press

\bibitem{UniversalityRescued}
{Y. Dokshitzer, A. Lucenti, G. Marchesini, G. Salam}, Nucl. Phys. B {\bf 511},
  396 (1998)

\bibitem{Universality}
{Y. Dokshitzer, A. Lucenti, G. Marchesini, G. Salam}, JHEP {\bf 05}, 003 (1998)

\bibitem{MovillaFernandez:OB259}
{P. A. Movilla Fern\'andez}, Nucl. Phys. Proc. Suppl. {\bf 74}, 384 (1999)

\bibitem{revisiting}
{Y. Dokshitzer, G. Marchesini, G. Salam}, Eur. Phys. J. direct C {\bf 1}, 3
  (1999)

\bibitem{OHab}
{O. Biebel}, Phys. Rept. {\bf 340}, 165 (2001)

\bibitem{Webber:1997zj}
{B. Webber}, Nucl.\ Phys.\ Proc.\ Suppl. {\bf 71}, 66 (1999)

\bibitem{DSE-paper}
{H.J. Lu, C.A.R. Sa de Melo}, Phys. Lett. B {\bf 273}, 260 (1991)

\bibitem{Gardi:privat}
{E. Gardi}, {Private communication}

\bibitem{vanRitbergen:1997va}
{T. van Ritbergen, J. Vermaseren, S. Larin}, Phys. Lett. B {\bf 400}, 379
  (1997)

\bibitem{Czakon}
M.~Czakon, Nucl. Phys. B {\bf 710}, 485 (2005)

\bibitem{OPALPR299}
JADE and OPAL Coll., {P. Pfeifenschneider et al}, Eur. Phys. J. C {\bf 17}, 19
  (2000)

\bibitem{pedrophd}
{P. A. Movilla Fern\'andez}, {Ph.D.} thesis, {RWTH Aachen} (2003). {\tt
  http://nbn-resolving.de/urn:nbn:de:hbz:82-\\opus-4836}

\bibitem{STKrev}
{S. Kluth}, Rept. Prog. Phys {\bf 69}, 1771 (2006)

\bibitem{hepdata}
{\tt http://durpdg.dur.ac.uk/HEPDATA/}

\bibitem{CHPphd}
{C.~Pahl}, {Ph.D.} thesis, {TU M\"unchen} (2007). {\tt
  http://nbn-resolving.de/urn:nbn:de:bvb:91-diss-\\20070906-627360-1-2}

\bibitem{bethke06}
{S. Bethke}, Prog. Part. Nucl. Phys. {\bf 58}, 351 (2006)

\end{thebibliography}
%\begin{thebibliography}{}
%Author, \textit{Book title} (Publisher, place year) page numbers
% etc
%\end{thebibliography}

\end{document}